\documentclass[oneside,pdflatex,sn-mathphys-num,referee,lineno]{sn-jnl}

\usepackage{graphicx}
\usepackage{multirow}
\usepackage{amsmath,amssymb,amsfonts}
\usepackage{amsthm}
\usepackage{mathrsfs}
\usepackage[title]{appendix}
\usepackage{xcolor}
\usepackage{textcomp}
\usepackage{manyfoot}
\usepackage{booktabs}
\usepackage{algorithm}
\usepackage{algorithmicx}
\usepackage{algpseudocode}
\usepackage{listings}
\usepackage{mathtools}

\usepackage[T2A]{fontenc}
\usepackage[utf8]{inputenc}
\usepackage[russian,english]{babel}

\usepackage{bm}
\usepackage{tabularx}
\usepackage{comment}
\makeatletter
\def\linenoon{}

\makeatother

\usepackage{lineno}
\theoremstyle{thmstyleone}

\theoremstyle{thmstyletwo}

\theoremstyle{thmstylethree}

\begin{document}

\title[Review]{On-chip optical isolators}

\author[1,2]{\fnm{Abanoub} \sur{Mikhail}}

\author[2]{\fnm{Nikolay} \sur{Solodovchenko}} 

\author[1]{\fnm{Sen} \sur{Xu}} 

\author[2]{\fnm{Fan} \sur{Zhang}} 

\author[1]{\fnm{Maksim} \sur{Tumashov}} 

\author[1]{\fnm{ Mingzhao} \sur{Song}} 

\author*[1,2]{\fnm{Sergei} \sur{Makarov}} \email{makarovsv@hrbeu.edu.cn}

\author*[1,2]{\fnm{Andrey} \sur{Bogdanov}} \email{a.bogdanov@hrbeu.edu.cn}

\affil[1]{\orgdiv{Qingdao Innovation and Development Center}, \orgname{Harbin Engineering University}, 
\city{Qingdao}, \postcode{266404},  
\country{China}}

\affil[2]{\orgdiv{Faculty of Physics}, \orgname{ITMO University}, 
\city{Saint Petersburg}, \postcode{ 197101}, 
\country{Russia}}

\abstract{
Optical isolation suppresses back-reflections that destabilize lasers, produce parasitic interference, and increase crosstalk in photonic systems. Although free-space and fibre-optic isolators are mature, no broadly adopted on-chip technology yet combines strong isolation, low insertion loss, broad bandwidth, compact footprint, and scalable fabrication. The difficulty is twofold: Lorentz reciprocity must be broken, and every relevant backward spatial, spectral, and polarization channel must be suppressed. This Review develops a common framework for assessing proposed solutions. We first distinguish pairwise nonreciprocity from intensity-asymmetric transmission and from true isolation using the constitutive relations and scattering matrix. We then compare three physical routes to an integrated nonreciprocal response: magneto-optical bias, externally driven space--time modulation, and nonlinear optical processes. Topological and non-Hermitian systems are treated as cross-cutting design frameworks that can route, protect, or enhance an underlying nonreciprocal response, but do not by themselves break reciprocity. Finally, we examine fundamental constraints and practical trade-offs among isolation, insertion loss, bandwidth, footprint, modal coverage, drive requirements, and fabrication compatibility. The resulting mechanism-based perspective connects reciprocity-breaking physics to the requirements of scalable photonic integrated circuits.}


\maketitle

\tableofcontents 

\section{Introduction} \label{Sect1}

Photonic integrated circuits (PICs) are evolving from assemblies of individual components into densely interconnected systems containing lasers, amplifiers, modulators, filters, processors, and detectors~\cite{Shekhar2024}. This development underpins optical communication and sensing as well as optical analogue computing and neuromorphic photonics, in which information is encoded and processed through optical amplitudes, phases, interference, and weighted interconnections~\cite{Shen2017,Fu2024}. In such systems, backward scattering is more than a local loss mechanism. Reflections from interfaces, waveguide transitions, splitters, and resonators create coherent feedback paths that can perturb the transfer function of an analogue processor, alter calibrated or trained weights in a photonic neural network, and increase inter-channel crosstalk~\cite{Komljenovic2015,Shen2017,Fu2024}. Integrated semiconductor lasers are particularly sensitive to optical feedback: returned light perturbs the intracavity field and can broaden the linewidth, degrade coherence, or trigger unstable dynamics~\cite{Komljenovic2015,Tkach1986,Liu2017}. Suppressing backward-propagating fields is therefore a key requirement for stable integrated sources and scalable photonic circuits.

Commercially available free-space and fibre-optic Faraday isolators already provide robust protection from optical feedback in communication, sensing, and precision-measurement systems. Closely related optical circulators are multiport nonreciprocal devices that route light sequentially between three ports, and can therefore separate forward and backward propagating signals by directing reflected light away from the source~\cite{Saleh2019,Mizumoto2018,Ghosh2013Circulator,Fujii1991Circulator}. Their magneto-optical (MO) materials, footprint, alignment requirements, and assembly, however, are poorly matched to standard PIC fabrication~\cite{Saleh2019,Mizumoto2018}. Moreover, a discrete isolator or circulator acts only where light leaves and re-enters a chip, whereas many harmful reflections originate inside the circuit and should ideally be suppressed close to an integrated source~\cite{Han2026}. The challenge of on-chip isolation extends well beyond the miniaturization of conventional bulk nonreciprocal components. It requires the simultaneous breaking of reciprocity, compatibility with stringent integration constraints, and effective control of polarization. The latter is particularly challenging in planar photonic waveguides, where strong transverse confinement and structural birefringence generally lift the polarization degeneracy that is present in free space and approximately preserved in conventional optical fibres. An optical isolator must provide a genuinely nonreciprocal response, which requires breaking Lorentz reciprocity~\cite{Jalas2013,Caloz2018,Asadchy2020}. Realizing strong isolation together with low insertion loss, broad bandwidth, compact footprint, and fabrication compatibility remains the central challenge of on-chip optical isolation.

We first clarify the terminology, since the terms \emph{optical isolator}, \emph{optical diode}, and \emph{asymmetric transmission} are not used consistently across the literature. An optical isolator is a nonreciprocal device designed to transmit a selected forward mode with low loss while strongly suppressing the corresponding backward excitation, ideally with transmission coefficients $T_{21}\simeq1$ and $T_{12}\ll1$~\cite{Jalas2013}. Thus, the asymmetry appears at the level of complex amplitude transmission coefficients.  The term optical diode is broader and less rigorously defined. In nonlinear photonics, it often refers to a spatially asymmetric device whose transmission contrast depends on the optical signal intensity, $I$, so that $T_{21}(I)\neq T_{12}(I)$ may be observed under separate forward and backward excitation~\cite{Callewaert2016,Liu2012,tokura2018nonreciprocal,Fernandes2018,Fernandes2022,Shi2015}. Such diode-like behaviour does not necessarily provide true isolation under simultaneous counterpropagating signals \cite{Shi2015}. However, in a broader sense, the term optical diode is also used for any device that preferentially transmits light in one direction, including genuinely nonreciprocal systems that satisfy $T_{21}\neq T_{12}$. In this Review, we reserve the term optical isolator for devices that provide genuine nonreciprocal suppression of backward propagation and use optical diode in the broader sense only when following the terminology of the original literature. \emph{Asymmetric transmission} refers more generally to unequal transmitted intensities for opposite propagation directions. It can occur in reciprocal multiport systems through interference between different spatial or polarization channels, as well as through mode conversion and selective detection. Such directional intensity contrast therefore does not by itself imply broken Lorentz reciprocity. These distinctions are discussed more rigorously in Section~\ref{Sec:Fundamentals}.

\begin{figure*}[t!]
\centering
\includegraphics[width=\textwidth]{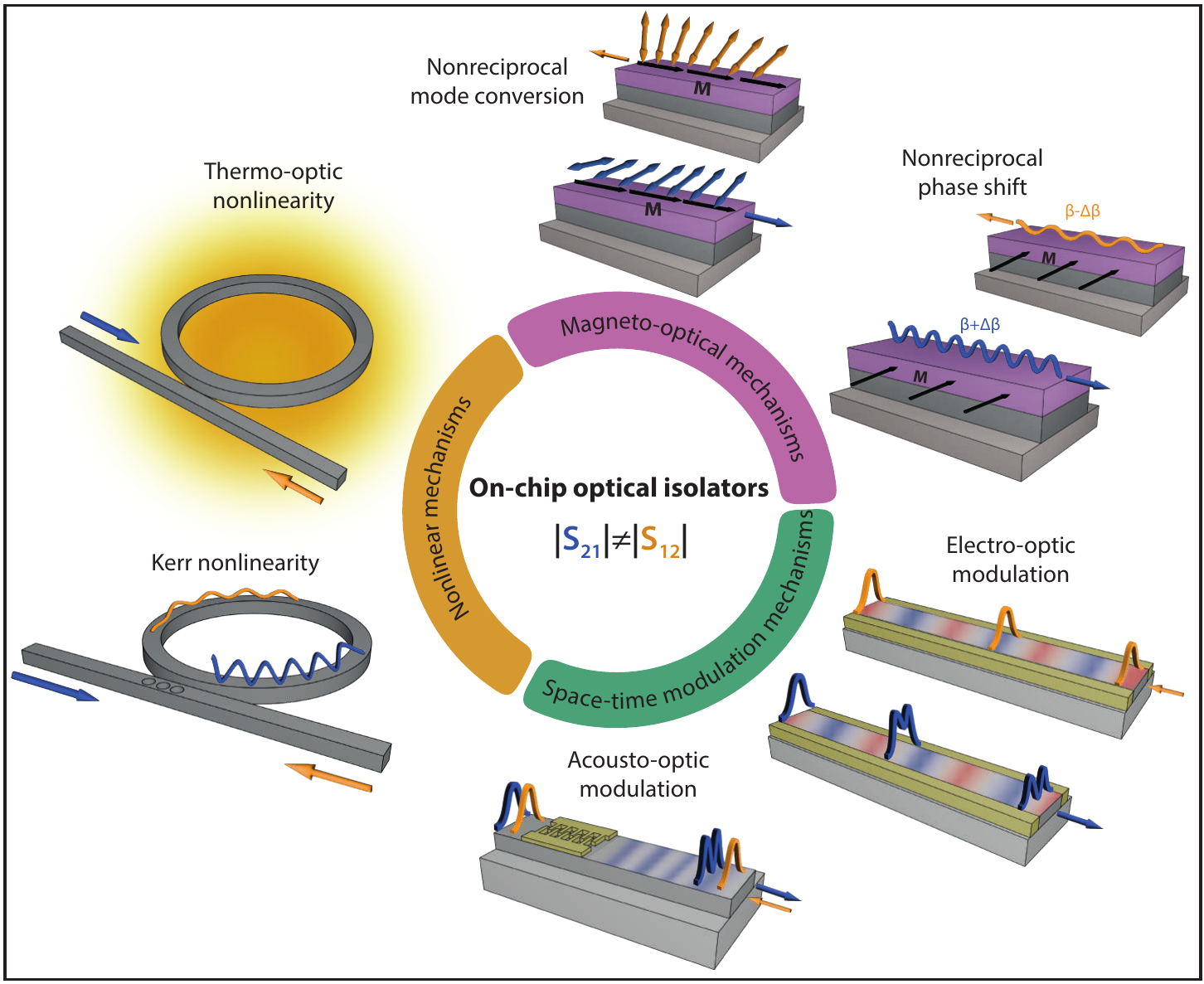}
\caption{Mechanism-based map of approaches to on-chip optical isolation. MO response provides a static time-odd material bias; externally driven space--time modulation provides a phase, momentum, or angular-momentum bias; and nonlinear schemes generate either signal-induced transmission contrast or a pump-defined response. Only the latter can provide genuine small-signal nonreciprocity when the pump establishes an independent directional bias. Topology and non-Hermiticity are cross-cutting design frameworks that can route, protect, or enhance an underlying response but do not by themselves break Lorentz reciprocity. The central two-port contrast is schematic: a complete isolator requires high forward transmission, strong backward suppression, and accounting for all relevant spatial, polarization, and frequency channels.}
\label{Fiq:graphical}
\end{figure*}

The Lorentz reciprocity theorem provides a natural starting point for classifying approaches to optical isolation~\cite{Lorentz1896,Carson1924,Potton2004,Asadchy2020}. Reciprocity can be broken in MO systems through static magnetization. This is the most established class of nonreciprocal optical devices. Particular implementations are based on the nonreciprocal phase shift (NRPS), nonreciprocal mode conversion (NRMC), the transverse magneto-optical Kerr effect (TMOKE), and magnetization-assisted spin-orbit coupling (SOC)~\cite{Ding2025,Saker2021,Asteraki2024,Abadian2021,Levy2017}. A second route is to break reciprocity through external time modulation. Appropriately designed traveling-wave, angular-momentum, or phase modulation can provide a directional bias and enable nonreciprocal mode conversion and synthetic gauge fields~\cite{YuFan2009,Cheng2025,Sounas2013,Ozawa2016,Orsel2023}. A third route combines Kerr-type nonlinearity with broken inversion symmetry. The resulting direction-dependent coupling to the cavity produces different intracavity intensities for opposite excitations and, consequently, different nonlinear resonance shifts and transmission~\cite{White2023,Fan2013Silicon}. Such signal-induced asymmetry can realize optical-diode behaviour, but does not by itself guarantee true isolation. Figure~\ref{Fiq:graphical} summarizes these physical routes and provides the organizing framework for this Review. MO bias, time modulation, and nonlinear optical processes form the three main mechanism classes considered below. Topological and non-Hermitian concepts are placed separately because they do not constitute independent sources of reciprocity breaking. Topology can guide light along robust directional channels, while coupling to external reservoirs enables non-Hermitian dynamics, gain and loss engineering, reservoir engineering, and exceptional-point physics~\cite{Xiao2025,Hurst2022,miri2019exceptional,Meng2024,Metelmann2015,Fang2017,Li2024,Grede2021,Ruan2025,Orsel2025}. Neither non-trivial topology nor non-Hermiticity alone violates Lorentz reciprocity~\cite{Carminati2000,Potton2004}. Instead, they provide additional degrees of freedom for enhancing, routing, or controlling a directional response generated by an underlying nonreciprocal mechanism~\cite{WangLu2025,Jia2025}. Each route presents a different balance between physical performance and technological complexity. MO materials can provide passive and potentially broadband nonreciprocity, but their integration remains limited by deposition or bonding, magnetic-domain control, device geometry, and foundry compatibility~\cite{Mizumoto2018,Srinivasan2022,Pintus2025}. Space-time modulation removes the need for static magnetic materials, but requires a sustained electrical, acoustic, or optical drive and may introduce sidebands, pump leakage, and control overhead~\cite{Han2026,Kim2026,Cui2026,Tan2026}. Nonlinear devices can be compact, but their transmission contrast, insertion loss, bandwidth, and response to simultaneous counterpropagating signals may depend strongly on optical power and dynamical state~\cite{Shi2015,White2023}. No present approach simultaneously optimizes all relevant metrics. In this Review, we compare the different mechanisms in terms of isolation, insertion loss, bandwidth, footprint, polarization and modal coverage, bias or drive requirements, simultaneous-excitation performance, and integration maturity.

The aim of the Review is to provide a unified perspective on the main physical routes to on-chip optical isolation by surveying the existing approaches, identifying the most promising concepts and platforms, and critically assessing their fundamental limitations, practical trade-offs, and prospects for scalable photonic integration. Previous reviews have focused primarily on integrated MO materials and devices~\cite{Srinivasan2018,Srinivasan2022} or on dynamically modulated photonic systems~\cite{Sounas2017Non-reciprocal,Williamson2020Integrated}. Building on the general theory of electromagnetic nonreciprocity~\cite{Caloz2018,Asadchy2020,Sigwarth2022} and the device-level criteria for optical isolation~\cite{Jalas2013}, we compare magnetic, time-modulated, and nonlinear approaches within a common reciprocity and scattering framework. Our focus is on monolithic and heterogeneous on-chip devices, as well as directly coupled or co-packaged components compatible with PIC architectures. Bulk, atomic-vapour, microwave, and free-space demonstrations are discussed only when they establish mechanisms that have not yet been realized optically on chip and are not counted as integrated isolators. Reported performance metrics are retained using the reference planes and operating conditions of the original studies.

Section~\ref{Sec:Fundamentals} first develops the hierarchy from Lorentz reciprocity to scattering nonreciprocity and finally to operational isolation, including the distinction from asymmetric transmission and optical-diode behaviour. The subsequent organization follows the three physical routes identified above: Section~\ref{Sec:magneto-optical} treats magnetic constitutive bias, Section~\ref{Sec:Space-time_modulation} treats time modulation, and Section~\ref{Sec:Nonlinear_Systems} treats Kerr-type and related nonlinear responses. Sections~\ref{sec:topological} and \ref{Sec:Dissipative_nonHermitian_Systems} then assess topological and non-Hermitian design as cross-cutting strategies. Section~\ref{Sec:Fundamental_limits} discusses general constraints, and Section~\ref{Sec:Conclusions} concludes with an outlook.

\section{Fundamentals of optical isolation} \label{Sec:Fundamentals}

Optical isolation fundamentally requires a nonreciprocal electromagnetic response. In linear and time-invariant systems, this requires breaking Lorentz reciprocity. However, nonreciprocity alone is not sufficient for isolation, since a device may exhibit a nonreciprocal phase or mode transformation without suppressing backward transmission. A complete description of optical isolation therefore requires both the fundamental reciprocity relations and their manifestation in the scattering response of the device. In this section, we first discuss electromagnetic reciprocity and then formulate the conditions for optical isolation in terms of the scattering matrix.

\subsection{Electromagnetic reciprocity} \label{Sec:reciprocity}

Consider two current distributions $\mathbf{J}_1$ and $\mathbf{J}_2$ that excite the same electromagnetic system at the same angular frequency $\omega$. The corresponding fields are denoted by $(\mathbf{E}_1,\mathbf{H}_1)$ and $(\mathbf{E}_2,\mathbf{H}_2)$. Throughout this section, we assume the harmonic convention $\exp(-i\omega t)$. Maxwell's equations then give the general Lorentz identity~\cite{Rumsey1954,Chang2026}

\begin{equation}
\begin{split}
\int_V \mathbf{J}_1\!\cdot\!\mathbf{E}_2\,\mathrm{d}V
-\int_V \mathbf{J}_2\!\cdot\!\mathbf{E}_1\,\mathrm{d}V
={}&
\oint_S
(\mathbf{E}_1\!\times\!\mathbf{H}_2
-\mathbf{E}_2\!\times\!\mathbf{H}_1)
\cdot\mathrm{d}\mathbf{S}
\\
&+i\omega
\int_V
\left(
\mathbf{E}_2\!\cdot\!\mathbf{D}_1
-\mathbf{E}_1\!\cdot\!\mathbf{D}_2
+\mathbf{H}_1\!\cdot\!\mathbf{B}_2
-\mathbf{H}_2\!\cdot\!\mathbf{B}_1
\right)
\mathrm{d}V .
\end{split}
\label{Eq:Nonreciprocity_general}
\end{equation}
Equation~(\ref{Eq:Nonreciprocity_general}) follows directly from Maxwell's equations. The two terms on the left compare the electromagnetic response when the roles of the two sources are interchanged. The surface integral describes the corresponding field flux through the boundary $S$, while the volume integral contains the contribution associated with the material response. For a linear, local, and time-invariant medium, the material relations can be written as~\cite{Honda2023}

\begin{equation}
\begin{pmatrix}
\mathbf{D} \\
\mathbf{B}
\end{pmatrix}
=
\begin{pmatrix}
\hat{\bm{\varepsilon}} & \hat{\bm{\xi}} \\
\hat{\bm{\zeta}} & \hat{\bm{\mu}}
\end{pmatrix}
\begin{pmatrix}
\mathbf{E} \\
\mathbf{H}
\end{pmatrix},
\label{eq:constitutive}
\end{equation}
where $\hat{\bm{\varepsilon}}$ and $\hat{\bm{\mu}}$ are the permittivity and permeability tensors, while $\hat{\bm{\xi}}$ and $\hat{\bm{\zeta}}$ describe magnetoelectric coupling between the electric and magnetic fields. Such coupling appears in bianisotropic media, including chiral and Tellegen media~\cite{Safaei2024,caloz2020ElectromagneticPart1}. Substitution of Eq.~(\ref{eq:constitutive}) into Eq.~(\ref{Eq:Nonreciprocity_general}) shows that the volume integral on the right-hand side vanishes for arbitrary fields when the material tensors satisfy~\cite{electromagnetic_nonreciprocity2018,Asadchy2020,Rado1973}

\begin{equation}
\hat{\bm{\varepsilon}}
=
\hat{\bm{\varepsilon}}^{T},
\qquad
\hat{\bm{\mu}}
=
\hat{\bm{\mu}}^{T},
\qquad
\hat{\bm{\xi}}
=
-\hat{\bm{\zeta}}^{T}.
\label{symmetric_constitutive_tensors}
\end{equation}
These symmetry relations define the conditions for Lorentz reciprocity. More generally, the material tensors may depend on an external time-odd ($\mathcal{T}$-odd) bias $\mathbf{F}$, such as velocity, electric current, magnetic field, or magnetization. Microscopic reversibility then leads to the Onsager-Casimir relations~\cite{Caloz2018}

\begin{equation}
\hat{\bm{\varepsilon}}(\mathbf{F})
=
\hat{\bm{\varepsilon}}^{T}(-\mathbf{F}),
\qquad
\hat{\bm{\mu}}(\mathbf{F})
=
\hat{\bm{\mu}}^{T}(-\mathbf{F}),
\qquad
\hat{\bm{\xi}}(\mathbf{F})
=
-\hat{\bm{\zeta}}^{T}(-\mathbf{F}).
\label{symmetric_constitutive_tensors_t}
\end{equation}
These relations connect the electromagnetic response at opposite directions of the external bias. At a fixed nonzero bias, the tensors need not satisfy Eq.~(\ref{symmetric_constitutive_tensors}), which allows Lorentz reciprocity to be broken. Under the reciprocity conditions of Eq.~(\ref{symmetric_constitutive_tensors}), Eq.~(\ref{Eq:Nonreciprocity_general}) reduces to

\begin{equation}
\int_V
\left(
\mathbf{J}_1\!\cdot\!\mathbf{E}_2
-
\mathbf{J}_2\!\cdot\!\mathbf{E}_1
\right)
\mathrm{d}V
=
\oint_S
\left(
\mathbf{E}_1\!\times\!\mathbf{H}_2
-
\mathbf{E}_2\!\times\!\mathbf{H}_1
\right)
\cdot\mathrm{d}\mathbf{S}.
\label{Eq:Lorentz_surface}
\end{equation}

For localized sources in an unbounded medium with outgoing-wave boundary conditions, or for other boundary conditions that make the surface contribution vanish, one obtains the familiar Lorentz reciprocity relation

\begin{equation}
\int_V
\mathbf{J}_{1}\cdot\mathbf{E}_{2}\,\mathrm{d}V
=
\int_V
\mathbf{J}_{2}\cdot\mathbf{E}_{1}\,\mathrm{d}V .
\label{Eq:Lorentz_reciprocity_theorem}
\end{equation}
An equivalent form of Lorentz reciprocity can be expressed in terms of the electric Green tensor

\begin{equation}
\mathbf{E}(\mathbf{r})
=
\int_V
\hat{\mathbf{G}}(\mathbf{r},\mathbf{r}')
\mathbf{J}(\mathbf{r}')
\,\mathrm{d}V'.
\end{equation}
Equation~(\ref{Eq:Lorentz_reciprocity_theorem}) then implies

\begin{equation}
\hat{\mathbf{G}}(\mathbf{r},\mathbf{r}')
=
\hat{\mathbf{G}}^{T}(\mathbf{r}',\mathbf{r}),
\label{Eq:Green_reciprocity}
\end{equation}
or equivalently $G_{ij}(\mathbf{r},\mathbf{r}')=G_{ji}(\mathbf{r}',\mathbf{r})$. Violation of this transpose symmetry therefore provides an equivalent Green-function criterion for nonreciprocity. It is important to emphasize that reciprocal media may be anisotropic and lossy. Complex material parameters, absorption, gain, and non-Hermiticity do not by themselves break Lorentz reciprocity~\cite{Caloz2018,Asadchy2020}.

Time modulation and optical nonlinearity fall outside the assumptions of the  Lorentz identity [Eq.~\eqref{Eq:Nonreciprocity_general}] and the standard reciprocity theorem, which applies to linear and time-invariant systems. However, leaving these assumptions does not automatically make a system nonreciprocal~\cite{Fernandes2018,Fernandes2022,Blanchard2018,Tai1992,Xie2009,Xie2009Clarification,Naguleswaran1998,Trzeciecki2000,Sotoodehfar2023}. Nonreciprocity must instead be identified from the device response itself, namely from unequal scattering amplitudes between reciprocal input and output channels. This can be described with the scattering matrix formalism.

\subsection{Scattering matrix, optical isolation, and asymmetric transmission}

\begin{figure*}[t]
\centering
\includegraphics[width=\linewidth]{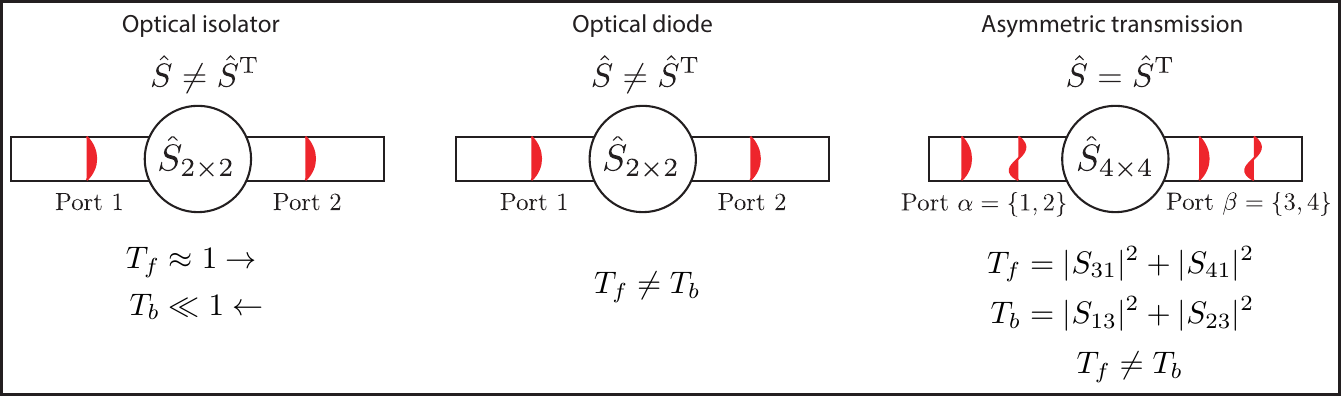}
\caption{Schematic comparison of an optical isolator, an optical diode, and asymmetric transmission. An optical isolator is characterized by a nonreciprocal scattering matrix and strong suppression of transmission in one direction while maintaining high transmission in the opposite direction. An optical diode exhibits direction-dependent transmission, although the degree and origin of the asymmetry may depend on the operating conditions. Asymmetric transmission can occur even for a reciprocal system with $\hat{S}=\hat{S}^{T}$ when the detected intensity contains contributions from several scattering channels. In the illustrated four-port example, reciprocity imposes pairwise equality of the scattering amplitudes, while the sums of the transmitted intensities for opposite excitation directions can remain different.}\label{Fig:isolator_diode_AT}
\end{figure*}

For on-chip photonic components, reciprocity is most conveniently formulated in terms of the scattering matrix. Let $\mathbf{a}$ and $\mathbf{b}$ denote the complex amplitudes of the incoming and outgoing power-normalized ports. They are related by

\begin{equation}
\mathbf{b}=\hat{\mathbf{S}}\mathbf{a}.
\label{Eq:S_matrix_general}
\end{equation}
For a reciprocal linear system with reciprocal port definitions, the complete scattering matrix is symmetric~\cite{Carminati2000,Nieto1986}
\begin{equation}
\hat{\mathbf{S}}=\hat{\mathbf{S}}^{T}.
\label{Eq:S_matrix_reciprocity}
\end{equation}
For a two-port system,

\begin{equation}
\begin{pmatrix}
b_1\\
b_2
\end{pmatrix}
=
\begin{pmatrix}
S_{11} & S_{12}\\
S_{21} & S_{22}
\end{pmatrix}
\begin{pmatrix}
a_1\\
a_2
\end{pmatrix},
\label{Eq:S_matrix_two_port}
\end{equation}
where $S_{21}$ and $S_{12}$ are the transmission amplitudes in the two opposite directions. Reciprocity requires
\begin{equation}
S_{21}=S_{12}.
\end{equation}
Breaking this equality is therefore a direct signature of nonreciprocity. However, nonreciprocity alone does not imply asymmetric transmission. For example, a gyrator with $S_{21}=-S_{12}$ is nonreciprocal in phase while transmitting equal power in both directions~\cite{Asadchy2020}. We therefore distinguish nonreciprocity ($\hat{\mathbf{S}}\neq\hat{\mathbf{S}}^{\rm T}$) from the asymmetry in transmission ($T_f\neq T_b$).

In this Review, we use the term \emph{optical diode} in this general sense to denote a nonreciprocal optical component that exhibits asymmetric transmission, as illustrated in Fig.~\ref{Fig:isolator_diode_AT}. Thus, an optical diode requires both $\hat{\mathbf{S}}\neq\hat{\mathbf{S}}^{\rm T}$ and unequal transmission in the two directions. In nonlinear photonics, the same term is often used more specifically for spatially asymmetric structures in which optical nonlinearity produces a direction-dependent transmission~\cite{Callewaert2016,Liu2012,Fan2013Silicon,tokura2018nonreciprocal}. In such systems, the internal optical state depends on the propagation direction and incident intensity, $I$, resulting in asymmetric transmission $T_f(I)\neq T_b(I)$.  Signal-induced nonlinear diodes can therefore exhibit a large transmission contrast while remaining subject to dynamic reciprocity, which may prevent suppression of an independently injected backward signal during simultaneous forward operation~\cite{Shi2015,Fernandes2018,Fernandes2022}.

An \emph{optical isolator} is a more restrictive realization of a nonreciprocal optical diode. In addition to asymmetric transmission, it must provide high transmission in the desired direction and strong suppression or redirection of light propagating toward the protected source, as shown schematically in Fig.~\ref{Fig:isolator_diode_AT}. Ideally,

\begin{equation}
T_f\approx 1,
\qquad
T_b\ll 1.
\end{equation}
The corresponding isolation ratio and forward insertion loss are 

\begin{equation}
\mathcal{I}
=
10\log_{10}\frac{T_{21}}{T_{12}},
\qquad
\mathrm{IL}
=
-10\log_{10}T_{21}.
\end{equation}
The nonreciprocal bias may originate from magnetization, external time modulation, or an independently maintained optical pump. Pumped nonlinear processes can therefore realize true isolation when the resulting small-signal scattering matrix is nonreciprocal~\cite{Shi2015}.

\emph{Asymmetric transmission} alone does not necessarily imply nonreciprocity as shown in Fig.~\ref{Fig:isolator_diode_AT} (right panel). This situation can arise when the detector measures the total intensity contributed by several scattering channels. In the four-port example shown in Fig.~\ref{Fig:isolator_diode_AT}, ports $1$ and $2$ correspond to two channels on one side of the structure, while ports $3$ and $4$ correspond to two channels on the opposite side. For excitation through ports $1$ and $3$ from opposite directions, the measured transmissions are

\begin{equation}
T_f
=\frac{|b_3|^2+|b_4|^2}{|a_1|^2} =
|S_{31}|^2+|S_{41}|^2,
\qquad
T_b
=\frac{|b_1|^2+|b_2|^2}{|a_3|^2} =
|S_{13}|^2+|S_{23}|^2.
\end{equation}
A reciprocal system satisfies $\hat{\mathbf{S}}=\hat{\mathbf{S}}^{\rm T}$ for every pair of reciprocal ports. Nevertheless, $T_f\neq T_b$ is still possible because the two measured intensities contain contributions from different scattering channels. The resulting directional contrast originates from redistribution of optical power among these channels rather than from broken Lorentz reciprocity. Moreover, the asymmetry in transmission depends on the modal composition of the incident signal. 

The distinction adopted throughout this Review is summarized in Fig.~\ref{Fig:isolator_diode_AT}. Asymmetric transmission refers to unequal measured intensities and may occur in a reciprocal multiport system. An optical diode is understood more generally as a nonreciprocal component with asymmetric transmission. An optical isolator additionally requires efficient transmission in the desired direction together with robust suppression or redirection of the backward signal.

\section{Magneto-optical mechanisms} \label{Sec:magneto-optical}
\begin{figure*}[t!]
\centering
\includegraphics[width = 170mm]{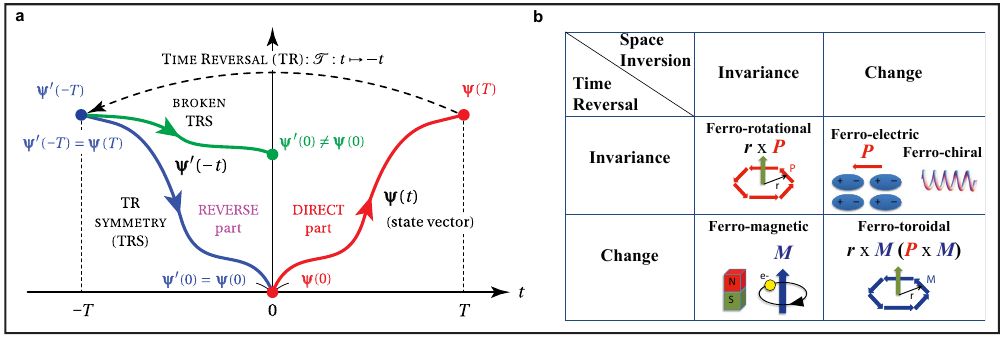}
\caption{Symmetry criteria for nonreciprocity based on time-reversal symmetry and $\mathcal{T}$-odd ferroic order parameters.
\textbf{a} Time-reversal test for nonreciprocity~\cite{electromagnetic_nonreciprocity2018}. The direct process (red) describes the evolution of the state vector $\boldsymbol{\psi}(t)$ from the initial state at $t=0$ to the final state at $t=T$. The time-reversal operation $\mathcal{T}\!:t\mapsto -t$ maps $\boldsymbol{\psi}(T)$ onto $\boldsymbol{\psi}'(-T)$, after which the reverse process evolves back to $t=0$. The system is time-reversal symmetric when the initial state is recovered, $\boldsymbol{\psi}'(0)=\boldsymbol{\psi}(0)$ (blue). If $\boldsymbol{\psi}'(0)\neq\boldsymbol{\psi}(0)$ (green), time-reversal symmetry is broken and a nonreciprocal response becomes possible.
\textbf{b} Ferroic orders classified according to their transformation under space inversion and time reversal~\cite{Cheong2018broken}. Ferro-rotational order is invariant under both transformations. Ferro-electric ($\boldsymbol{P}$) and ferro-chiral orders change sign under space inversion only, ferro-magnetic order ($\boldsymbol{M}$) changes sign under time reversal only, and ferro-toroidal order ($\boldsymbol{r}\times\boldsymbol{M}\propto\boldsymbol{P}\times\boldsymbol{M}$) changes sign under both. In linear time-invariant media, a $\mathcal{T}$-odd bias or order parameter can provide the symmetry breaking required for nonreciprocity.
\textbf{a} Reproduced with permission from Ref.~\cite{electromagnetic_nonreciprocity2018}. Copyright 2018 by the American Physical Society. \textbf{b} Reproduced from Ref.~\cite{Cheong2018broken} under the terms of the Creative Commons Attribution 4.0 International License.}
\label{Fig:Magneto_optics_reciprocity}
\end{figure*}

MO media provide the most established route to breaking Lorentz reciprocity in linear and time-invariant optical systems. An external magnetic field or spontaneous magnetization introduces a $\mathcal{T}$-odd bias and gives rise to antisymmetric components of the material tensors. This enables nonreciprocal effects such as Faraday rotation, the MO Kerr effect, and their guided-wave counterparts. For a ferromagnetic dielectric, the leading gyrotropic contribution to the relative permittivity tensor can be written as~\cite{Popkov1998,Bahlmann2005}

\begin{equation}
\hat{\bm{\epsilon}}
=
\frac{\hat{\bm{\varepsilon}}}{\varepsilon_0}
=
n_0^2 \hat{\mathbf{I}}
+
K
\begin{pmatrix}
0 & M_z & -M_y \\
-M_z & 0 & M_x \\
M_y & -M_x & 0
\end{pmatrix},
\end{equation}
where $\varepsilon_0$ is the vacuum permittivity, $n_0$ is the refractive index of the unperturbed medium, $\hat{\mathbf{I}}$ is the identity tensor, $\mathbf{M}=(M_x,M_y,M_z)$ is the magnetization vector, and $K=K'+iK''$ is the complex MO coupling coefficient. The antisymmetric contribution changes sign upon reversal of $\mathbf{M}$ and therefore directly reflects the $\mathcal{T}$-odd character of the magnetic bias (Fig. \ref{Fig:Magneto_optics_reciprocity}a).

A related route to nonreciprocity is provided by ferrotoroidal order, whose toroidal moment breaks both time-reversal (TR) symmetry  and spatial-inversion symmetry and can produce directional anisotropy through the optical magnetoelectric effect~\cite{Cheong2018broken,szaller2013symmetry,arima2008magneto}. Within the material relations of Eq.~(\ref{eq:constitutive}), the magnetoelectric tensors $\hat{\bm{\xi}}$ and $\hat{\bm{\zeta}}$ may contain both $\mathcal{T}$-even and $\mathcal{T}$-odd contributions. The time-even chiral coupling satisfies the reciprocity condition
$\hat{\bm{\xi}}=-\hat{\bm{\zeta}}^{T}$ of Eq.~(\ref{symmetric_constitutive_tensors}) and therefore remains reciprocal. In contrast, the $\mathcal{T}$-odd Tellegen-type coupling can violate this condition and requires a $\mathcal{T}$-odd bias or order parameter such as magnetization or a toroidal moment~\cite{Asadchy2020,Safaei2024,Asadchy2018}. Directional anisotropy therefore originates from the $\mathcal{T}$-odd magnetoelectric response rather than from chirality alone. When chirality and magnetic order coexist, their interplay can additionally produce magneto-chiral anisotropy~\cite{tomita2018metamaterials}. Such effects can lead to nonreciprocal propagation over a broad range of polarization states and have been observed in materials with finite ferrotoroidal order, as illustrated in Fig.~\ref{Fig:Magneto_optics_reciprocity}b~\cite{Cheong2018broken,arima2008magneto}.

The $\mathcal{T}$-odd bias does not enter the permittivity and permeability tensors on an equivalent footing, and the tensor that carries the gyrotropic response determines how the nonreciprocal problem is formulated. This choice reflects the effective-material-parameter representation adopted for the medium \cite{Amert2012,Lee2016}. Maxwell's equations constrain the total induced current, whereas its decomposition into polarization and magnetization currents is not unique. At low frequencies, magnetization is defined as the magnetic dipole moment per unit volume. At optical frequencies, however, this separation becomes less fundamental, as both magnetization and polarization currents can contribute to the induced magnetic moment \cite{Stancil2009}. Consequently, a medium described by both $\hat{\bm{\varepsilon}}$ and $\hat{\bm{\mu}}$ can, under an appropriate effective-medium description, be represented equivalently by a spatially dispersive electric response \cite{Stancil2009}. In the microwave regime, magnetization precesses near ferromagnetic resonance, giving rise to the Polder tensor, which makes $\hat{\bm{\mu}}$ strongly dispersive and gyrotropic through microscopic spin precession \cite{Polder1949,Lax1963,Gurevich2020}. By contrast, in magnetized plasmas and doped semiconductors, gyrotropy appears in $\hat{\bm{\varepsilon}}$ through the cyclotron motion of free carriers \cite{Zvezdin1997}. At optical frequencies, the relative permeability of conventional natural materials is generally close to unity \cite{Landau1984,Merlin2009}, and the response is therefore conventionally described in terms of $\hat{\bm{\varepsilon}}$ alone. The antisymmetric part of $\hat{\bm{\varepsilon}}$ arises from spin--orbit coupling acting on exchange-split states, transferring the spin polarization to the electric-dipole response \cite{Argyres1955,Zvezdin1997}. Electric and magnetic gyrotropy at optical frequencies are therefore effective-response quantities that can encompass multiple microscopic contributions, rather than unique signatures of distinct mechanisms within the effective-material-parameter description \cite{Zhiyu2023,Mnasri2029}.

However, there are electromagnetic problems which require the use of more complicated constitutive relations due to the breakdown of the effective material description \cite{SIMOVSKI2010,Menzel2010,Andrei2016}. Furthermore, in some nonreciprocal problems, symmetry may require the use of both antisymmetric permittivity and permeability tensors in order to get all possible solutions related to TE modes \cite{Dmitriev2013,Gurevich2020}. In addition, the permeability tensor can be engineered in resonant metamaterials \cite{cui2025metamaterials,Yang2022}. The constitutive relations $\mathbf{D}=\hat{\bm{\varepsilon}} \mathbf{E}$ and $\mathbf{B}=\hat{\bm{\mu}}\mathbf{H}$, together with the Voigt parameters $Q_e$ and $Q_m$ quantify electric- and magnetic-dipole contributions to nonreciprocity \cite{Dmitriev2013,Safaei2024}. While conventional integrated isolators are dominated by $Q_e$, artificial magneto-plasmonic metamaterials exhibit measurable $Q_m\sim10^{-3}$, enabling permeability-driven nonreciprocity inaccessible to $\hat{\bm{\varepsilon}}$-only platforms \cite{Shiu2020,Zhiyu2023,Li2024Magneto-optical,Zhang2024Broadband,Lei2022,Jiang2024}. In practice, $\hat{\bm{\varepsilon}}$ alone suffices when $|Q_e|\gg|Q_m|$ (MO garnets), $\hat{\bm{\mu}} $ alone when $|Q_m|\gg|Q_e|$ (ferrite-based devices), whereas hybrid MO/metamaterial systems with $|Q_e|\sim|Q_m|$ require simultaneous inclusion of both tensors to correctly capture mode coupling and isolation physics in photonic isolators and circulators \cite{Yang2022,Dmitriev2013,Gabriel1966}. Table \ref{tab:MOisolators} summarizes representative state-of-the-art metrics for MO isolators. 

\begin{table*}[t]
\caption{Representative magneto-optical isolators}\label{tab:MOisolators}%
\centering
\footnotesize
\setlength{\tabcolsep}{4pt}
\begin{tabular*}{170mm}{@{\extracolsep{\fill}}lccccccc@{}}
\toprule
Mechanism & Platform & \shortstack{Isolation (dB)} & \shortstack{Bandwidth (nm)}
& \shortstack{Insertion loss (dB)} & Polarization & \shortstack{Footprint ($\mu$m)} & ref. \\
\midrule
NRPS (MZI) & Ce:YIG/YIG/SiN & 28 & 29  & 2.7
& TM & $1300 \times 300$ & \cite{Yan2024} \\
%NRPS (MRR) \textsuperscript{\dag} & CuCrP$_2$S$_6$/Si & 28 & 0.4 & 0.15--1.8  & TE & ${\sim}90 \times 90$ & \cite{Dushaq2025} \\
NRPS (MMI) & Ce:YIG/SOI & 20.1 & 3.5 & 2.6
& TE & $1050 \times 1.15$ & \cite{Wang2025Low-loss} \\
Faraday rotation & BIG, hybrid & $>25$ & $>90$ & $<1.5$
& Linear & $50 \times 50 \times 540$ & \cite{Lapointe2025} \\
SOC (WGM) & YIG/Si$_3$N$_4$ & 16.9 & ${\sim}0.005$ & 1.4
& NR & $1000$ & \cite{Qiao2025} \\
TMOKE\textsuperscript{*} & Ce:YIG/SOI & 58 & 35  & ${\sim}1.8$
& TE & $3 \times 500$ & \cite{Chao2025} \\
\botrule
\end{tabular*}
\par\vspace{4pt}
\begin{minipage}{170mm}
\setlength{\parindent}{0pt}\raggedright\scriptsize
\textsuperscript{*}Simulation/theory only (not experimentally realized design). 
\end{minipage}
\end{table*}

\subsection{Nonreciprocal phase shift}  
In MO waveguides, NRPS arises from the gyrotropic perturbation of the propagation constant induced by  static magnetization.  
This mechanism has been exploited in several integrated isolation architectures, including Mach–Zehnder interferometers (MZIs) \cite{Chen2024,Liang2023,Ding2025}, multimode interferometers (MMIs) \cite{Wang2025Low-loss}, and microring resonators (MRRs) \cite{Dushaq2025,Liu2022TE-Mode,Bi2011}. In MO media, the forward propagation constant $\beta_f$ and backward propagation constant $\beta_b$ of an optical mode are not identical. The magnetization induces a direction-dependent perturbation $\delta \beta$ of the propagation constant $\beta$. As a result, the propagation constants in the forward and backward direction become \cite{Popkov1998}
\begin{equation}
\beta_f = \beta + \delta \beta, \qquad
\beta_b = \beta - \delta \beta .
\end{equation}
The sign reversal is the source of the direction-dependent accumulated phase \cite{Zhou2012,Yokoi2008}. 

\begin{figure*}[t!]
\centering
\includegraphics[width = 170 mm]{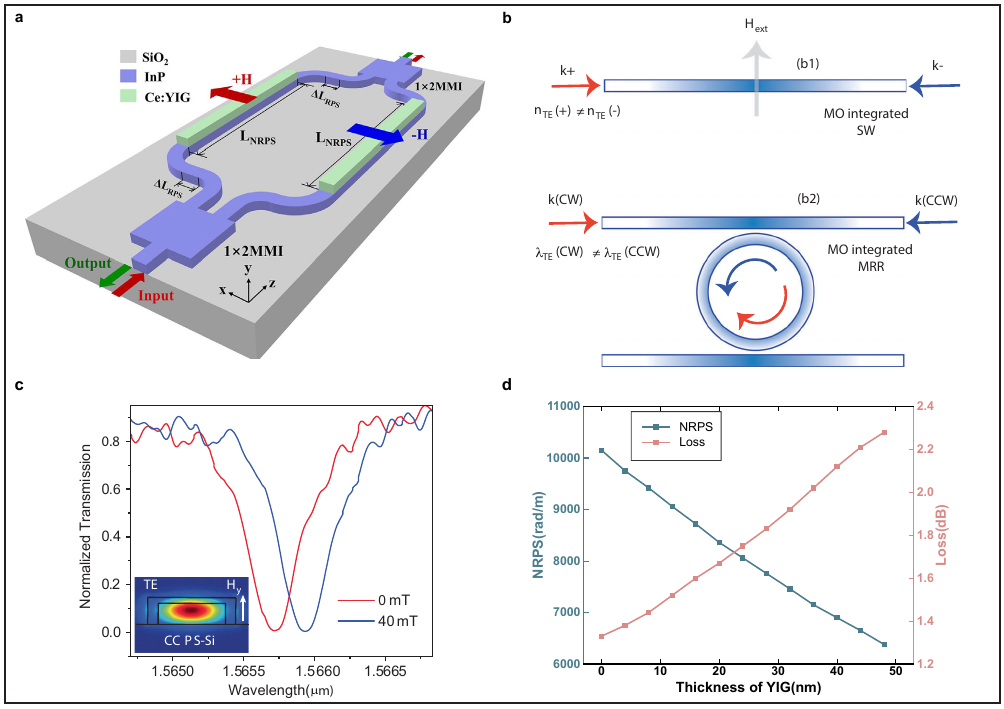}
  \caption{Working principle of isolators based on nonreciprocal phase shift.
  \textbf{a} Sketch of a Mach--Zehnder isolator formed by Ce:YIG bonded to
  InP-on-insulator waveguides. The arms are magnetised antiparallel and transverse
  to propagation ($\pm H$): the nonreciprocal sections ($L_\mathrm{NRPS}$) reverse
  the sign of their $\pi/2$ contribution with propagation direction, while the
  reciprocal length offset ($\Delta L_\mathrm{RPS}$) adds a fixed $\pi/2$, so the
  arms interfere constructively in the forward direction and destructively in the
  backward direction.
  \textbf{b} The same phase shift in two geometries: a magneto-optically loaded
  straight waveguide (b1), in which the forward and backward effective indices
  differ, and a microring resonator (b2), in which the clockwise and
  counterclockwise resonances are no longer degenerate.
  \textbf{c} Measured normalised transmission of the TE mode of a
  CuCrP$_2$S$_6$/Si microring at 0 and 40~mT, showing the field-induced resonance
  detuning. Inset: electric-field profile with the applied field.
  \textbf{d} Simulated nonreciprocal phase shift and propagation loss versus YIG
  seed-layer thickness, illustrating the trade-off between the two.
  \textbf{a} Reproduced from ref.~\cite{Chen2024} and \textbf{b}, \textbf{c} from
  ref.~\cite{Dushaq2025}, under the terms of the Creative Commons Attribution 4.0
  International License. \textbf{d} Reproduced with permission from
  ref.~\cite{Liang2023}. Copyright 2023 Optica Publishing Group}
\label{Fig:Magneto_optics_NRPS_1}
\end{figure*}

Because gyrotropy is small, perturbation theory is used to evaluate the NRPS \cite{Popkov1998,Yamamoto1974,Bahlmann2005,Yokoi2008,Zhou2012,Yan2021}.  Throughout this section the mode propagates along $z$, $y$ is the normal
to the film plane, and $x$ is the in-plane transverse direction
(Fig.~\ref{Fig:Magneto_optics_NRPS_1}a) \cite{Chen2024}. Two magnetization configurations produce NRPS.
For the TM mode, the magnetization lies in the film plane, transverse to
the propagation direction ($\mathbf{M}\parallel\hat{x}$, the Voigt
configuration of Fig.~\ref{Fig:Magneto_optics_NRPS_1}a) \cite{Chen2024}. For the TE mode, the
magnetization points along the film normal
($\mathbf{M}\parallel\hat{\mathbf{y}}$) \cite{Bahlmann2005,Popkov1998,Zhou2012}: 
\begin{equation}
\Delta \beta^{\mathrm{TE}}
=
\frac{2 \omega \varepsilon_0}{\beta^{\mathrm{TE}} N}
\iint \left( K'' M_y \right) E_x \,\partial_x E_x \, dx\, dy ,
\end{equation}
where $N$ denotes the power flux along the light-propagation direction (z-axis) and the fields evolve as $e^{-i\omega t}$. When the magnetization is parallel to the $x$ axis ($\mathbf{M}\parallel\hat{\mathbf{x}}$), the NRPS of the TM mode is \cite{Bahlmann2005,Yamamoto1974,Zhou2012}
\begin{equation}
\Delta \beta^{\mathrm{TM}}
=
-\,\frac{2 \beta^{\mathrm{TM}}}{\omega \,\varepsilon_0 N}
\iint \frac{K'' M_x}{n_0^{4}}\, H_x \,\partial_y H_x \, dx\, dy ,
\end{equation}
According to this formulation, the TE-mode NRPS is governed by the lateral variation of $E_x$ along $x$, whereas the TM-mode NRPS is governed by the vertical variation of $H_x$ along $y$.

Figure~\ref{Fig:Magneto_optics_NRPS_1}a illustrates this mechanism for a MZI isolator based on TM-mode \cite{Chen2024}. Isolation is realized by controlling the total phase difference $\Delta \phi$ between the two interferometric waveguide arms through the combined use of a reciprocal phase shifter (RPS) and NRPS. For forward propagation along the positive $z$-direction, the total phase difference $\Delta \phi = \phi_{\mathrm{RPS}} + \phi_{\mathrm{NRPS}} = 2m\pi$, where $m$ is an integer, leading to constructive interference and efficient coupling of light into the output waveguide. In the backward direction, the total phase difference is $\Delta \phi = (2m+1)\pi$, which causes destructive interference. Consequently, coupling into the input waveguide mode is suppressed and most of the optical power is radiated away. 

When incorporated into a ring resonator, the NRPS lifts the degeneracy between clockwise (CW) and counter-clockwise (CCW) circulating modes (Fig. \ref{Fig:Magneto_optics_NRPS_1}b) \cite{Dushaq2025}. This effect manifests as a direction-dependent modification of the effective refractive index resulting in distinct propagation constants for forward- and backward-propagating modes and spectral splitting between CW and CCW resonances (Fig. \ref{Fig:Magneto_optics_NRPS_1}c) \cite{Dushaq2025,Bi2011}. As illustrated in Fig. \ref{Fig:Magneto_optics_NRPS_1}d, the NRPS decreases while the propagation loss increases as the thickness of YIG seed layer increases \cite{Liang2023}. Interferometric broadband MO isolators based on NRPS have been demonstrated on several integrated photonics platforms, including silicon-nitride, InP-on-insulator, and InGaAs-on-insulator \cite{Yan2020,Chen2024,Ding2025}.  
Polarization-independent interferometric isolators have also been realized \cite{Zhang2019,Liu2023,Ma2021}, while TM-mode implementations remain common in multiple material platforms \cite{Yan2020,Yan2024}.

\subsection{Nonreciprocal polarization mode conversion}
NRMC is the waveguide analogue of Faraday rotation~\cite{Stadler2018,Liu2021}. In the Faraday geometry, the magnetization is parallel to the direction of light propagation. For propagation along the $z$ axis, the relative permittivity tensor tensor can be written as 

\begin{equation} \begin{split} \hat{\boldsymbol{\epsilon}} & = \, n_0^2 \hat{\mathbf{I}} + K \begin{pmatrix} 0 & M_z & 0 \\ - M_z & 0 & 0 \\ 0 & 0 & 0 \end{pmatrix} . \\ \end{split} \label{Dielectric_tensor_NRMC} \end{equation}
The corresponding phase mismatch between the TE and TM modes is
\begin{equation}
\Delta\beta
=
\beta^{\mathrm{TE}}-\beta^{\mathrm{TM}}
=
\frac{2\pi}{\lambda}\Delta n,
\end{equation}
where $\Delta n=n_{\mathrm{TE}}-n_{\mathrm{TM}}$ is the difference between their effective refractive indices~\cite{Matyavs1988,Tien1974}. The off-diagonal elements of the dielectric tensor couple the TE and TM modes, such that a TE input can acquire a TM component during propagation, enabling magneto-optical mode conversion~\cite{Tien1974}. The maximum efficiency of TE-TM mode conversion is \cite{Tien1974,Choueikani2009,Zayets2004,Saker2021}
\begin{equation}
R_{m} = \frac{{\theta_{F}^{2} }}{{\theta_{F}^{2} + \left( {{\Delta }\beta /2} \right)^{2} }}, 
\end{equation}
where $\theta_{F}$ is Faraday rotation per unit length given by \cite{Popkov1998}
\begin{equation}
    \theta_{F} = - k_0\frac{K'' M_s}{2n_0},
\end{equation} 
$M_s$ is the saturation magnetization and $k_0$ is the vacuum wavenumber. If a TE mode is launched at the input of the waveguide, the MO interaction causes part of its energy to be progressively converted into the TM mode during propagation. With further propagation the optical power is transferred back from the TM
mode to the TE mode, resulting in a periodic exchange of energy between the two modes. The coupling length of the waveguide is \cite{Saker2021}
\begin{equation}
    L_{C} = \frac{\pi }{{\sqrt {4\theta_{F}^{2} + {\Delta }\beta^{2} } }} .
\end{equation}

\begin{figure*}[t!]
\centering
\includegraphics[width=170mm, height=65mm]{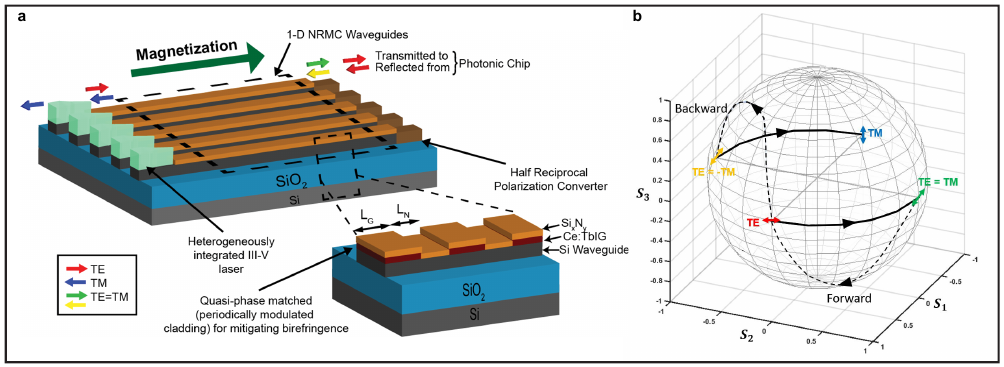}
\caption{Working principle of isolators based on nonreciprocal mode conversion.
  \textbf{a} Monolithic Ce:TbIG/SOI isolator placed directly after a heterogeneously
  integrated III--V laser. Longitudinal magnetisation drives nonreciprocal mode
  conversion (NRMC), the waveguide analogue of Faraday rotation, in a segmented
  Ce:TbIG/Si$_x$N$_y$ cladding that quasi-phase-matches the TE and TM modes against
  waveguide birefringence; a half reciprocal polarisation converter (H-RPC) completes
  the device. Inset: waveguide cross-section and cladding period.
  \textbf{b} Poincar\'e-sphere evolution of the polarisation state. In the forward
  direction NRMC rotates TE into the equal superposition TE\,=\,TM (solid) and the
  H-RPC rotates it back to TE (dashed); in the backward direction the H-RPC gives
  TE\,=\,$-$TM (dashed) and NRMC adds a further $45^\circ$ (solid), so returning light
  arrives at the laser as TM, to which the laser is weakly coupled.
  \textbf{a}, \textbf{b} Reprinted with permission from ref.~\cite{Srinivasan2019}.
  Copyright 2019 American Chemical Society}
\label{Fig:Magneto_optical_NRMC}
\end{figure*}

Figure~\ref{Fig:Magneto_optical_NRMC}a illustrates the concept of NRMC isolator \cite{Srinivasan2019}. A static magnetic bias applied longitudinally induces NRMC in the waveguide. The mode-conversion sequence for forward and backward propagation is shown in Fig.~\ref{Fig:Magneto_optical_NRMC}b. In the forward direction an input TE mode is nonreciprocally converted to an equal TE=TM superposition by NRMC (solid line). A half-reciprocal polarization converter (H–RPC, dashed line) at the waveguide end then completes the conversion back to the TE basis required by the remainder of the circuit, so that the transmitted light exits in the original polarization state. In the backward direction the order of conversions is reversed: the H–RPC first maps the backward-propagating TE into a superposed TE $=-$ TM state, which is then converted by NRMC into a polarization state that is not accepted by the source \cite{Srinivasan2019}. A polarization-selective source or analyser is therefore part of the isolation function; without it, the structure is a nonreciprocal mode converter rather than a complete two-port isolator. As planar waveguides are birefringent, quasi-phase matching (QPM) or tailored dispersion engineering is generally required to maintain efficient NRMC across the device length \cite{Zhang2017}. 

A significant recent development is eliminating the need for external bias magnets through the realization of one-dimensional NRMC devices that exploit intrinsic remanent magnetization in high-gyrotropy Ce:TbIG films \cite{Srinivasan2019} and in bismuth-doped iron garnet slabs (BIG) \cite{Lapointe2025}. Enhancement of Faraday rotation can also be achieved by hybridizing plasmonic structures with MO materials \cite{Chin2013}.  
Nonreciprocal effects in MO dielectric waveguides with metal–insulator–metal and insulator–metal–insulator configurations have also been investigated, where the degree of nonreciprocity can be tuned by varying the thickness of the intermediate layer \cite{Lafmejani2021,Khatir2012,Khatir2013}. Despite this, many studies of integrated isolators based on MO plasmonic waveguides remain at the design or theoretical stage, with relatively limited experimental demonstrations \cite{Nabiei2024,Mansouri2025}. Optomechanically induced Faraday effects have been proposed as a promising route to realize all-optical isolation capable of operating over the entire Poincaré sphere \cite{Lafmejani2021}. Nevertheless, an integrated on-chip optical isolator based on this mechanism has not yet been experimentally demonstrated.

\subsection{Spin-orbit coupling} 

SOC denotes a class of couplings in which light's (or particle's) spin angular momentum becomes intrinsically linked to its spatial degrees of freedom, including orbital angular momentum, propagation direction, or mechanical rotation \cite{Aharony2025,Bliokh2015,Petersen2014,Luo2021,Levy2017,Maayani2018}. In nanophotonic structures, transverse confinement generates longitudinal field components and transverse SOC that is locked to the propagation direction, enabling chiral and unidirectional light–matter coupling \cite{Qian2025,Suarez2025,Wang2019}. In MO media, SOC is modified by the magnetization, producing asymmetric spin-to-orbital angular momentum conversion for counter-propagating modes and thereby enhancing MO effects and optical isolation \cite{Levy2017,Varela2024,Hu2021Coin}. Moreover, SOC enables polarization-controlled lifting of degeneracies between counter-propagating modes in confined geometries 
\cite{Ma2016,Chai2020}.

\begin{figure*}[t!]
\centering
\includegraphics[width=170mm]{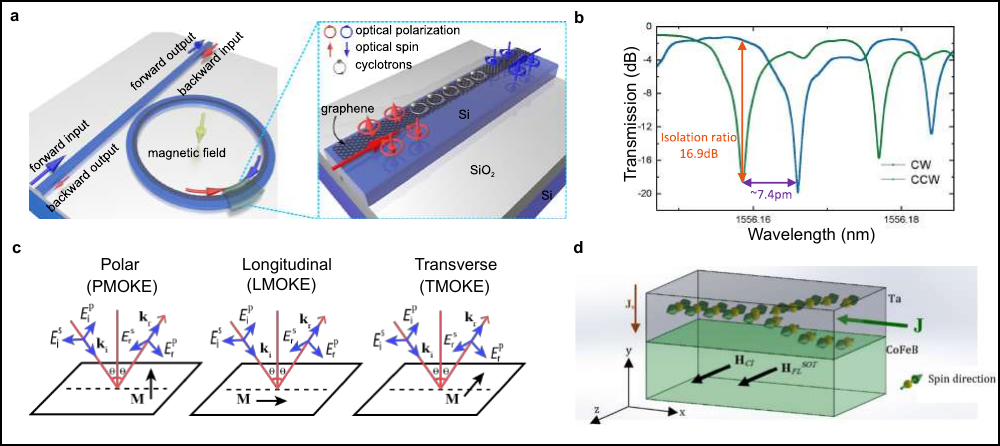}
\caption{ Spin-orbit coupling and magneto-optical Kerr effect. \textbf{a} Hybrid graphene/silicon microring isolator. A graphene nanoribbon covers the inner top surface of the ring; under a vertical magnetic field the induced cyclotrons experience opposite photonic spins for counterpropagating light, giving different forward and backward responses. \textbf{b} Measured transmission of the Si$_3$N$_4$ bus waveguide coupled to a 1~mm YIG microsphere in a static bias field. Photonic spin--orbit coupling and the Faraday effect split the CW and CCW whispering-gallery resonances.  \textbf{c} The three magneto-optical Kerr geometries, set by the orientation of the
  magnetisation $\mathbf{M}$: polar (PMOKE, $\mathbf{M}$ normal to the surface),
  longitudinal (LMOKE, $\mathbf{M}$ in-plane and in the plane of incidence) and
  transverse (TMOKE, $\mathbf{M}$ in-plane and perpendicular to the plane of
  incidence). PMOKE and LMOKE rotate the polarisation; TMOKE instead changes the
  reflected intensity and phase of p-polarised light only.
  \textbf{d} Spin--orbit fields generated by a lateral charge current density
  $\mathbf{J}$ in a Ta/CoFeB bilayer. Spin--orbit coupling in the heavy-metal layer
  converts $\mathbf{J}$ into a transverse spin current $\mathbf{J}_{\mathrm{s}}$, so
  spins of opposite sign accumulate at the two interfaces. The black arrows give the
  directions of the field-like spin--orbit field $\mathbf{H}^{\mathrm{SOT}}_{\mathrm{FL}}$
  and the current-induced Oersted field $\mathbf{H}_{\mathrm{CI}}$, which are collinear
  and add, $\mathbf{H}^{\mathrm{SOT}}_{\mathrm{FL}}$ being the larger. \textbf{a} Reproduced from ref.~\cite{Ma2016}, with the permission of AIP
  Publishing. \textbf{b} Reproduced with permission from ref.~\cite{Qiao2025}.
  Copyright 2025 Chinese Optics Letters. \textbf{d} Reproduced with permission from ref.~\cite{Asteraki2024} under the terms of the Creative
Commons Attribution-NonCommercial-NoDerivatives 4.0 International License. }
\label{Fig:Magneto_optical_SOC}
\end{figure*}

SOC itself preserves TR symmetry \cite{Aharony2025}. Nonreciprocity therefore requires an additional $\mathcal{T}$-odd perturbation. This symmetry breaking can be introduced via magnetization, which couples asymmetrically to spin-polarized optical fields inducing asymmetric frequency splitting (Fig.~\ref{Fig:Magneto_optical_SOC}a and b) \cite{Qiao2025,Chai2020,Ma2016}. Alternatively, mechanical rotation of a nanoparticle or resonator induces direction-dependent coupling to the transverse spin of guided modes, generating distinct chiral resonances for opposite propagation directions and unidirectional absorption \cite{Shi2021,Qian2025,Yang2025Nonreciprocal}. Thus, nonreciprocity in SOC-based isolators arises from the interplay between spin-selective light–matter interactions and a $\mathcal{T}$-odd mechanism \cite{Sayrin2015,Revah2019,Hallett2022}. 

\subsection{Transverse magneto-optical Kerr effect}
The MOKE refers to the influence of the magnetization of the medium on the reflected light. Three distinct MO effects in reflection are identified depending on the alignment of the magnetization vector with respect to the reflecting surface and the plane of incidence: the polar Kerr effect (PMOKE), the longitudinal Kerr effect (LMOKE), and the transverse Kerr effect (TMOKE)  as shown in Fig. \ref{Fig:Magneto_optical_SOC}c \cite{Sato2022}. In the TMOKE, the magnetization is perpendicular to the plane of incidence and the projection of the light wave vector in the magnetization direction is zero, resulting in a change in the intensity and phase of the reflected light \cite{Zvezdin1997}. In the PMOKE, the magnetization is perpendicular to the reflective surface and parallel to the plane of incidence with nonzero component of the wavevector along the magnetization direction. In the LMOKE, the magnetization is parallel to the reflective surface with nonzero component of the wavevector along the magnetization direction. In both PMOKE and LMOKE, the influence of magnetization is a rotation of the plane of polarization and the
appearance of ellipticity of the reflected linearly polarized light \cite{Zvezdin1997,Floess2018}. 

The  relative magnitude of  Faraday effect and MOKE depend on the material, geometry, resonance conditions and interaction length. Faraday rotation can accumulate with propagation length, whereas the Kerr response is determined by the magnetization-dependent reflection and can be strongly enhanced by resonant optical structures. Magneto-plasmonics exploit such resonances to enhance MOKE \cite{borovkova2016transverse,kreilkamp2013waveguide}. SOC provides an important microscopic contribution to MO responses. One prominent mechanism uses spin-orbital fields generated inside plasmonic heterostructures by an applied lateral current density to enhance TMOKE. As shown in Fig. \ref{Fig:Magneto_optical_SOC}d, owing to the strong spin-orbit interaction, the spin-Hall effect is induced and the generated internal magnetic fields change the magnetization and hence enhance the TMOKE \cite{Asteraki2024}. Although the TMOKE is mainly a reflection effect occurring at the MO material interface without modifying the light polarization \cite{Abadian2021}, an enhanced TMOKE in transmission geometry was reported in ref. \cite{Belotelov2011}. Through the excitation of surface plasmon polariton, the $\delta_{T}$ parameter of the TMOKE increases from approximately $ 10^{-5}$ in case of bare BIG up to approximately $10^{-3}$ when the film is covered with a continuous gold layer \cite{Belotelov2011}. 

TMOKE has been theoretically identified to be a more compatible solution to planar integrated isolators in both magneto-plasmonic and magneto-dielectric waveguides \cite{Abadian2021,Chao2025}. The basic working principle is the loss of modal symmetry when plasmonic (dielectric) waveguides are integrated with magnetized MO material. 
An integrated TE optical isolator that exploits nonreciprocal modal interaction in a pair of evanescently coupled silicon waveguides coated with MO material was recently studied \cite{Chao2025}. The TMOKE breaks the symmetry of the guided modes so that the modal beating between even and odd supermodes depends on the direction of light propagation. This direction-dependent perturbation of the propagation constants produces nonreciprocal energy transfer between the coupled waveguides resulting in strong isolation for one propagation direction. 
Contrary to isolators using NRMC, TMOKE-based isolators do not require QPM \cite{Chao2025,Fujita2000}.

Isolators utilizing TMOKE remain theoretical and simulation-based, with no reported experimental demonstrations of two-port isolation to date \cite{Chao2025,Abadian2021,Hu2021Coin}. 
Experimental work on TMOKE in nanostructures and magneto-plasmonic systems has shown measurable TMOKE signals, but has not been translated into a two-port isolator with measured isolation ratios on-chip,  underscoring a current gap between TMOKE theory and experimental optical isolation \cite{Abadian2021,Huang2026}.  

\section{Space-time modulation} \label{Sec:Space-time_modulation} 
Integrated photonic devices rely on controlled perturbations of the refractive index or permittivity to manipulate optical signals. Such modulation enables phase \cite{Eltes2019,Wang2018}, amplitude \cite{Le2022,Zhang2021}, and frequency control \cite{Jiang2025,Zhu2022} and forms the foundation of active PICs. Refractive index modulation mechanisms include electro-optic (EO) \cite{Kashi2020,Li2024Linearity,Ma2024}, acousto-optic (AO) \cite{Tadesse2014,Shi2026,Yu2024,Beller2022}, opto-mechanical \cite{Pruessner2021,Eichenfield2009}, and thermo-optic \cite{Jayatilleka2015,Zhou2014}. Each modulation mechanism originates from a distinct physical interaction between electromagnetic fields, lattice dynamics, or charge carriers \cite{Ogawa2024}. 

\subsection{Reciprocity theorem for time-varying media} 
 Regardless of the modulation mechanism, the generalized TR asymmetry is a necessary but not a sufficient condition for nonreciprocity \cite{Williamson2020Integrated,Caloz2018}
 \begin{equation}
 \varepsilon (\mathbf {r}, t - t_{0}) \neq \varepsilon (\mathbf {r}, -t - t_{0})
 \end{equation}
for all choices of the initial time $t_0$. A spatially uniform, TR symmetric harmonic modulation of the permittivity does not break reciprocity. Nonreciprocity requires that the modulation carry a signed direction, which is supplied either by a spatiotemporal (traveling-wave) profile or by a modulation function that is itself asymmetric in time \cite{Asadchy2020}. The reciprocity theorem for time-varying media in the frequency domain is \cite{Shi2016,Asadchy2020} 
\begin{equation}
\int_{V_A}
\sum_n
\frac{\mathbf{E}^{(B)}_n \cdot \mathbf{J}^{(A)}_n}{\omega_n} \, dV_A
=
\int_{V_B}
\sum_n
\frac{\mathbf{E}^{(A)}_n \cdot \mathbf{J}^{(B)}_n }{\omega_n}\, dV_B.
\end{equation}
Reciprocity in time-varying media implies restriction on the evolution of photon numbers in the forward and backward directions, rather than intensities of waves as is the case of time-invariant systems \cite{Asadchy2020,Caloz2018}.  

\subsection{Electro-optic modulation}
EO modulation  relies on the EO effect, whereby the optical properties of specific materials are altered by an applied electric field. In \textit{electro-refractive modulators}, EO interaction changes the refractive index. This is the case of the Pockels effect, the DC Kerr effect, and free-carrier dispersion (plasma effect) \cite{Liu2015}. Electro-refractive modulators generally rely on interferometric configurations, most commonly Mach–Zehnder interferometers or MRRs \cite{Sinatkas2021,Yu2025}.  By contrast, when the applied electric field alters the imaginary part of the permittivity (absorption coefficient), the device operates as \textit{an electro-absorption modulator}. This functionality is typically achieved through the Franz–Keldysh effect, quantum-confined Stark effect, or free-carrier absorption \cite{Xu2023,Wu2024Survey}.

The dependence of the permittivity on the applied electric field can be expressed as
\begin{equation}
\hat{\bm{\varepsilon}}\mathbf{E}
\equiv
\mathbf{D},
\qquad
D_i = \varepsilon_0 \left[ \delta_{ij} + \chi_{ij}^{(1)}
+ \mkern-32mu \underbrace{\chi^{(2)}_{ijk}}_{ \text{Linear EO modulation}} \mkern-32mu E_k
+ \chi^{(3)}_{ijkl} E_k E_l + \cdot \cdot \cdot  \right] E_j ,
\label{Eq:permittivity_tensor_time_modulation}
\end{equation} 
where $\varepsilon_{ij}^{\mathrm{L}} = \varepsilon_0 \left[ \delta_{ij} + \chi_{ij}^{(1)} \right] $ is the linear permittivity tensor. The $\chi^{(2)}_{ijk}$ term gives rise to the Pockels effect, also known as the linear EO effect, and is allowed only in media lacking inversion symmetry \cite{Boyd2020}. The $\chi^{(3)}_{ijkl}$ term gives rise to the DC Kerr effect, or quadratic EO effect, under a static electric-field bias. %Although both effects originate from nonlinear material susceptibilities, the resulting response can remain linear with respect to a weak optical field for a fixed external bias.
%Space–time modulation breaks the time-invariance assumption underlying the conventional Lorentz reciprocity theorem. 
Although the electro-optic permittivity tensor [Eq.~(\ref{Eq:permittivity_tensor_time_modulation})] remains symmetric, its temporal modulation can violate the Lorentz reciprocity assumption of the time-invariant media.  In time-periodic media, reciprocity instead requires the modulation harmonics to satisfy $\hat{\bm{\varepsilon}}_{m}=\hat{\bm{\varepsilon}}_{-m}^{T}$. Thus, the relevant reciprocity constraint involves the transpose symmetry relation between Fourier components rather than the full permittivity tensor~\cite{Sotoodehfar2023}. In particular, suitable spatiotemporal modulation can emulate an effective gyrotropic response with off-diagonal components~\cite{Boshgazi2025,Zhiyu2023}.

\subsection{Acousto-optic modulation} 

AO modulation arises from changes in the refractive index of a material induced by strain associated with acoustic waves. Acoustic waves can interact with and manipulate light in multiple ways, including shifting the optical frequency, deflecting light into different spatial modes, and rotating the polarization state. Because the velocity of acoustic waves is much smaller than that of optical waves, the acoustic field can be treated as quasi-static in the dielectric function \cite{Wolff2021}.

The AO perturbation in permittivity $\Delta \bm{\varepsilon}$ can be divided into a photoelastic contribution $\Delta \bm{\varepsilon}^{p}$ and  a birefringence contribution $\Delta \bm{\varepsilon}^{b}$ as \cite{Engan1996,Smith1980,Nelson1970}
\begin{equation}
\Delta \mathbf{\varepsilon}_{ij}
= \Delta \mathbf{\varepsilon}^{p}_{ij}
+ \Delta \mathbf{\varepsilon}^{b}_{ij}
=-n^{4}\left( p_{ijkl}S_{kl} + p'_{ijkl}\mathsf{R}_{kl} \right),
\label{Eq:acousto_optic_permittivity}
\end{equation}
where $S_{kl}$ and $\mathsf{R}_{kl}$ are the symmetric and antisymmetric parts of the displacement gradient, respectively; $p_{ijkl}$ is the photoelastic tensor, and $p'_{ijkl}$ is the rotational tensor introduced by Nelson and Lax, which is antisymmetric in the acoustic indices $(kl)$ and vanishes in isotropic and cubic media~\cite{Nelson1970}. Both tensors are symmetric in the optical indices $(ij)$, so the acousto-optic permittivity perturbation remains symmetric. In absence of rotation, $\Delta \varepsilon$ is strictly symmetric as well \cite{Engan1996}. As in the electro-optic case, however, the travelling acoustic wave renders the optical medium time dependent violating the time-invariance assumption of Lorentz reciprocity. %The distinction is particularly clear for torsional acoustic waves, which rotate the local optical axes and couple the two polarization modes of a fibre through the rotational contribution in Eq.~(\ref{Eq:acousto_optic_permittivity})~\cite{Engan1996,Smith1980}. The coupling between the two modes remains symmetric, whereas the directionality of the interaction is determined by momentum matching to the travelling acoustic wave~\cite{Engan1996}. 

In the linear regime, the optical refractive index is modulated linearly by acoustic strain through mechanisms such as the \emph{photoelastic effect} and the \emph{moving boundary effect} \cite{Johnson2002}. In the former, strain within the waveguide core modifies the local refractive index, leading to optical phase modulation. In the latter, acoustic motion displaces the waveguide boundaries or alters its cross-section, thereby changing the optical mode confinement and effective refractive index \cite{Xu2024,Kittlaus2021,Wolff2015}.

Nonlinear AO modulation arises from intensity-dependent photon–phonon interactions, most notably Brillouin scattering, which constitutes a third-order optical nonlinearity \cite{Wolff2021,Eggleton2019}. In this regime, the optical fields themselves drive coherent acoustic vibrations. Fundamentally, the Brillouin interaction originates from the interference between optical modes in a waveguide, which generates a time-varying force density within the material through  radiation pressure and electrostriction \cite{Wolff2021}.
 
In recent integrated platforms, the distinction between AO and optomechanical modulation has become increasingly blurred. Both processes originate from the same underlying photon–phonon interactions, namely photoelastic coupling and radiation-pressure-induced boundary motion, and differ primarily in how the mechanical degrees of freedom are excited and confined \cite{Wolff2021,Eggleton2019,Ruesink2016,Liu2020,Dostart2020,Kittlaus2021}. Table \ref{tab:space_time_modulation} contains representative state-of-the-art metrics for optical isolators based on space-time modulation.

\begin{table*}[t]
\caption{Optical isolators based on space-time modulation}\label{tab:space_time_modulation}%
\centering
\footnotesize
\setlength{\tabcolsep}{4pt}
\begin{tabular*}{170mm}{@{\extracolsep{\fill}}lccccccc@{}}
\toprule
Mechanism & Platform & \shortstack{Isolation (dB)} & \shortstack{Bandwidth (nm)}
& \shortstack{Insertion loss (dB)} & Polarization & \shortstack{Footprint ($\mu$m)} & ref. \\
\midrule
Momentum bias           & AlN/Si         & 24.5   & 16          & 8.09  & TE  & $\sim$7640 & \cite{Cheng2025}    \\
Angular momentum bias         & Si/SiO$_2$     & 16     & NR \textsuperscript{a}          & 0.6  & TE & $40 \times 40$ & \cite{Zarif2024}    \\
Asymmetric bandgap\textsuperscript{$\ast$} & Graphene & $>20$ & NR & $<3$ & TE & $\sim$1500 & \cite{Correas2018}  \\
ATS (EO)                & TFLN/SiO$_2$   & 47.7   & 0.00032 & 1.45 & TE & $\approx$900 & \cite{Kim2026} \\
Rabi splitting (AO)     & AlN/SOI        & 14.4   & 0.47  & 0.7 & TE  & $\sim$9950 & \cite{Zhou2024} \\
AC Stark shift          & LN/Al$_2$O$_3$ & 45     & 1           & 8.2   & TE  & NR   & \cite{Xu2025}       \\
Synthetic fields (AO)   & AlN/Si         & 3      & NR          & NR    & TE  & NR & \cite{Kim2021} \\
Polarization conversion \textsuperscript{$\ast$} & Dielectric & 29.95 & $\sim$40 & 0.05 & TE & $\sim$11600 & \cite{Wang2020} \\
Phase modulation (EO)   & TFLN           & 48     & 120         & 0.5   & TE  & $\sim$17500 & \cite{Yu2023}   \\
Phase modulation (DRDI) & SiN/TFLN       & 30     & 30          & 8.05  & TE  & $\sim$15000 & \cite{Han2026}  \\
\botrule
\end{tabular*}
\par\vspace{4pt}
\begin{minipage}{170mm}
\setlength{\parindent}{0pt}\raggedright\scriptsize
\textsuperscript{$\ast$}Simulation/theory only; not experimentally realized. \textsuperscript{a} NR denotes data not reported in the original reference. 
\end{minipage}
\end{table*}

\subsection{Momentum bias} 
\begin{figure*}[t!]
\centering
\includegraphics[width =170mm]{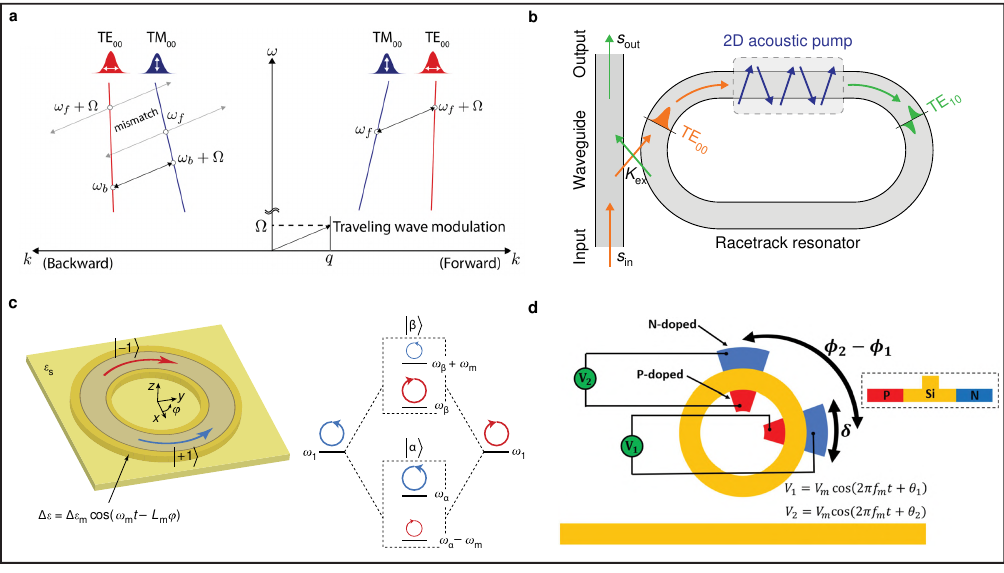}
\caption{Non-reciprocity from traveling wave modulation, indirect interband scattering and angular momentum bias. \textbf{a} Frequency-momentum diagram for indirect interband scattering between the fundamental TE$_{00}$ and TM$_{00}$ modes of a waveguide. A traveling wave modulation of frequency $\Omega$ and momentum $q$ phase-matches the transition in the forward direction only; the backward transition is mismatched. \textbf{b} Acoustically pumped non-reciprocal modulator: a racetrack resonator supporting TE$_{00}$ and TE$_{10}$ modes, evanescently coupled to a bus waveguide. The two-dimensional acoustic pump simultaneously breaks the orthogonality of the two optical modes and satisfies the phase-matching condition, so intermodal scattering occurs for one propagation direction only. \textbf{c} Azimuthally symmetric ring resonator carrying a spatiotemporal permittivity modulation $\Delta\varepsilon=\Delta\varepsilon_m\cos(\omega_m t - L_m\varphi)$, an azimuthally traveling wave of orbital angular momentum $L_m$. Without modulation the ring supports two degenerate counter-propagating states $|\pm1\rangle$. The modulation lifts that degeneracy, acting as a magnet-free counterpart of the static magnetic bias that Zeeman-splits atomic orbitals. To the right shown is the frequency diagram of the ring eigenstates without (outer columns) and with (centre column) modulation, for $L_m=2$. The eigenstates $|\alpha\rangle$ and $|\beta\rangle$ of the modulated ring are hybridisations of $|+1\rangle$ and $|-1\rangle$, each comprising a dominant (thick line) and a secondary (thin line) sub-state separated by $\omega_m$. \textbf{d} Silicon-on-insulator ring resonator side-coupled to a bus waveguide, modulated at two point perturbations formed by p--i--n junctions acting through the plasma-dispersion effect. Their angular separation $\varphi_2-\varphi_1$, width $\delta$ and modulation phase difference $\theta_2-\theta_1$ set the non-reciprocal response. Inset: junction cross-section. \textbf{a} Reproduced from ref.~\cite{Orsel2023} under the terms of the Creative  Commons Attribution 4.0 International License. \textbf{b} Reproduced with permission from ref.~\cite{Sohn2018}. Copyright 2018 Springer Nature Limited. \textbf{c} Reproduced with permission from ref.~\cite{Sounas2013}. Copyright 2013 Macmillan Publishers Limited. \textbf{d} Reproduced from ref.~\cite{Zarif2022} under the terms of the Creative Commons Attribution 4.0 International License} 
\label{Fig:SpaceTime_Modulation_Fig_6}
\end{figure*}

 Interband photonic transition is the photonic analog of electronic interband transition \cite{Winn1999}. Such a transition results in \emph{nonreciprocal frequency conversion} 
 via spatiotemporal modulation of permittivity \cite{YuFan2009,Yu2009Optical}. Consider two optical modes with different longitudinal wavevectors $k_1 \ne k_2$  and assume the permittivity is modulated by 
\begin{equation}
    \Delta\varepsilon (x, z,t) = \delta (x) \cos (\Omega t - qz) ,
\end{equation}
where $\delta (x)$ is the modulation amplitude (transverse to the propagation direction),   $\Omega$ is the modulation frequency, and $q$ is the modulation wavenumber. In the forward direction, an interband photonic transition of the modes occurs when the phase matching condition $\Delta k = k_2 - k_1 -q \approx0$ is satisfied. In the backward direction, such a transition is not possible for the chosen modes, frequency, and modulation momentum (Fig. \ref{Fig:SpaceTime_Modulation_Fig_6}a). Space-time modulation via interband transition can be thought of as biasing the photonic structure via linear momentum which is a  $\mathcal{T}$-odd physical quantity under TR. Combined with filters in the backward direction, complete isolation of  spatially asymmetric modes is possible \cite{YuFan2009,Yu2009Integrated,Yu2009Optical}.

On-chip traveling wave EO implementation requires a sophisticated network of electrical junctions \cite{Lira2012}. Optomechanical systems are used recently to break reciprocity via optical pumping \cite{Shen2016}. However, by inverting the roles and using acoustic pumping instead of optical pumping to break TR symmetry, the bandwidth of the effect is no longer determined by the mechanics, but by the photonic modes instead \cite{Sohn2018,sohn2019}. The photoelastic perturbations of the medium by the driven acoustic wave enable indirect interband scattering (Fig. \ref{Fig:SpaceTime_Modulation_Fig_6}a and b) \cite{Sohn2018,Kittlaus2021}. In addition to phase-matching, the overlap integral between the acoustic mode and optical modes must be non-zero.

Broadband nonreciprocity based on \textit{indirect interband scattering} has been reported in several studies \cite{Sohn2018,Cheng2025,Zhou2024}. One notable recent implementation is a terahertz-bandwidth isolator comprising two multiport AO modulators that serve as beam splitters in a nonreciprocal multimode interferometer \cite{Cheng2025}. The AO modulators are used to partially convert the forward-propagating symmetric mode at frequency $\omega_1$ into an antisymmetric mode at frequency $\omega_2 = \omega_1 + \Omega$, and likewise partially convert the backward-propagating antisymmetric mode at frequency $\omega_2$ into a symmetric mode at frequency $\omega_3 = \omega_2 + \Omega$. The beam-splitting action is realized through an electrically driven intermodal scattering process. The phase-matching condition associated with this scattering gives rise to nonreciprocal frequency conversion. Within the interferometer, this directional frequency shift leads to constructive interference for forward-propagating light and destructive interference for backward light \cite{Cheng2025}.

%Two acousto-optic modulators acting as non-reciprocal beamsplitters ina multimode interferometer. Each AOM partially scatters between the symmetric andantisymmetric modes while shifting the frequency by the phonon frequency $\Omega_0$ (left). Balancing the group delay of the two arms, here by mapping the modes onto length-matched single-mode waveguides, sets the isolation bandwidth.

\subsection{Angular momentum bias}
Nonreciprocity in ferromagnetic materials originates at the microscopic level from the unidirectional precession of electron spins or from the Zeeman splitting of atomic orbitals caused by an applied static magnetic field \cite{Dionne2005}. Angular-momentum bias is a form of spatiotemporal symmetry breaking that imparts an effective \emph{azimuthal} (rotational) momentum to a resonant photonic structure lifting the degeneracy between CW and CCW modes $|\pm \ell\rangle$: see Fig \ref{Fig:SpaceTime_Modulation_Fig_6}c. In integrated MRRs, this bias is initially implemented through an azimuthally traveling wave modulation of permittivity. The approach was originally developed in metamaterial and microwave circulator platforms and later translated to integrated photonics \cite{Sounas2013,Estep2016,Estep2014}. In its general form, the modulation can be written as \cite{Sounas2013,Estep2014}
\begin{equation}
\Delta\varepsilon(\varphi,t)=\Delta\varepsilon_m
\cos\!\left(\omega_m t - \ell_m \varphi\right),
\end{equation}
where $\omega_m$ is the modulation frequency and $\ell_m$ is the modulation orbital angular momentum. This modulation carries orbital angular momentum and corresponds to an effective angular velocity $\Omega_m=\omega_m/\ell_m$.   
Angular-momentum bias breaks reciprocity because it introduces a bias quantity that is \emph{$\mathcal{T}$-odd under TR}, thereby violating the conditions of the Onsager-Casimir relations \cite{Estep2014,Fang2012}. Under TR, the sign of angular momentum reverses, playing a role directly analogous to a magnetic field in MO systems. 

The physics of angular-momentum bias in photonic resonators is well captured by temporal coupled-mode theory, where the CW and CCW modal amplitudes $a_{\mathrm{cw}}$ and $a_{\mathrm{ccw}}$ evolve under a time-periodic coupling matrix \cite{Zarif2019,Suh2004,Fang2012}
\begin{equation}
\frac{d\mathbf{a}}{dt}
= i\,\mathbf{M}(t)\,\mathbf{a},
\qquad
\mathbf{a}=
\begin{bmatrix}
a_{\mathrm{cw}} \\
a_{\mathrm{ccw}}
\end{bmatrix}.
\end{equation}
Here $\mathbf{M}(t)$ contains off-diagonal terms oscillating as $e^{\pm i\omega_m t}$, which mediate frequency- and momentum-selective coupling between counter-rotating modes and their Floquet sidebands. When the modulation is discretized into multiple spatial segments with progressive phase delays, such as stepwise  modulation with different steps corresponding to different pn junctions (Fig. \ref{Fig:SpaceTime_Modulation_Fig_6}d) \cite{Sounas2014} or three resonators driven with $120^\circ$ phase offsets  \cite{Estep2016,Yang2025}, the modulation emulates a unidirectionally rotating bias. In this regime, CW and CCW eigenfrequencies split, the eigenmodes become directionally asymmetric, and the resulting scattering matrix becomes non-symmetric, enabling linear optical isolation within the resonator linewidth \cite{Zarif2024,zarif2020o}.

\subsection{Asymmetric photonic bandgap} 

\begin{figure*}[t!]
\centering
\includegraphics[width = 170mm]{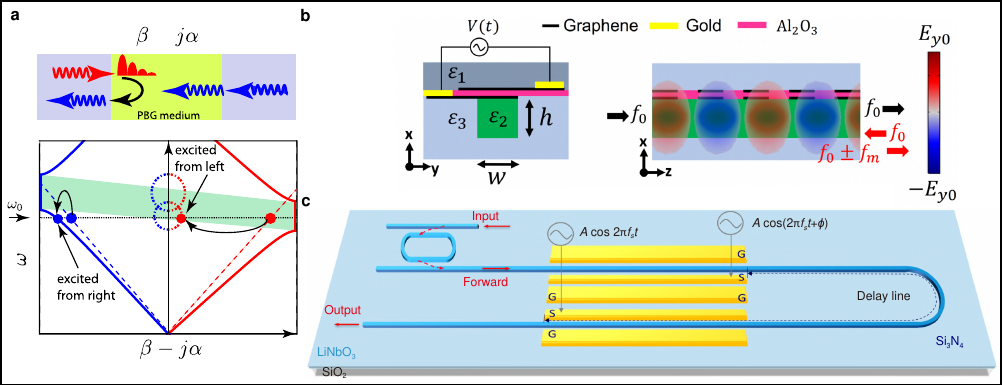}
\caption{\textbf{a} Non-reciprocal Bragg reflection in a space-time modulated slab. The
  modulation tilts the photonic band gap in the $\beta-j\alpha$ plane, so at the
  excitation frequency $\omega_0$ a wave incident from the left (red) is evanescent
  and totally reflected, whereas one incident from the right (blue) remains
  propagating and is fully transmitted.
  \textbf{b} Hybrid dielectric/graphene realisation of the same principle: a
  Si/SiO$_2$ waveguide loaded with a series of time-modulated graphene capacitors
  driven by $V(t)$. Forward light at $f_0$ is transmitted, whereas backward light is
  frequency-converted to $f_0\pm f_m$ and reflected. \textbf{c} Non-reciprocity from traveling-wave electro-optic phase modulation.  Spatio-temporal isolator in Si$_3$N$_4$-loaded lithium niobate on
insulator. Light passes twice through traveling-wave modulator sections driven at
$A\cos(2\pi f_s t)$ and $A\cos(2\pi f_s t+\phi)$, separated by a folded delay
line. Forward light emerges at the carrier frequency, whereas backward light is
converted into sidebands, which the add--drop racetrack resonator at the input
rejects. \textbf{a} Adapted with permission from ref.~\cite{Chamanara2016}. Copyright 2017
  American Physical Society. \textbf{b} Adapted with permission from
  ref.~\cite{Correas2018}. Copyright 2018 American Physical Society. \textbf{c} Adapted from ref.~\cite{Huang2022} under the terms of the Creative Commons Attribution 4.0 International License  }
\label{Fig:Asymmetric_photonic_bandgap}
\end{figure*}

Asymmetric photonic bandgaps rely on engineering the dispersion diagrams so that the Bloch–Floquet band structure is \emph{asymmetric} in momentum-frequency space $(\beta,\omega)$. A space-time \emph{periodic perturbation} of the permittivity is \cite{Chamanara2016} 
\begin{equation}
\Delta \varepsilon(z,t) =  M\cos(\omega_m t-k_m z),
\end{equation}
where $M$ is the modulation depth. Such periodic modulation couples the harmonics \((\beta\pm n k_m,\ \omega\pm n\omega_m)\) and yields an oblique Brillouin zone, i.e., the resulting band edges are shifted differently for forward and backward propagation as shown in Fig. \ref{Fig:Asymmetric_photonic_bandgap}a. When the operating frequency $\omega_0$ falls inside the bandgap for one direction of propagation but lies within a band for the opposite propagation direction, a wave incident from one port is evanescent (reflected) while the opposite incidence excites a propagating Bloch mode (transmitted), producing isolation in a linear magnet-free medium \cite{Chamanara2016}. Contrary to interband photonic transition, the modulation is uniform in the cross-section of the waveguide and relies on \emph{intraband transition}.

The operation mechanism of the isolators based on asymmetric band gaps can also be understood as nonreciprocal coupling of an incident mode at $\omega_0$ and waveguide modes at $\omega_0 \pm \omega_m$. As in the case of interband transition, for any two modes to exchange power the phase-matching condition must hold, $\Delta k = k_z^{(2)}(\omega_0\pm\omega_m) - k_z^{(1)}(\omega_0) - k_m =0$, so that the fields scattered along the structure’s length add constructively \cite{Correas2018}. When the scattered field does not correspond to a mode of the waveguide, the corresponding wave passes through the space-time modulated
region largely unaffected (Fig. \ref{Fig:Asymmetric_photonic_bandgap}b). This operating principle underpins both nonreciprocal metasurfaces and hybrid graphene/dielectric implementations \cite{Chen2023,Correas2018,Pakniyat2024}.

The modulation creates an effective moving space-time perturbation without motion of matter. A wave reflected from this bandgap experiences a frequency shift analogous to the Doppler shift of a wave reflected from a moving medium \cite{Chamanara2016,Sounas2017Non-reciprocal}. Generally, traveling wave modulation mimics certain aspects of physical motion including synthetic Doppler shift \cite{Serra2026,Mazor2019} and synthetic Fresnel drag \cite{Prudencio2023,Huidobro2019,Huidobro2021}. 

\subsection{Traveling-wave phase modulation}
The direction-dependent electro-optic phase accumulation produced by a traveling microwave drive at frequency \(f_{\mathrm{RF}}\) is given by \cite{Yu2023,Huang2022}:
\begin{equation}
\label{eq:phi_integral}
\phi(f_{\mathrm{RF}},t)
= \frac{V_{\mathrm{pk}}}{V_{\pi}L}\,\pi
\int_{0}^{L} e^{-\alpha z/2}
\cos\!\Bigg[2\pi f_{\mathrm{RF}}\Big(\frac{(n_{\mathrm{opt}}\pm n_{\mathrm{RF}})\,z}{c}-t\Big)\Bigg]\mathrm{d}z,
\end{equation}
where \(V_{\mathrm{pk}}\) is the RF peak voltage, \(V_{\pi}\) the half-wave voltage, \(\alpha\) the microwave attenuation constant, \(n_{\mathrm{opt}}\) the optical group index, \(n_{\mathrm{RF}}\) the microwave phase index, \(c\) the vacuum speed of light and $L$ is the modulation length. The sign \(+\) (\(-\)) denotes counter-propagating (co-propagating) optical and microwave waves, respectively.

For negligible microwave attenuation the forward (counter-propagating) accumulated phase vanishes whenever the integrand completes an integer number of half-periods over the modulation length $f_{\mathrm{RF}} = \frac{m\,c}{2L\,n_{\mathrm{opt}}}$ with $m$ being an integer, so the forward-propagating optical wave is transmitted unmodulated. In the backward (co-propagating) direction the phase accumulates monotonically with \(z\), producing frequency sidebands at \(f_0\pm n f_{\mathrm{RF}}\). Complete carrier depletion occurs when the effective modulation index \(\beta\) reaches the first root of the zeroth-order Bessel function $J_{0}(\beta)=0$ with $\beta \approx 0.765\pi$ \cite{Yu2023,Doerr2011}, such that essentially all optical power is transferred to sidebands. This architecture has been demonstrated both in a low-loss TFLN device (Fig. \ref{Fig:Asymmetric_photonic_bandgap}c) \cite{Huang2022}  and as a cascaded-modulator plus a racetrack approach in LNOI \cite{Yu2023}.

Han et al. demonstrated a four-channel traveling-wave electro-optic isolator on a SiN/TFLN hybrid waveguide that drives parallel Mach–Zehnder arms with quarter-period–staggered RF waveforms to implement dynamic rotating destructive interference, achieving continuous destructive interference under time-periodic
phase modulation. The device achieves approximately $ 30$ dB peak isolation and maintain more than $24$ dB across approximately $30$ nm   \cite{Han2026}. 
Unlike tandem-phase \cite{Doerr2011} or other traveling-wave schemes \cite{Doerr2014,Gall2013,Yang2014,Lin2019}, which rely on interferometric delays or carrier-to-sideband conversion plus spectral filtering, dynamic rotating destructive interference achieves broadband suppression with no sidebands detected above the reported noise level by continuously time-rotating pairwise destructive interference across balanced channels. 

\subsection{Synthetic electric and magnetic fields} \label{sec:synthetic}
The Aharonov-Bohm effect is a quantum mechanical phenomenon in which the phase of a charged particle is affected by the vector potential of electromagnetic fields even in regions where the electric and magnetic fields vanish \cite{Aharonov1959,Wilczek1984,Gao2025}. Analogously in photonics, \emph{the phase of the temporal modulation} can be used to create an effective gauge potential for photons leading to phenomena such as an effective Lorentz force on photons \cite{song2025artificial,Fang2012Photonic}.

\begin{figure*}[t!]
\includegraphics[width = 170mm]{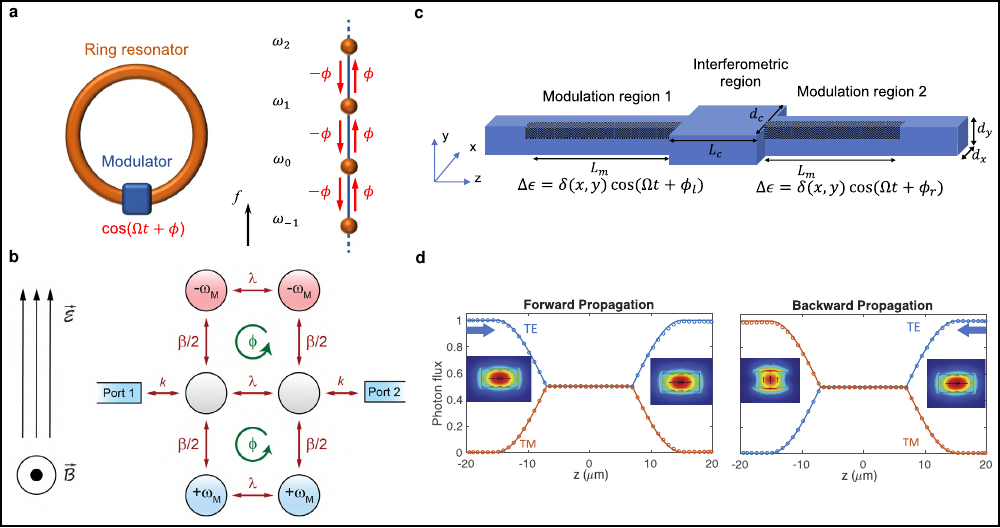}
\caption{Direct photonic transition via synthetic gauge fields for photons and dynamic refractive index modulation.
  \textbf{a} Synthetic frequency dimension. An intracavity modulator driven at
  $\cos(\Omega t+\phi)$ couples the resonator's frequency modes into a
  one-dimensional lattice along frequency, with the modulation phase imprinted on
  the couplings as a gauge potential ($+\phi$ upward, $-\phi$ downward).
  \textbf{b} Pictorial representation of a modulated two-resonator chain (ports 1
  and 2, coupling rate $k$) as a synthetic lattice carrying both fields. The
  synthetic electric field generates a potential of $\mp\omega_M$ on the upper and
  lower chains, and the synthetic magnetic field generates a direction-dependent
  phase shift $\phi$ around each closed loop.  \textbf{c} Isolator design. A static waveguide (blue) carries two modulated regions
  (dark) of length $L_m$, separated by an interferometric region of length $L_c$ and
  width $d_c$. Only one quadrant of the waveguide cross-section is modulated in each
  region, with permittivity modulation $\delta(x,y)\cos(\Omega t+\phi)$ and
  modulation phases $\phi_l$ and $\phi_r$.
  \textbf{d} Photon flux of the TE and TM modes versus propagation distance, from
  finite-element simulation (circles) and vectorial coupled-mode theory (lines).
  Forward (left), TE injected from the left leaves the device in the TE mode;
  backward (right), TE injected from the right is fully converted to TM. Insets:
  mode profiles at the input and output.
  \textbf{a} Adapted from ref.~\cite{yuan2021synthetic} under the terms of the
  Creative Commons Attribution 4.0 International License. \textbf{b} Reproduced with
  permission from ref.~\cite{Peterson2019}. Copyright 2019 American Physical Society.   \textbf{c}, \textbf{d} Adapted with permission from ref.~\cite{Wang2020}.
  Copyright 2020 Optical Society of America} 
\label{Fig:phase_modulation}
\end{figure*}

When two photonic modes having opposite spatial symmetries undergo \emph{direct interband transition}, the coupled-mode theory gives \cite{Fang2012Photonic}
\begin{equation}
i\frac{d}{dz}\begin{pmatrix}a_1\\a_2\end{pmatrix}
=\begin{pmatrix}0 & C\,e^{-i\phi}\\[4pt] C^*\,e^{i\phi} & 0\end{pmatrix}
\begin{pmatrix}a_1\\a_2\end{pmatrix},
\label{eq:coupled}
\end{equation}
with \(C\propto\int \delta(x) E_1^*E_2\,d\mathbf x\).  The modulation phase \(\phi\) appears multiplicatively on the off-diagonal coupling and is therefore the photonic analogue of the Peierls phase \(C\mapsto C e^{i\int\mathbf{A}\cdot d\mathbf{l}}\). A photon that follows one conversion path picks up \(+\phi\) while the time-reversed path picks up \(-\phi\) leading to a direction-dependent phase shift  and a gauge potential (Fig. \ref{Fig:phase_modulation}a) \cite{Fang2012Photonic}. %The modulation strength must be a nonuniform function as uniform modulation leads to an ambiguity in the phase. 

Synthetic magnetic fields alone do not guarantee nonreciprocity in time-modulated linear media.  
Nonreciprocity requires breaking of both TR symmetry (here via \(\phi\)) and the mirror symmetry in the \emph{frequency} dimension \cite{Kim2021}. 
Modulation with frequency \(\omega_M\) builds a ladder of sideband chains separated by \(\omega_M\), see Fig. \ref{Fig:phase_modulation}b. The spatial gradient of \(\phi\) gives the synthetic magnetic field %\(B_{\rm syn}\)
while \(\omega_M\) plays the role of a synthetic electric potential (energy offset between chains). 

To lowest nontrivial order in the modulation amplitude \(\beta\), the forward and backward linear scattering amplitudes at the same input frequency can be written as \cite{Peterson2019,Kim2021}
\begin{align}
S_{21}(\omega) &\approx \tau(\omega)
+\frac{\beta^2}{4}\Big[(\tau_+ + \tau_-)\cos\phi
- i(\tau_+ - \tau_-)\sin\phi\Big],\label{eq:S21}\\
S_{12}(\omega) &\approx \tau(\omega)
+\frac{\beta^2}{4}\Big[(\tau_+ + \tau_-)\cos\phi
+ i(\tau_+ - \tau_-)\sin\phi\Big],\label{eq:S12}
\end{align}
with \(\tau_\pm\equiv\tau(\omega\pm\omega_M)\) the (reciprocal) responses of the frequency-shifted channels. The symmetric part (the \(\cos\phi\) term and the base \(\tau\)) is identical for both directions while the antisymmetric term proportional to \(\left( \tau_+ - \tau_-\right)\sin\phi\) flips sign between \(S_{21}\) and \(S_{12}\).  Thus nonreciprocity requires (i) a nonzero synthetic flux (TR breaking) and (ii) a frequency-dimension asymmetry \(\tau_+\neq\tau_-\) (mirror symmetry breaking), which can be provided by choosing \(\omega_M\) so that the upper and lower conversion paths sample different dispersive (or resonant) phases. For example, \(\phi\approx\pm\pi/2\) and \(\omega_M\) tuned to bring one sideband near resonance will make the antisymmetric term dominant and can suppress transmission in one direction while leaving the reverse largely unaffected \cite{Tzuang2014,Peterson2019,Li2014}. 

\subsection{Nonreciprocal polarization conversion}

Nonreciprocal polarization conversion using \emph{direct photonic bandgap transition}  can be realized on chip for waveguides supporting quasi-TE and quasi-TM modes. The spatiotemporal modulation of the permittivity couples quasi-TE and quasi-TM modes in a direction-dependent manner while preserving the longitudinal wavevector. This mechanism emulates the functionality of a MO Faraday rotator followed by a reciprocal polarizer, but eliminates the need for MO materials and external magnetic bias \cite{Wang2020}.

An isolator can be realized using two phase-controlled modulated regions separated by a passive interferometric section (Fig.  \ref{Fig:phase_modulation}c). By choosing the modulation phases $\phi_\ell - \phi_r = \pi/2$ and engineering a relative interferometric phase shift $\Delta\varphi = \pi/2$, forward-propagating light remains in the input TE polarization mode, whereas backward-propagating light is converted to the orthogonal polarization TM mode and can be rejected by a polarizer (Fig. \ref{Fig:phase_modulation}d). An asymmetric transverse modulation profile maximizes the longitudinal-field overlap required for efficient TE-TM coupling \cite{Wang2020}. 

EO nonreciprocal polarization rotation in lithium niobate using indirect interband transition has been experimentally demonstrated  \cite{Orsel2023}. However, its counterpart using direct interband transition and phase modulation of permittivity remains in the design stage \cite{Wang2020}.

\subsection{Direction-dependent attenuation mechanisms} 
\begin{figure*}[t!]
\centering
\includegraphics[width=170mm]{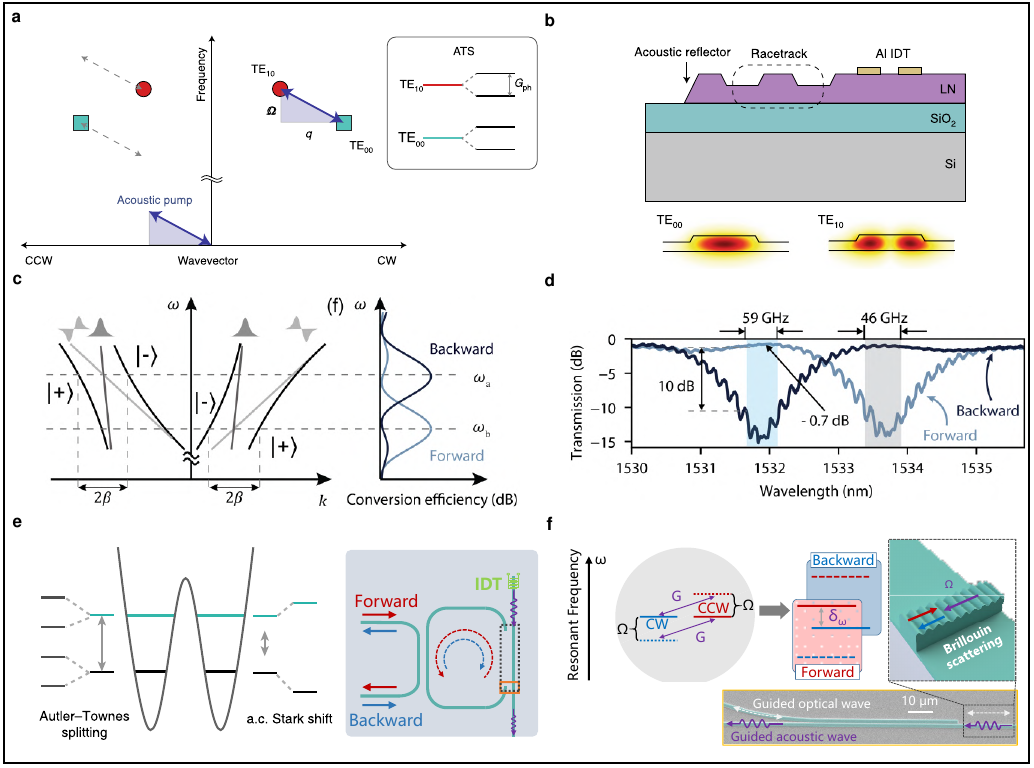}
\caption{Non-reciprocity from acousto-optic Autler--Townes splitting and AC Stark
  shifts.
  \textbf{a} Phonon-mediated Autler--Townes splitting between the symmetric
  TE$_{00}$ and antisymmetric TE$_{10}$ modes of a whispering-gallery resonator.
  Acousto-optic phase matching admits the splitting only for clockwise circulation,
  making the resonator chiral; the phonon-enhanced coupling rate $G_{\mathrm{ph}}$
  sets the splitting.
  \textbf{b} Cross-section of the acousto-optic interaction region: an aluminium
  interdigital transducer and an acoustic reflector on a lithium niobate racetrack
  over SiO$_2$/Si, with the TE$_{00}$ and TE$_{10}$ mode profiles below.
  \textbf{c} Non-reciprocal photonic band hybridisation in the rotating frame.
  Interband optomechanical scattering is phase matched at different frequencies in
  the two directions, hybridising the bands at $\omega_b$ forward and $\omega_a$
  backward with wavevector splitting $2\beta$, and giving direction-dependent
  conversion (right).
  \textbf{d} Measured transmission, with 0.7~dB insertion loss and a 59~GHz 10-dB
  isolation bandwidth; the blue and grey windows are the 10-dB isolation windows at
  $\omega_a$ and $\omega_b$.
  \textbf{e} Resonant Autler--Townes splitting contrasted with the off-resonant AC
  Stark shift (left), and a microring in which a traveling acoustic wave launched by
  an interdigital transducer couples the clockwise and counter-clockwise modes
  off-resonantly (right).
  \textbf{f} Energy diagram of the phonon-mediated coupling: the acoustic field
  couples the CW and CCW modes asymmetrically, lifting their degeneracy by
  $\delta_\omega$.
  \textbf{a}, \textbf{b} Reproduced with permission from ref.~\cite{Sohn2021}.
  Copyright 2021 Springer Nature Limited. \textbf{c}, \textbf{d} Reproduced from
  ref.~\cite{Zhou2024} and \textbf{e} right, \textbf{f} adapted from
  ref.~\cite{Xu2025}, under the terms of the Creative Commons Attribution 4.0
  International License. \textbf{e} left, Reproduced with permission from
  ref.~\cite{ZhangElectronically_Archive}. Copyright 2019 Springer Nature Limited}
\label{Fig:Rabi_and_Autler_Townes_splitings}
\end{figure*}

Autler--Townes splitting (ATS) provides another mechanism to break reciprocity via on-resonance attenuation \cite{Kim2026,Peng2014What}. In atomic physics, Autler--Townes splitting corresponds to stationary states of the coupled atom-field system \cite{Boyd2020}. In phonon-mediated (p-ATS) photonic implementations, a launched traveling acoustic wave provides a unidirectional coupling between nondegenerate resonator modes. The acoustic wave must possess odd spatial symmetry in order to break the orthogonality of the resonator modes and enable a non-zero overlap integral between the acoustic and optical fields \cite{Kim2026}. When this overlap is nonzero, the bare optical levels split into an Autler–Townes doublet (see Fig.~\ref{Fig:Rabi_and_Autler_Townes_splitings}a).

The strong-coupling (p-ATS) regime is reached when the phonon-enhanced optomechanical coupling rate $G_{\mathrm{ph}}$ exceeds the optical loss rate, $G_{\mathrm{ph}}>\sqrt{\kappa_1\kappa_2}$ \cite{Sohn2021}, where $\kappa_{1,2}$ are the loss rates of the two optical modes. As a result, light at the unperturbed resonance frequencies $\omega_1$ and $\omega_2$ of the modes TE$_{00}$ and TE$_{10}$ experiences transparency in the phase-matched propagation direction, but complete on-resonance absorption in the counter-propagating direction. Figure ~\ref{Fig:Rabi_and_Autler_Townes_splitings}b shows a cross-section of the acousto-optic interaction region and the racetrack resonator  with an aluminium IDT actuator on a lithium-niobate-on-insulator platform \cite{Sohn2021}. The acoustic reflector ensures the transverse standing-wave characteristic for the acoustic excitation. 

\medskip
\noindent
Rabi-type schemes exploit nonreciprocal dissipation engineering by inducing strong, direction-dependent coupling between photonic bands that each support a continuum of modes \cite{Zhou2024}. In integrated implementations, an electromechanically driven, counter-propagating acoustic wave provides the coherent modulation that couples a spatially symmetric optical band to an antisymmetric band along the propagation direction. When the phase-matching condition is satisfied, this phonon-mediated interaction hybridizes the bands. The resulting eigenmodes exhibit an anticrossing with a characteristic splitting $2\beta$ (Fig.~\ref{Fig:Rabi_and_Autler_Townes_splitings}c). In the coupled-mode picture, optical power oscillates between the bands according to \cite{Zhou2024}
\begin{equation}
P_{1}^{\mathrm{out}} \approx P_{1}^{\mathrm{in}}\,e^{-\alpha L}\cos^{2}(\beta L),
\qquad
P_{2}^{\mathrm{out}} \approx P_{1}^{\mathrm{in}}\,e^{-\alpha L}\sin^{2}(\beta L),
\end{equation}
so that complete energy transfer from the first to the second band is achieved when $\beta L=\pi/2$, where $L$ is the device length. Because phase matching for this coherent conversion depends on the sign of the traveling-wave momentum, full interband transfer, and subsequent routing into a deliberately lossy channel, can be engineered for one propagation direction but not the other (Fig.~\ref{Fig:Rabi_and_Autler_Townes_splitings}d) \cite{Zhou2024}. An equivalent interpretation can be obtained from Floquet analysis, where dynamic modulation induces a direction-dependent Rabi splitting between photonic modes \cite{Shi2018}. As a result, incident light can be critically coupled to the lossy photonic modes only for one propagation direction. 

The Stark shift is the electric-field analogue of the magnetic Zeeman effect \cite{Autler1995}. A Stark effect of phonons was discovered recently \cite{Huang2024}. The AC Stark shift occurs when the applied microwave frequency that excites acoustic field is detuned far from the transition frequency of the photonic energy levels leading to a dispersive effect (Fig. \ref{Fig:Rabi_and_Autler_Townes_splitings}e) \cite{Zhang2019Electronically,Autler1995}. AC Stark shift  relies on phonon-mediated asymmetric coherent coupling, through Brillouin interaction, between CW and CCW modes in microring resonators (Fig. \ref{Fig:Rabi_and_Autler_Townes_splitings}e). The traveling acoustic wave produces off-resonant interactions that give rise to propagation-direction-dependent dissipation as described in detail in ref.\cite{Xu2025}. In contrast to intermodal conversion schemes, the off-resonant interaction acts as an effective magnetic field, leading to antisymmetric behavior for opposite propagation directions and inducing differential energy shifts that lift modal degeneracy (Fig.~\ref{Fig:Rabi_and_Autler_Townes_splitings}f). As a result, TR symmetry is broken. 
In contrast to interband transition, Autler–Townes splitting, and Rabi splitting, no intermodal phase-matching is required \cite{Xu2025}. 

\section{Nonlinear systems} \label{Sec:Nonlinear_Systems}

We have previously seen that EO space-time modulation of permittivity can be viewed as second order optical nonlinearity $\chi^{(2)}_{ijk}$ [Eq. (\ref{Eq:permittivity_tensor_time_modulation})]. AT can also be achieved using other second order optical processes including second-harmonic generation  
\cite{Boroviks2023,Toyoda2021,Sun2019,Stankov1992,Trevino1996,Lu2017}, sum/difference frequency generation 
\cite{Abdelsalam2020Linear,Rangelov2017}, and optical parametric amplification 
\cite{Liu2023Parametric}. Third order optical nonlinearities may also lead to AT including third-harmonic generation  
\cite{Dongrui2023,Guo2022,Kruk2017}, four-wave mixing 
\cite{Lin2025Optical,Shui2022,Wang2016}, Kerr effect \cite{Del2018,Sounas2018,Ren2020,White2023}, and Brillouin scattering \cite{Kim2017,Kim2015,Kittlaus2018,Ma2020}. 

Integrated optical isolation schemes based on third-order optical nonlinearities have attracted considerable interest because a non-zero $\chi^{(3)}_{ijkl}$ response can arise in both centrosymmetric and non-centrosymmetric materials. In particular, silicon has a centrosymmetric crystal lattice, so EO modulation based on second order optical nonlinearity in silicon photonics is highly limited \cite{Sirleto2023}; however, its Kerr nonlinearity is substantial \cite{Jin2020}. By contrast, lithium niobate and aluminium nitride exhibit robust $\chi^{(2)}_{ijk}$ nonlinearity \cite{Geng2024,Leidinger2015,Liu2019}, making them well suited for EO modulation.

For the class of local third-order nonlinear responses considered by Maker et al.~\cite{Maker1964} and Kaplan~\cite{Kaplan1983}, the nonlinear contribution to the permittivity can be written as
\begin{equation}
\Delta\varepsilon_{ik}
=
\varepsilon^{\mathrm{L}}\chi
\left[
\mathsf{B}\left(E_iE_k^*+E_i^*E_k\right)
+
(\mathsf{A}-\mathsf{B})I_{\Sigma}\delta_{ik}
\right],
\label{Eq:third_order_permittivity}
\end{equation} 
where $\varepsilon^{\mathrm{L}}$ is the linear permittivity of the medium, $\chi$ is a constant describing the nonlinear interaction for a single linearly polarized plane wave, $I_{\Sigma}=|E_x|^2+|E_y|^2+|E_z|^2$, and $\mathsf{A}$ and $\mathsf{B}$ are dimensionless constants characteristic of the particular mechanism of nonlinearity, real and satisfying $\mathsf{A}+\mathsf{B}=1$ in a lossless medium. The tensor is symmetric, so a third order nonlinearity does not violate the constitutive-symmetry condition of the Lorentz reciprocity theorem. It violates linearity instead since the permittivity depends on the field acting on it. The anisotropy of this response, governed by $\mathsf{B}$, produces light-induced birefringence and gyrotropy, the latter observed as self-rotation of the polarization ellipse of an elliptically polarized beam~\cite{Maker1964,Kaplan1983}. Kaplan identified four mutual arrangements of the polarizations of two counterpropagating beams for which the nonlienar electric displacement $\mathbf{D}_j^{\mathrm{NL}}$ remains parallel to $\mathbf{E}_j$, so that the polarization of each beam is unchanged on propagation: both linearly polarized and parallel, both linearly polarized and orthogonal, and both circularly polarized with co-rotating or counter-rotating fields. For beams of intensities $I_1$ and $I_2$ in any of these arrangements, the partial nonlinear susceptibilities are $\chi_j^{\mathrm{NL}}=\chi[\mathsf{C} I_j +(\mathsf{C}+\mathsf{D})I_{3-j}] \,\,\text{for} \, j=1,2$, and the nonlinear nonreciprocity as defined by Kaplan is~\cite{Kaplan1983}
\begin{equation}
\chi^{\mathrm{NR}}
=
\chi_1^{\mathrm{NL}}-\chi_2^{\mathrm{NL}}
=
\chi\mathsf{D}\left(I_2-I_1\right),
\label{Eq:nonlinear_nonreciprocity}
\end{equation}
where $\mathsf{C}$ and $\mathsf{D}$ are set by the mutual arrangement of the two polarizations. The nonreciprocity is therefore proportional to the intensity difference of the counterpropagating beams and vanishes when they are equal. Equation~(\ref{Eq:nonlinear_nonreciprocity}) is a bulk measure of a single material quantity, defined for two counterpropagating plane waves and manifested as a nonreciprocal phase difference $\Delta\phi^{\mathrm{NR}}=\chi^{\mathrm{NR}}Lk$~\cite{Ezekiel1982,Bergh1982}. In an integrated device the corresponding quantity is the transmission in each direction, which is not fixed by the material alone but depends on the structure and on how the ports are driven. Table \ref{tab:nonlinear_systems} summarizes representative state-of-the-art metrics for optical diodes.

\begin{table*}[t]
\caption{On-chip optical diodes based on optical nonlinearities}\label{tab:nonlinear_systems}%
\centering
\footnotesize
\setlength{\tabcolsep}{4pt}
\begin{tabular*}{170mm}{@{\extracolsep{\fill}}lccccccc@{}}
\toprule
Scheme & Platform & \shortstack{Contrast (dB)} & \shortstack{Bandwidth (nm)}
& \shortstack{Insertion loss (dB)} & Polarization & \shortstack{Footprint ($\mu$m)} & Ref.  \\
\midrule
Asymmetric Kerr effect                     & Si/SiO$_2$   & 20.3 & NR \textsuperscript{a} & 1.1  & TE           & NR      & \cite{Yang2020} \\
XPM (single ring)\textsuperscript{$\ast$}  & Si           & 15   
& NR & 1.96 & NR           & NR                         & \cite{Pan2022} \\
XPM (multiple ring)                        & Si$_3$N$_4$  & 35   & NR          & $\sim$5 & NR        & 200& \cite{White2023} \\
Saturable absorption\textsuperscript{$\ast$}  & Graphene/Si & 9 & 70 & $\le$ 6  & TE & 600            & \cite{Pitilakis2021} \\
Thermo-optic 
& Si/SiO$_2$ & 33.6 & 0.11  & NR  & NR & $400\times100$ & \cite{Liu2018On-chip} \\

\botrule
\end{tabular*}
\par\vspace{4pt}
\begin{minipage}{170mm}
\setlength{\parindent}{0pt}\raggedright\scriptsize
\textsuperscript{$\ast$}Simulation/theory only; not experimentally realized. \textsuperscript{a} NR denotes data not reported in the original reference.
\end{minipage}
\end{table*}

\subsection{Kerr nonlinearity-induced symmetry breaking}

\begin{figure*}[t!]
\centering
\includegraphics[width = 170mm]{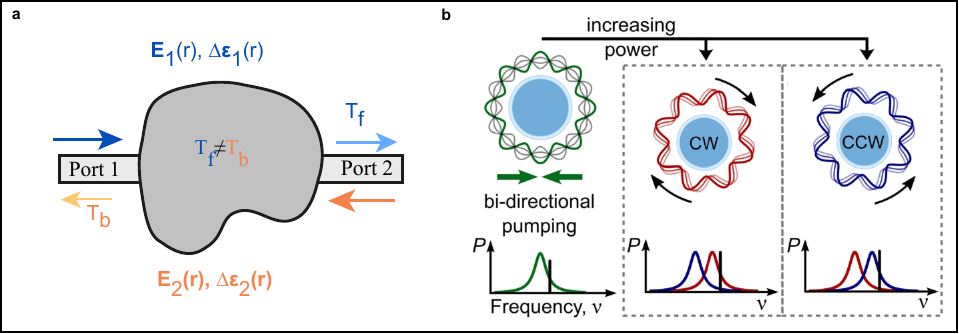}
\caption{Nonlinear non-reciprocity: structural asymmetry and spontaneous symmetry breaking.
\textbf{a} Nonlinear non-reciprocity in a geometrically asymmetric structure.
Forward and backward excitation establish different field distributions
$\mathbf{E}1(\mathbf{r})$ and $\mathbf{E}2(\mathbf{r})$, which through the
intensity-dependent index produce different permittivity profiles
$\Delta\varepsilon_1(\mathbf{r})$ and $\Delta\varepsilon_2(\mathbf{r})$, and hence
unequal transmittances $T\mathrm{f}\neq T\mathrm{b}$.
\textbf{b} Spontaneous symmetry breaking between the counterpropagating modes of
a Kerr resonator. For degenerate, co-polarized counterpropagating modes in an
isotropic, instantaneous Kerr medium, cross-phase modulation is twice as strong
as self-phase modulation. Bidirectional pumping therefore sustains a symmetric
standing-wave state only below a threshold power; above it, the state becomes
unstable and the system spontaneously selects clockwise or counter-clockwise
circulation, with the CW and CCW resonances splitting about the pump frequency
(black line, lower plots). Adapted from ref.~\cite{Del2017} under the terms of the
Creative Commons Attribution 4.0 International License.}
\label{Fig:Nonlinear_Systems_Fig_1}
\end{figure*}

In centrosymmetric dielectrics, the absence of a second-order susceptibility makes the third-order response the leading optical nonlinearity. The Kerr effect describes the resulting dependence of the refractive index on the local optical intensity \cite{Boyd2020,Shen1984}
\begin{equation}
\Delta \epsilon(\mathbf{r}) = \epsilon_0 \chi^{(3)} |E(\mathbf{r})|^2 .
\end{equation}
To leading order, this cubic response gives rise to an intensity-dependent refractive-index change,
\begin{equation}
    \Delta n = n_2 I, \qquad
    n_2 = \frac{3\chi^{(3)}}{4 n_0' n_0 \varepsilon_0 c},
\end{equation}
where $n_0 = n_0^{\prime} + i n_0^{\prime\prime}$ is the complex linear refractive index. In geometrically asymmetric nonlinear structures, self-phase modulation (SPM) can produce different internal field intensities under opposite-side illumination, leading to distinct permittivity profiles and AT (Fig.~\ref{Fig:Nonlinear_Systems_Fig_1}a) \cite{Yang2020,Xu2014,Sergei2011,Fan2018}. 

A related symmetry-breaking mechanism arises in dielectric MRRs through Kerr-mediated interactions between counterpropagating modes~\cite{Del2017}. For degenerate, co-polarized counterpropagating modes in an isotropic, instantaneous Kerr medium, cross-phase modulation (XPM) contributes twice the self-phase modulation (SPM) coefficient. The nonlinear refractive-index change experienced by each mode therefore depends on both its own power and that of the counterpropagating mode~\cite{Del2017,Kaplan1981}:
\begin{equation}
\Delta n_A = \frac{n_2}{A_{\mathrm{eff}}}\left(P_A + 2P_B\right), \qquad
\Delta n_B = \frac{n_2}{A_{\mathrm{eff}}}\left(P_B + 2P_A\right).
\end{equation}
Here, $A_{\mathrm{eff}}$ is the effective mode area, $n_2$ is the Kerr nonlinear-index coefficient, and $P_A$ and $P_B$ denote the powers of the two counterpropagating modes. Thus, within this model, a counterpropagating mode produces twice the Kerr-induced resonance shift of the corresponding mode produced by the same power in the mode itself~\cite{Kaplan1981}. At low excitation power, the CW and CCW modes remain degenerate and can interfere to form a standing wave, as illustrated in the left panel of Fig.~\ref{Fig:Nonlinear_Systems_Fig_1}b. As the power increases, the symmetric state with equal CW and CCW powers becomes unstable, and the system spontaneously selects either clockwise or counter-clockwise circulation, thereby breaking the symmetry between the two directions~\cite{Del2017}.

\begin{figure*}[t!]
\centering
\includegraphics[width = 170mm]{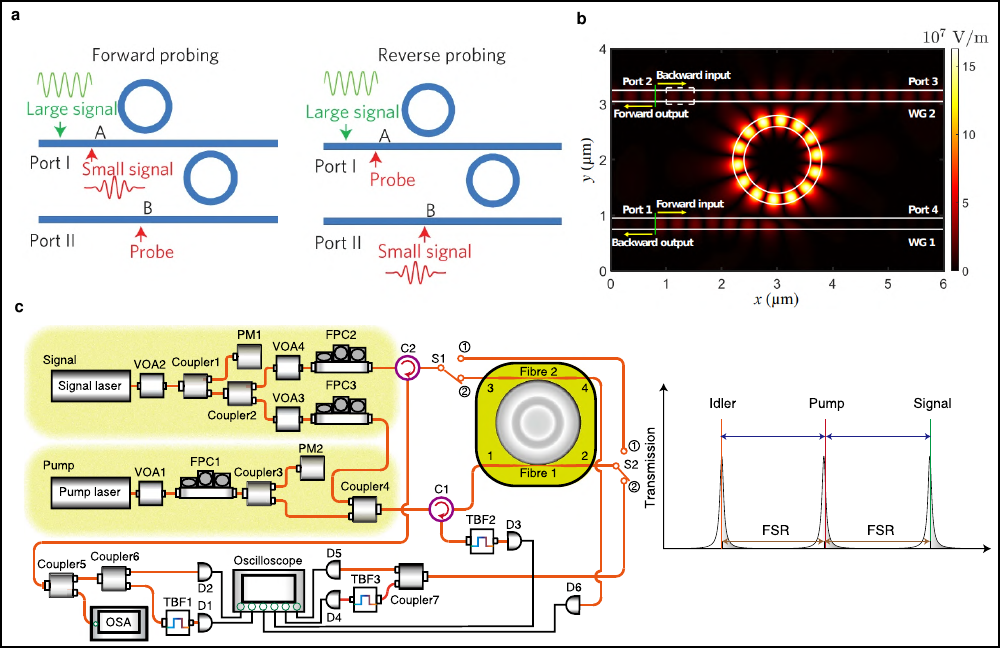}
  \caption{Dynamic reciprocity and its circumvention in nonlinear isolators.
  \textbf{a} Reciprocity analysis for a small probe in the presence of a large
  forward signal. The small-signal transmission from A to B (forward probing) equals
  that from B to A (reverse probing), so a nonlinear isolator cannot block a weak
  backward wave that coexists with a strong forward one.
  \textbf{b} Theoretical finite-difference
time-domain simulation of the proposal by Pan et al. with continuous light incident simultaneously on ports 1
  and 2 of a Kerr microring. Forward light entering port 1 is transmitted while
  backward light from port 2 is suppressed; green lines mark the source positions
  and the dashed rectangle an optical attenuator.
  \textbf{c} Experimental setup: signal and pump lasers are combined and coupled to
  the microtoroid through two tapered fibres, with forward and backward transmission
  monitored on an oscilloscope and an optical spectrum analyser. 
  \textbf{a} Reproduced with permission from ref.~\cite{Shi2015}. Copyright 2015
  Springer Nature Limited. \textbf{b} Adapted with permission from ref.~\cite{pan2022dynamic_archieve}, under the terms
of the Creative Commons Attribution 4.0 International License. \textbf{c} Adapted from ref.~\cite{Hua2016} under the terms of the Creative Commons Attribution 4.0 International License}
\label{Fig:Nonlinear_System_Fig_2}
\end{figure*}

A fundamental limitation of Kerr-based diodes is known as \emph{dynamic reciprocity}. Shi \emph{et al.} demonstrated that passive Kerr (or Kerr-like) ``isolators" obey a reciprocity constraint for small probe signals: in the presence of a strong forward signal, an arbitrarily weak backward-propagating perturbation cannot be completely suppressed (Fig. \ref{Fig:Nonlinear_System_Fig_2}a) ~\cite{Shi2015}. In other words, a Kerr-based isolator cannot fully reject backward noise when the forward field is continuously present. To overcome this constraint, practical schemes ensure that the backward wave does not behave as a small perturbation at the same frequency as the forward signal. The evidence is coupled-mode/FDTD simulation rather than an experimental demonstration, and robustness to an arbitrary weak backward perturbation, coherent phase, detuning, and backscattering remains to be tested.

Various strategies seeking to circumvent dynamic reciprocity have emerged. Pan \emph{et al.}  theoretically proposed that a Kerr nonlinear MRR fabricated from a material with both SPM and XPM 
can exhibit a form of \emph{dynamic nonreciprocity}~\cite{Pan2022}. In such a ring the SPM and XPM induced refractive-index changes differ depending on the light’s direction of propagation. Because co-propagating and counter-propagating fields contribute unequally to the total Kerr-induced resonance shift, the ring effectively acquires a \emph{chiral} nonlinear response~\cite{Pan2022}. In practice this implies that a strong forward wave shifts the resonance by a different amount than a backward wave of identical frequency and power. As a consequence, a continuous-wave input in one direction can remain on resonance and transmit efficiently, whereas an input in the opposite direction experiences detuning and is suppressed (Fig. \ref{Fig:Nonlinear_System_Fig_2}b). Pan \emph{et al.} propose that this mechanism enables on-chip optical diodes and even three-port circulators based on a single MRR operating under bidirectional excitation~\cite{Pan2022}. %Velocity-selective nonlinearity bypasses the constraint in atomic media \cite{Hu2021Passive}. 

Another route employs active or parametric nonlinear processes \cite{Hua2016,Wang2016}. For instance, Hua \emph{et al.} demonstrated a silicon-compatible diode based on \emph{four-wave mixing} in a high-$Q$ silica microtoroid resonator~\cite{Hua2016}. Because of the inversion symmetry of silica, the dominant parametric process is four-wave mixing through third-order nonlinearity. In this process, two pump photons with angular frequency $\omega_p$ and wave vector $\mathbf{k}_p$ are converted into one signal photon $(\omega_s, \mathbf{k}_s)$ and one idler photon $(\omega_i, \mathbf{k}_i)$ by satisfying the phase-matching conditions $2\omega_p = \omega_s + \omega_i$ (energy conservation) and $2\mathbf{k}_p = \mathbf{k}_s + \mathbf{k}_i$ (momentum conservation). The strong pump induces phase-matched parametric gain exclusively in the forward direction (Fig. \ref{Fig:Nonlinear_System_Fig_2}c). The forward signal is amplified inside the cavity, whereas the backward signal at the same frequency does not experience gain and instead undergoes attenuation. The directional pump together with phase-matching breaks TR symmetry and enables nonreciprocal transmission even under simultaneous bidirectional excitation. Because the backward wave does not act as a small-signal perturbation of the same nonlinear state, the device circumvents the dynamic reciprocity constraint. Experimentally, tens of decibels of isolation were achieved, demonstrating continuous-wave nonreciprocity enabled by nondegenerate frequency conversion~\cite{Hua2016}.

\subsection{Temperature-enhanced asymmetric transmission}
\begin{figure*}[t!]
\centering
\includegraphics[width = 170mm]{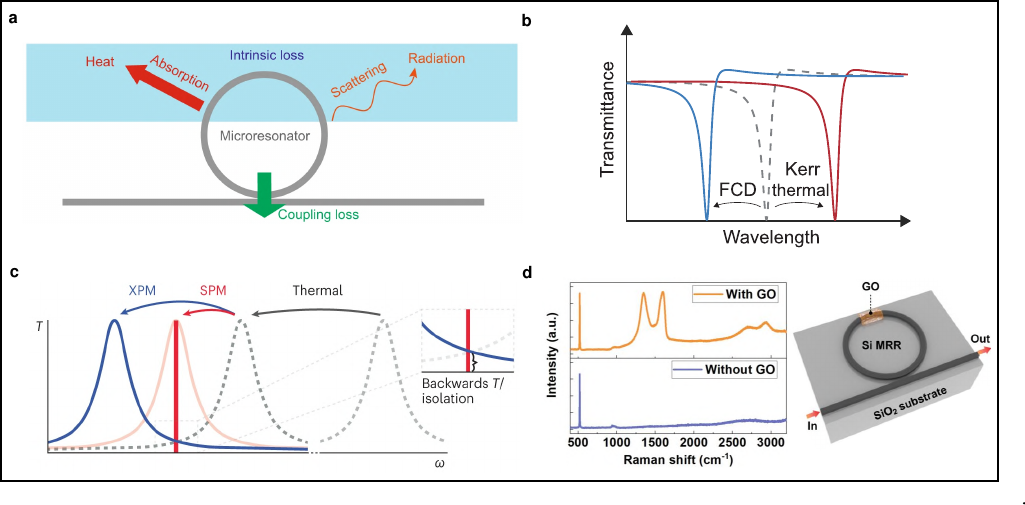}
  \caption{Loss channels and competing nonlinearities in microresonator isolators.
  \textbf{a} Optical loss channels of a high-$Q$ integrated microresonator: surface
  and bulk scattering, material absorption and radiation set the intrinsic loss rate
  (intrinsic quality factor $Q_0$), while bus-waveguide coupling adds the external
  loss characterised by $Q_\mathrm{e}$.
  \textbf{b} Competition between free-carrier dispersion (FCD), the Kerr effect and
  the thermo-optic effect in silicon: FCD blueshifts the resonance whereas the Kerr
  and thermo-optic contributions redshift it.
  \textbf{c} A strong unidirectional pump imposes self-phase modulation on the
  co-propagating mode and cross-phase modulation, twice as large, on the
  counter-propagating mode. The resulting resonance splitting, together with the
  thermo-optic shift, leaves near-unity forward transmission while suppressing
  backward transmission at the pump frequency (inset).
  \textbf{d} Silicon microring resonator coated with a two-dimensional graphene-oxide
  (GO) film (right), and Raman spectra of the chip before and after coating (left);
  the D and G peaks confirm successful integration of the GO film.
  \textbf{a} Reproduced from ref.~\cite{Gao2022} under the terms of the Creative
  Commons Attribution 4.0 International License. \textbf{b} Adapted with permission
  from ref.~\cite{Hong2024}. Copyright 2025 Wiley-VCH GmbH. \textbf{c} Adapted with
  permission from ref.~\cite{White2023}. Copyright 2022 Springer
  Nature Limited. \textbf{d} Adapted with permission from ref.~\cite{Moss2025_Preprint}.
Copyright 2025 Optica Publishing Group}
\label{Fig:Nonlinear_Systems_Fig3}
\end{figure*}

In silicon photonic resonators, diverse third-order nonlinearities arise including the Kerr effect, two-photon absorption (TPA), free-carrier effect, and the thermo-optic effect due to TPA. The Kerr coefficient $n_2$, TPA coefficient $\beta_{\rm{TPA}}$, and third-order susceptibility are related by \cite{Osgood2009,Lin2007,Wang2012Theoretical} \begin{equation}
\frac{\omega}{c} n_{2} - i \frac{\beta_{\rm{TPA}}}{2} = \frac{3 \omega}{4 n_0^{2} \varepsilon_{0} c^{2}} \chi^{(3)}
\end{equation} 
The imaginary part of $\chi^{(3)}$ represents TPA. In the optical band, TPA generates electron--hole pairs that modify both the refractive index and material loss, while the absorbed optical power and the associated free-carrier dissipation are converted into heat (Fig. \ref{Fig:Nonlinear_Systems_Fig3}a). The resulting temperature rise produces a slower, reciprocal thermo-optic refractive index change. In high Q-resonators, these mechanisms often compete: the free carrier effect typically blue-shifts the resonance frequency, whereas thermo-optic effect red-shifts it (Fig. \ref{Fig:Nonlinear_Systems_Fig3}b) \cite{Liu2018On-chip}. The net resonance shift is therefore determined by the balance between the free carriers dispersion and thermal contributions \cite{Wang2012Theoretical,Gao2022,White2023}. 

The local temperature rise can be written as \cite{Wang2012Theoretical}
\begin{equation}
\frac{d}{dt}\Delta T
= - \frac{\Delta T}{\tau_{\theta}}
+ \frac{1}{C\rho}
\left| \frac{dP_{\mathrm{TPA}}}{dV} \right|
\end{equation}
where $\tau_{\theta}$ is the thermal relaxation time, $C$ is the specific heat capacity, $\rho$ is the mass density, and $P_{\mathrm{TPA}}$ is the absorbed optical power density. In steady state  $\Delta T = \frac{\tau_{\theta}}{C\rho}
\left| \frac{dP_{\mathrm{TPA}}}{dV} \right|$, the change in refractive index due to thermo-optic effect is \cite{Komma2012} 
\begin{equation}
\Delta n_{\theta} = \kappa_{\theta}\Delta T,
\end{equation}
with $\kappa_{\theta} = \partial n/\partial T$ being the thermo-optic coefficient. As thermal relaxation is much slower than Kerr SPM, the two contributions can be separated experimentally by quasi-static transmission measurements and modulation-response spectroscopy \cite{Gao2022}. 

A strong pump signal launched into the resonator heats the cavity, producing a reciprocal thermo-optic increase in the refractive index and a corresponding decrease in the resonance frequency. At the same time, the intracavity Kerr nonlinearity produces SPM of the CW pump mode and XPM of the CCW mode. For the same intracavity pump intensity, the Kerr-induced resonance-frequency shift of the CCW mode is therefore twice the shift of the CW pump mode~\cite{White2023,Wang2012Theoretical}. The two directions therefore experience different total resonance shifts (Fig. \ref{Fig:Nonlinear_Systems_Fig3}c), allowing high transmission in the pump direction while substantially reducing transmission at the same frequency in the reverse direction. 

Over the last two decades, significant progress in on-chip integration of advanced
thermo-optic two-dimensional materials 
has opened up new avenues for efficient heat control and manipulation in integrated
photonic devices \cite{Nika2012two,hu20242d,li2020high}. Two-dimensional graphene
oxide integrated on silicon MRR exhibit more significant resonance wavelength
shifts in the response of both environmental and input power temperature variations as compared to the
uncoated MRRs (Fig. \ref{Fig:Nonlinear_Systems_Fig3}d)   \cite{Moss2025_Published}. An enhanced nonreciprocal AT in the hybrid MRRs is experimentally observed under much less input power as compared with the uncoated design highlighting the potential of films integrated onto silicon photonic platforms to implement high-performance thermo-optic devices \cite{Moss2025_Published}.

\section{Topological isolators}\label{sec:topological}

Topology studies conserved properties under continuous deformation \cite{Yin2025}. Topological photonics emerged from concepts originally developed to describe topological phases of matter in solid-state systems. Beyond the energy-band dispersion, the geometry of Bloch eigenstates in momentum space plays a crucial role in determining the properties of photonic systems. This geometry is characterized by topological invariants associated with the bands~\cite{Ozawa2019}. For a two-dimensional band that is separated from neighbouring bands, the relevant invariant is the band Chern number, $C_n$, obtained by integrating the Berry curvature $\Omega_n(\mathbf{k})$ over the Brillouin zone $C_n=\frac{1}{2\pi}\int_{\mathrm{BZ}}\Omega_n(\mathbf{k}),\mathrm{d}^2k\in\mathbb{Z}$, where $\Omega_n(\mathbf{k})$ acts as an effective magnetic field in momentum space~\cite{Raghu2008,Ozawa2019,Bisharat2021}. Time-reversal symmetry makes the Berry curvature odd in momentum, $\Omega_n(\mathbf{k})=-\Omega_n(-\mathbf{k})$, and therefore forces $C_n=0$ for each band~\cite{Wang2008PRL}. When time-reversal symmetry is preserved, other symmetry-protected or symmetry-related descriptors, such as pseudo-spin and valley degrees of freedom, can instead characterize topological phases~\cite{Lu2014,Ozawa2019}. For a given band gap, the relevant topological quantity for the edge spectrum is the gap Chern number, defined as the sum of the band Chern numbers of all bands below that gap. Bulk--boundary correspondence then relates the net number of chiral edge modes traversing the gap to the difference between the gap Chern numbers on the two sides of an interface~\cite{Wang2008PRL,Lu2014}.

\begin{figure*}[t!]
\centering
\includegraphics[width = 170mm]{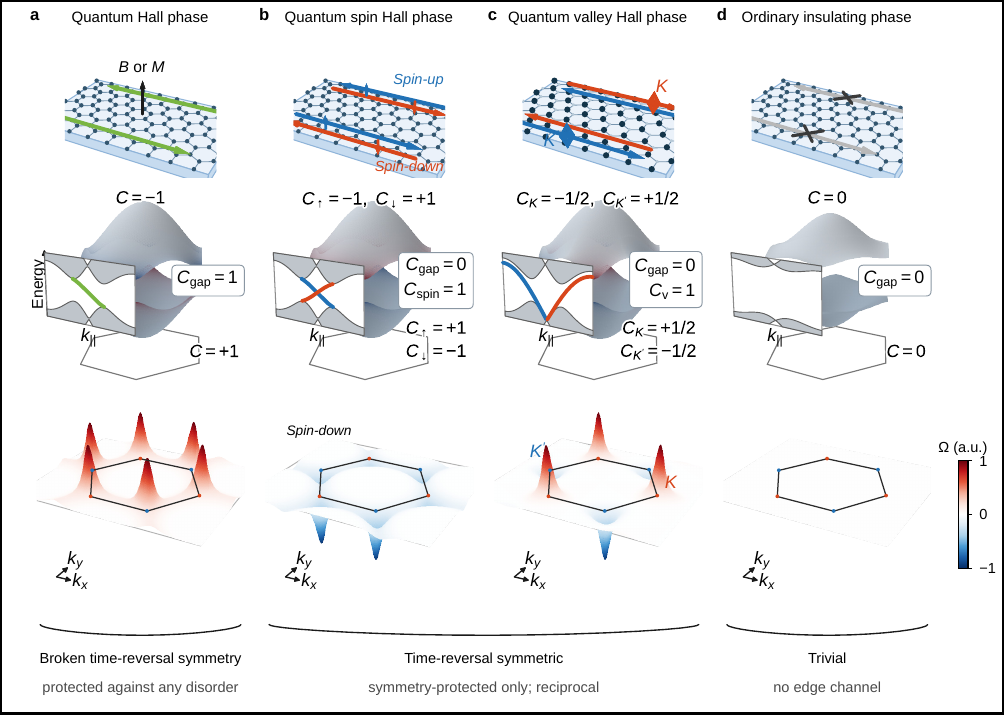}
\caption{\textbf{Topological phases that can and cannot carry light one way.}
Top row, honeycomb slab and edge channels; middle, bulk bands over the first
Brillouin zone; bottom, Berry curvature $\Omega$ of the lower band, hexagon
marking the zone.
\textbf{a} Quantum Hall phase. A magnetic bias breaks time-reversal symmetry;
$\Omega$ has the same sign at both valleys. The bands carry $C=+1$ below the gap
and $-1$ above, so $C_{\mathrm{gap}}=+1$ and one chiral channel crosses each edge.
\textbf{b} Quantum spin Hall phase. Each surface is a pseudo-spin doublet, one
Chern number per sector. Time reversal makes the sectors opposite,
$C_\uparrow=-C_\downarrow$, so $C_{\mathrm{gap}}=C_\uparrow+C_\downarrow=0$ while
$C_{\mathrm{spin}}=(C_\uparrow-C_\downarrow)/2=1$ gives one counter-propagating
pair per edge. The spin-down curvature is plotted.
\textbf{c} Quantum valley Hall phase, valley-locked kink states at a mass domain
wall. $C_K$ and $C_{K'}$ are $\Omega$ integrated around each valley, not band
Chern numbers; opposite in sign, reaching $\mp1/2$ only in the continuum limit.
Their difference $C_{\mathrm{v}}=C_K-C_{K'}=1$ counts the kink pair.
\textbf{d} Ordinary insulator. Inversion and time reversal force
$\Omega\equiv0$; no in-gap channel.
Labels give each band's invariants; the boxed number is their sum over the bands
below. Only the Chern number counts one-way channels: $C_{\mathrm{spin}}$ and
$C_{\mathrm{v}}$ count counter-propagating pairs and require their labelling
symmetry, so \textbf{b} and \textbf{c} stay reciprocal. Surfaces are tinted by
the sign of $\Omega$. Green, chiral channel; blue and red, negative and positive
group velocity, locked in \textbf{c} to $K'$ and $K$.}
\label{fig:phases}
\end{figure*}

Figure~\ref{fig:phases} contrasts the three routes to a one-way channel with
the trivial case. In the quantum Hall phase (Fig.~\ref{fig:phases}a) a
magnetic bias breaks TR symmetry; the two bands acquire equal and
opposite Chern numbers, and a single chiral channel with one sign of group
velocity traverses the gap on each edge. The quantum spin Hall phase
(Fig.~\ref{fig:phases}b) preserves TR symmetry and substitutes a
pseudo-spin: the two pseudo-spin sectors carry opposite Chern numbers, so
the total vanishes and each edge supports a counter-propagating,
spin-locked pair rather than a single channel. The quantum valley Hall
phase (Fig.~\ref{fig:phases}c) instead breaks inversion symmetry; the Berry
curvature forms hot spots of opposite sign at the two valleys $K$ and $K'$,
and a domain wall in the inversion-breaking term binds a pair of
valley-locked counter-propagating kink states. The ordinary insulator
(Fig.~\ref{fig:phases}d) supports neither in-gap channels nor a nonzero
Berry curvature. The bulk dispersions of the three nontrivial cases are nearly indistinguishable, and are exactly identical for the two pseudo-spin sectors of Fig.~\ref{fig:phases}b, because the phases differ in the sign of the Dirac mass at the two valleys and not in its magnitude. Only the
geometry of the eigenstates, plotted in the bottom row, separates them.

\subsection{Magneto-optical topological systems}
\begin{figure*}[t!]
\centering
\includegraphics[width = 170mm]{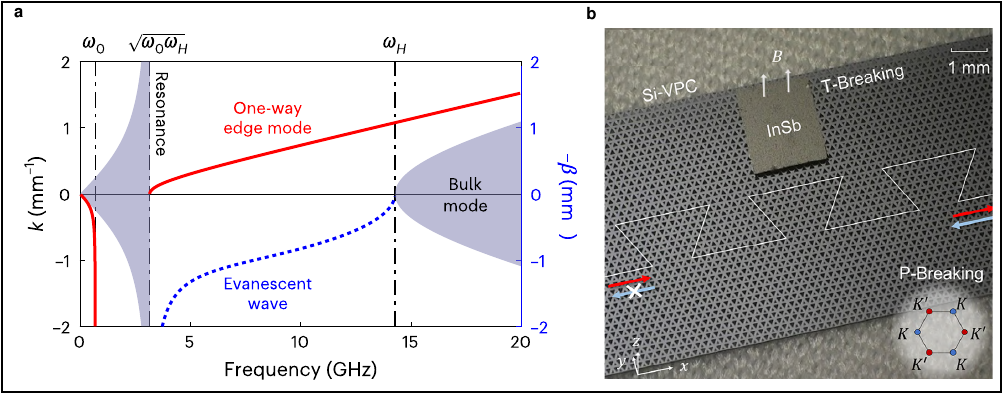}
\caption{\textbf{Topological non-reciprocity from ferrite microwave isolators to
on-chip terahertz photonics.}
\textbf{a} One-way edge-mode band diagram of a magnetized ferrite. The topological
bandgap spans the magnon resonance $\sqrt{\omega_{0}\omega_{H}}$ to the bias frequency
$\omega_{H}$ ($\omega_{0}$: Larmor frequency of the internal field); the in-gap edge
mode (red, left axis) is one-way, leaving only an evanescent backward channel (blue
dotted, right axis). 
\textbf{b} Terahertz silicon valley photonic crystal (Si-VPC) chip with magneto-optical
InSb. The field $B$ (T-breaking) and unequal triangular holes (P-breaking) render the
forward (red) and backward (blue, crossed) valley edge modes non-reciprocal. Inset:
inequivalent $K$ and $K'$ valleys. \textbf{a} Reproduced with permission from ref.~\cite{WangLu2025},
Copyright 2025 Springer Nature. \textbf{b} Reproduced under the CC BY-NC-ND 4.0 licence from
ref.~\cite{Jia2025}, Copyright 2025 published by Wiley-VCH GmbH.
}
\label{fig:topological_MO}
\end{figure*}

In MO photonic crystals, a magnetic bias can open bandgaps with a nonzero \emph{gap Chern} number~\cite{Raghu2008,Wang2008PRL}. At the interface between topologically distinct media, this band topology
gives rise to edge modes that propagate in a single direction. These modes
are robust against back-reflection from disorder as long as the bandgap
remains open, no counter-propagating edge channel is available, and the
disorder does not couple the edge mode to radiative continua outside the
crystal. This principle was proposed for photonic analogs of quantum-Hall
edge states~\cite{Raghu2008,Wang2008PRL} and was experimentally demonstrated
in microwave gyromagnetic photonic crystals~\cite{Wang2009Nature}. A one-way edge channel is not automatically a two-port isolator. Topology can protect propagation along a selected edge, but isolation also requires backward-injected light to be removed from or redirected away from the source port. This can be achieved by combining topological mode selection with loss, filtering, or additional routing channels. The clearest demonstration of this principle is a ferrite stripline in which the backward direction supports no propagating edge mode at all, so an absorber can be distributed along the return path without changing the forward signal. The residual backward signal is then set only by the evanescent tail across the topological bandgap (Fig.~\ref{fig:topological_MO}a) \cite{WangLu2025}. Wang and Lu showed analytically that this limits the topological isolation only by the decay constant and the length of the waveguide, giving $200\,\mathrm{dB}$ within a single-wavelength-sized device, and measured more than $100\,\mathrm{dB}$ of isolation with $1\,\mathrm{dB}$ insertion loss in the microwave region \cite{WangLu2025}.
El-Ganainy and Levy proposed a related mechanism in a finite, dissipative Su--Schrieffer--Heeger waveguide array \cite{ElGanainyl2015Optical}.

A recent approach combines valley-Hall photonics with MO bias. Valley-Hall waveguides break spatial inversion symmetry and support counter-propagating edge modes associated with opposite valleys (Fig. \ref{fig:phases}c). These modes can suppress intervalley backscattering at properly designed interfaces and bends, but the structure remains reciprocal unless TR symmetry is also broken. The strength of that suppression at optical frequencies is contested~\cite{Rosiek2023}. Jia \textit{et al.} integrated an InSb film with a silicon valley-Hall platform operating in the terahertz band and applied an out-of-plane magnetic bias of $0.632\,\mathrm{T}$ (Fig. \ref{fig:topological_MO}b)~\cite{Jia2025}. Without the MO film, the CW and CCW cavity modes are reciprocal. Under magnetic bias, their resonance frequencies and loss rates become different~\cite{Jia2025}. As a result, the \emph{critical-coupling condition} is reached for one circulation direction but not for the other, producing strong asymmetric transmission at the cavity resonance. The experiment reported a maximum isolation of $64.3\,\mathrm{dB}$ with $2.6\,\mathrm{dB}$ chip loss~\cite{Jia2025}. % from a cavity of footprint $6.4\times2.5\,\lambda^{2}$, and non-reciprocity across the whole topological band from $0.325$ to $0.345\,\mathrm{THz}$; an optical pump at $525\,\mathrm{nm}$ was used to tune the isolation ratio over a range of up to $48\,\mathrm{dB}$.

\subsection{Time-modulated topological systems} \label{sec:topolocial_time_modulated}
\begin{figure*}[t!]
\centering
\includegraphics[width=170mm]{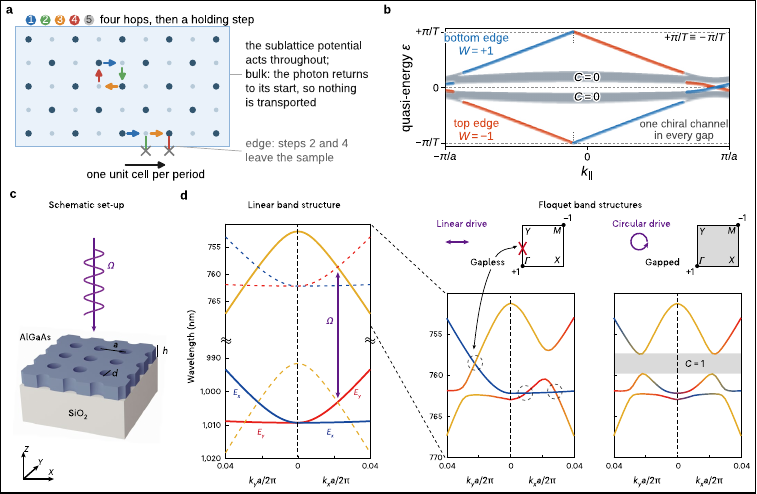}
\caption{Winding number and Floquet Chern insulators of light.
\textbf{a} One cycle of a periodically driven lattice, in which the
couplings are switched on in turn around each plaquette. A photon in the
bulk completes a closed loop and returns to the site it started from, so
nothing is transported and every band Chern number vanishes, while at the
boundary two hops are blocked and the photon advances by one unit cell per
period. \textbf{b} Quasi-energy spectrum of a strip in the anomalous Floquet
phase. The quasi-energy zone closes on itself, so $+\pi/T$ and $-\pi/T$ are
the same quasi-energy, and the chiral edge branch wraps it once as
$k_\parallel$ crosses the Brillouin zone: the winding number is $W=1$ even
though both bulk bands have $C=0$. Grey marks the bulk bands, and red and
blue mark the channels bound to the two boundaries of the strip.
\textbf{c} An AlGaAs photonic crystal on SiO$_2$ driven at frequency
$\Omega$. \textbf{d} Its linear band structure, and the Floquet bands under
a linearly polarized drive, which remain gapless, and under a circular
drive, which opens a gap of Chern number $C=1$. \textbf{c}, \textbf{d}
Adapted with permission from ref.~\cite{Jin2025}, copyright 2025, Springer
Nature}
\label{fig:time_mod}
\end{figure*}

Modulating a structure in time is the same tool that produced the synthetic gauge fields of Section~\ref{sec:synthetic}. A modulation whose phase varies from site to site produces a synthetic magnetic flux on the photons \cite{Fang2012}. Periodic driving changes the structure of the band diagram as real magnetic fields would. Under a drive of period $T$, the analogue of the energy spectrum is the Floquet spectrum and eigenstates are no longer labeled by an energy but by a quasi-energy $\varepsilon$ which is unique up to an integer multiple of $\omega = 2 \pi/T$. The quasi-energy zone closes on itself: $+\pi/T$ and $-\pi/T$ are the same quasi-energy (Fig.~\ref{fig:time_mod}b).  Despite the similarities between static and periodically driven systems, Chern numbers employed in the static case do not give a full characterization of the topological properties of periodically driven systems. Because the zone is a circle, an edge channel can disperse upwards, leave the top of the zone, re-enter at the bottom and so traverse \emph{every} gap in the spectrum. %---behaviour with no static counterpart. 
The number of times the edge branch wraps the zone as the momentum crosses the Brillouin zone is the \emph{winding number} $W$, and unlike a Chern number it is a property of the full evolution operator over one drive period rather than of any single band \cite{Rudner2013}. Its origin is easiest to see in real space (Fig.~\ref{fig:time_mod}a): when the drive cycles the couplings around each plaquette in turn, a photon in the bulk completes a closed loop and returns to the site it started from, so nothing is transported and every band Chern number vanishes. At a boundary, part of that loop is missing, and the photon instead advances by one unit cell per period. The result is the anomalous Floquet phase: $C=0$ in every bulk band, yet $W=1$ and a one-way channel in every gap (Fig.~\ref{fig:time_mod}b). In the case of static Hamiltonians,  such edge modes are only produced at the boundaries of systems characterized by nonzero Chern numbers (Fig. \ref{fig:phases}). 

The most advanced chip-scale realization is electronic rather than photonic. Nagulu \textit{et al.} implemented a $4\times4$ lattice of quasi-electrostatic, helicoidally rotating switched-capacitor circulators in a $45\,\mathrm{nm}$ silicon on-insulator CMOS process, occupying $3.5\times3.4\,\mathrm{mm^2}$ and clocked at $f_m=500\,\mathrm{MHz}$ \cite{Nagulu2022}. Isolation accumulates at roughly $20\,\mathrm{dB}$ per unit cell along the edge and saturates at a chip-level parasitic floor of $60$-$70\,\mathrm{dB}$. Carrying this to optical frequencies is hard because the modulation rate must be a significant fraction of the topological gap, which for a photonic crystal means coherent terahertz-rate driving distributed over many subwavelength elements. Two partial routes are emerging. In the first, the drive is itself optical. Jin \textit{et al.} drove an AlGaAs photonic crystal with an intense mid-infrared field near $3\,\mathrm{\mu m}$, coupling its photonic bands through the second-order nonlinearity of the material, and reconstructed the Floquet bands from transient sum-frequency generation (Fig.~\ref{fig:time_mod}c,d) \cite{Jin2025}. The spectrum stays gapless under a linearly polarized drive and acquires a $247\,\mathrm{GHz}$ gap under a circular one, which the theory assigns a Chern number of unity. %---the signature of a drive that breaks TR symmetry. 
This topological phase exists only while the pump is on. No isolation has been reported. In the second route the modulation is a traveling wave rather than a rotation, which is far easier to distribute. Serra and Silveirinha showed that a linear-momentum bias can still produce a gap Chern number of $-1$, provided each unit cell carries an internal angular-momentum texture, while cautioning that the unbounded spectrum of a photonic crystal makes Floquet gaps ill-defined in general \cite{Serra2026}.  Among propagation-as-time Floquet analogues, only genuine temporal modulation can break Lorentz reciprocity. Photonic Floquet insulators built from helical waveguide arrays, in which the propagation axis substitutes for time, are reciprocal and cannot isolate \cite{Rechtsman2013,Nagulu2022}. For now, time-modulated topological isolators remain a microwave and acoustic technology; the outstanding obstacles are phase-synchronized distribution of the drive across the lattice, harmonic conversion loss, and modulation power \cite{Ma2025}.

\subsection{Nonlinear topological systems}
\begin{figure*}[t!]
\centering
\includegraphics[width=170mm]{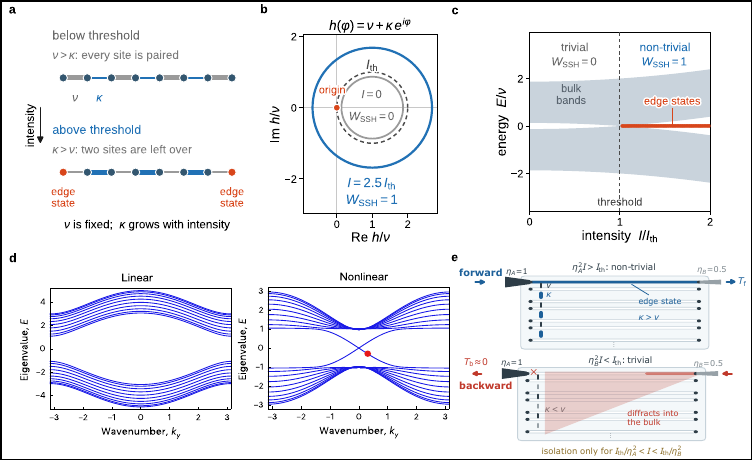}
\caption{Self-induced topological transitions in nonlinear lattices.
\textbf{a} A Su--Schrieffer--Heeger chain, drawn in the dimerised limit. The
intracell coupling $\nu$ is fixed and the intercell coupling $\kappa$ grows
with intensity: below threshold every site is paired, above it the two end
sites are left over and each carries an edge state. \textbf{b} The same
statement as a winding number. The curve $h(\varphi)=\nu+\kappa e^{i\varphi}$
leaves the origin outside itself below threshold, passes through it at
threshold, and encircles it above, so $W$ steps from $0$ to $1$. \textbf{c}
Bulk bands of the Bloch Hamiltonian frozen at each intensity, with the in-gap
levels of a finite chain: the gap closes at the threshold and reopens with two
zero-energy edge states. \textbf{d} Ribbon spectra of a nonlinear Chern
insulator below and above threshold, the in-gap branches appearing only in the
nonlinear regime. \textbf{e} The nonlinear topological diode~\cite{Zhou2017}: unequal port
couplers deliver an above-threshold intensity $I$ into the dimerised array
only in the forward direction, where a self-induced edge state transmits the
signal, while backward the lattice stays trivial and the light diffracts into
the bulk. \textbf{d} Adapted
with permission from ref.~\cite{Sone2024}, copyright 2024, Springer Nature.}
\label{fig:nonlinear_topological}
\end{figure*}

Which topological phase a lattice occupies is decided by a competition between couplings. In the Su--Schrieffer--Heeger (SSH) chain, the simplest case, each unit cell holds two sites joined by an intracell coupling $\nu$, and neighbouring cells are joined by an intercell coupling $\kappa$. If $\nu>\kappa$ every site is paired inside its own cell and the chain is trivial. When $\kappa>\nu$, the stronger intercell coupling produces an unpaired site at each boundary, giving rise to a pair of zero-energy edge states and signalling the topologically nontrivial phase (Fig.~\ref{fig:nonlinear_topological}a). The off-diagonal element of the Bloch Hamiltonian can be written as $h(\varphi)=\nu+\kappa e^{i\varphi}$, where $W$ is the number of times $h$ encircles the origin as $\varphi$ crosses the Brillouin zone, and it equals $0$ or $1$ according to which coupling is the larger (Fig.~\ref{fig:nonlinear_topological}b). 

%In a driven finite chain the transition is local, spreading outwards from the excitation point rather than converting the lattice at once. They then realized the effect in a radio-frequency ladder of coupled $LC$ resonators near $100\,\mathrm{MHz}$, in which back-to-back varactor diodes make the intracell coupling saturable so that it falls \emph{through} the fixed intercell value as the drive rises---the same transition reached from the opposite side (Fig.~\ref{fig:nonlinear_topological}e) \cite{Hadad2018}. The edge state emerges above a measured input power of $19\,\mathrm{dBm}$ ($80\,\mathrm{mW}$).
%Maczewsky \textit{et al.} realized the optical analogue in a laser written, Floquet-driven Kerr waveguide lattice in fused silica: raising the input peak power past $0.8\,\mathrm{MW}$ lifts the per step transfer ratio from $44\%$ through the $50\%$ topological criterion and on to $73\%$, driving the lattice into an anomalous Floquet phase---winding number $W=1$ with every Chern number zero---and transiently switching on a chiral edge channel, which was observed to round a corner of the lattice without scattering \cite{Maczewsky2020}.
 
A Kerr medium makes one of the two couplings a function of intensity, so the chain can start trivial, close its gap at the power where the couplings become equal, and reopen it with an edge state inside (Fig.~\ref{fig:nonlinear_topological}c). %Nothing about the structure changes; only the power flowing through it does. 
Hadad \textit{et al.} predicted this self-induced transition in an SSH chain
whose \emph{intercell} coupling grows with intensity through a Kerr term, so
that $W$ switches from $0$ to $1$ at a threshold modal amplitude
%$I_{\mathrm{th}}=\sqrt{(\nu-\kappa_0)/\alpha}$ \cite{Hadad2016} 
$I_{\mathrm{th}}$. The nonlinear Chern insulator is reciprocal as the edge states appear in both directions  (Fig.~\ref{fig:nonlinear_topological}d). Turning a power threshold into a diode requires an asymmetry that only the
forward direction crosses. Zhou \textit{et al.} analysed three such designs: a
nonlinear SSH waveguide array with unequal input and output couplings, a
nonlinear Haldane array in which the input and output sit on different sites
so that the lattice itself supplies the asymmetry, and an all-on-chip lattice
of coupled rings \cite{Zhou2017}. For the SSH array they bounded the
attainable isolation ratio by the inverse of the linear transmittance
$\mathrm{IR}_{\max}\approx1/T(0)$ (Fig.~\ref{fig:nonlinear_topological}e). This means that a lattice that transmits well
cannot isolate strongly. %, because the contrast is the difference between thelinear and the nonlinear state. 
In the steady-state ring lattice the transition produces a genuine discontinuity rather than a smooth threshold in the power-dependent transmittance, and the primary jump survives ensemble averaging over weak coupler-phase disorder. %Full-wave designs are now appearing. %Kilic \textit{et al.} loaded an all-dielectric checkerboard topological interface with deliberately asymmetric defects, shifting its leaky edge resonance by about $0.08\,\mathrm{THz}$ between the two directions at $1.32\,\mathrm{GW\,cm^{-2}}$, though at only ${\sim}45\%$ peak transmission; halving the defect raises the transmission to $96\%$ but narrows the forward--backward splitting to roughly $12.5\,\mathrm{GHz}$, an explicit trade-off between contrast and insertion loss \cite{Kilic2025}.
 
There are two important limitations when interpreting these results. Firstly, band topology is a property of a linear band structure, and the
invariants quoted above are evaluated on a Bloch Hamiltonian frozen at a given
intensity. The authors of the original proposal were explicit that a proper
definition for a nonlinear lattice was an open question
\cite{Hadad2016,Hadad2018}. That gap has since been partly closed. Sone
\textit{et al.} defined a nonlinear Chern number from the Berry curvature of
the fixed-amplitude solutions of the nonlinear eigenvalue problem. It reduces
to the ordinary Chern number in the linear limit and predicts phases that are
adiabatically disconnected from it, with edge states that appear only above a
threshold amplitude (Fig.~\ref{fig:nonlinear_topological}d) \cite{Sone2024}.
%It is not a general resolution: it presumes a conservative system with $U(1)$ and translation symmetry, an open gap, and Bloch solutions whose amplitude is independent of wavenumber, and bulk--boundary correspondence is recovered only in the joint continuum and thermodynamic limits. 
Secondly, nonlinearity can degrade the protection it creates. Michen and Budich showed that a wave packet
traveling in a chiral edge channel, meeting a second, locally circulating
packet of individually sub-threshold amplitude, can be irreversibly deflected
into the bulk when the two interfere constructively past the threshold that
sustains the topological phase; their relative phase tunes between unimpeded
transmission, deflection, and even amplification of the edge current
\cite{Michen2023}. %The circulating packet acts as an effectively time-dependent impurity, against which topological protection carries no guarantee---a scattering channel with no linear counterpart.

Topology does not lift dynamic reciprocity 
  \cite{Shi2015}.  Band topology fixes where and at what power the self-induced transition occurs; it does not make the linearized problem nonreciprocal. The one topological device tested under simultaneous bidirectional excitation shows that the forward and backward output contrast survives only when the counter-propagating field is three orders of magnitude weaker \cite{Kilic2025}. 
  
A distinct and more promising route is to let the nonlinearity create the bias rather than the asymmetry, leaving the signal itself linear. Duggan \textit{et al.} proposed pumping a silicon photonic crystal with two counter handed, frequency-detuned circularly polarized beams, whose interference is a linearly polarized field whose axis rotates at half the detuning; through $\chi^{(3)}$ this rotates the principal axes of the permittivity, imparting a uniform synthetic angular-momentum bias that breaks TR symmetry. It gaps the three defect-band degeneracies --two at $\Gamma$ and one at $K$, each a degeneracy of the two circular polarizations on a site-- leaving Chern numbers $(-1,0,0,1)$ \cite{Duggan2020}. Such a device escapes dynamic reciprocity because the permittivity the signal sees is not a scalar but a tensor whose axes rotate in time, so the Lorentz reciprocity theorem does not hold---the same reason Brillouin-based isolators are exempt \cite{Shi2015}. Absent one of these mechanisms, a self-induced topological device is more accurately described as a power-dependent limiter or switch than as a diode.
 
\section{Non-Hermitian systems}
\label{Sec:Dissipative_nonHermitian_Systems}
\begin{figure*}[t!]
\centering
\includegraphics[width=170mm]{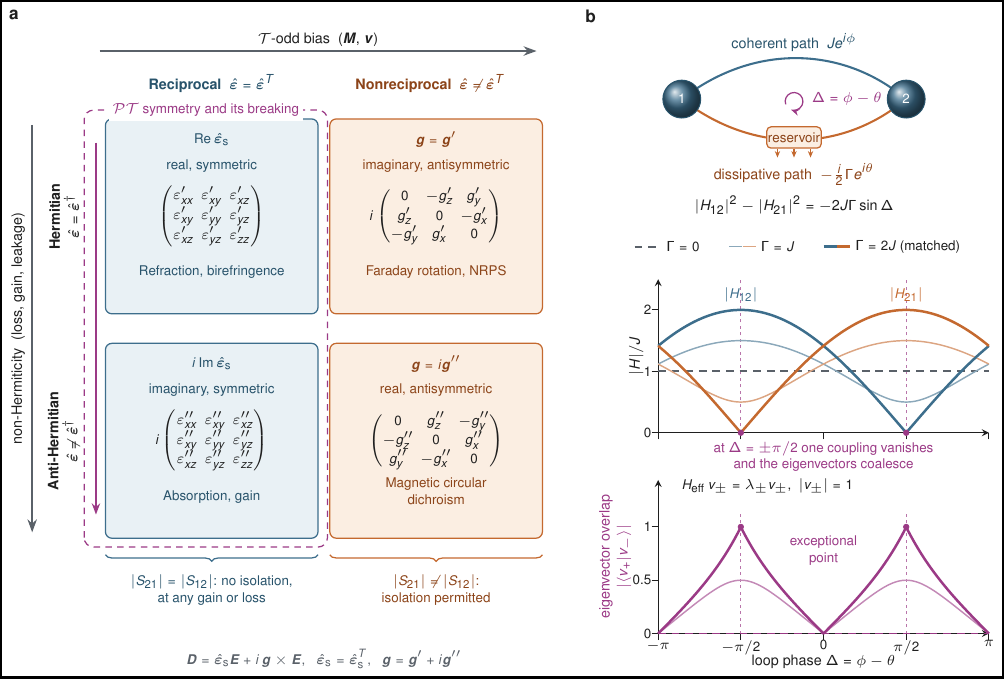}
\caption{\textbf{Non-Hermiticity and nonreciprocity are set by different symmetries of the same tensor.} \textbf{a} Permittivity of a linear, time-invariant, non-magnetic medium, written as $\mathbf{D}=\hat{\bm{\varepsilon}}_{\mathrm{s}}\bm{E}+i\bm{g}\times\mathbf{E}$ with $\hat{\bm{\varepsilon}}_{\mathrm{s}}=\hat{\bm{\varepsilon}}_{\mathrm{s}}^{T}$. Hermiticity (rows) and reciprocity (columns) are independent conditions; both blocks reduce to diagonal form in the principal axes of an isotropic or uniaxial medium. Loss and gain occupy only $i\operatorname{Im}\hat{\bm{\varepsilon}}_{\mathrm{s}}$, which is symmetric, so $|S_{21}|=|S_{12}|$ however strong they are; only a $\mathcal{T}$-odd bias reaches the right-hand column. Because $\mathcal{PT}$ symmetry constrains the spatial distribution of $\hat{\bm{\varepsilon}}$ rather than its structure at a point, it acts within the left-hand column alone (dashed). \textbf{b} Two modes coupled coherently, $Je^{i\phi}$, and dissipatively through a lossy reservoir, $-\tfrac{i}{2}\Gamma e^{i\theta}$; only the loop phase $\Delta=\phi-\theta$ enters. Upper, $|H_{12}|$ and $|H_{21}|$ from $|H_{12}|^{2}-|H_{21}|^{2}=-2J\Gamma\sin\Delta$, which coincide for every $\Delta$ when $\Gamma=0$. Lower, the overlap of the normalised eigenvectors $v_{\pm}$ of $H_{\mathrm{eff}}$, reaching unity at $\Delta=\pm\pi/2$ for $\Gamma=2J$; for degenerate, equally damped modes these are second-order exceptional points, where $H_{\mathrm{eff}}$ ceases to be diagonalisable.} \label{fig:nonhermitian_symmetry_concept}
\end{figure*}

Integrated photonic devices are open systems. Radiative leakage, material absorption and optical gain can render the effective operator governing the modal spectrum non-Hermitian~\cite{ElGanainy2018,Dong2025,Chang2014}. It is therefore natural to ask whether non-Hermiticity can make a device nonreciprocal. It cannot by itself, because Hermiticity and reciprocity impose distinct symmetry conditions. For the local, linear, time-invariant media considered here, material non-Hermiticity corresponds to $\hat{\bm{\varepsilon}}\neq\hat{\bm{\varepsilon}}^{\dagger}$, whereas nonreciprocity requires $\hat{\bm{\varepsilon}}\neq\hat{\bm{\varepsilon}}^{T}$~\cite{Landau1984,Caloz2018,Asadchy2020}. Separating $\hat{\bm{\varepsilon}}$ into its symmetric and antisymmetric parts,
\begin{equation}
\bm{D}=\hat{\bm{\varepsilon}}_{\mathrm{s}}\bm{E}+i\,\bm{g}\times\bm{E},
\qquad
\hat{\bm{\varepsilon}}_{\mathrm{s}}=\hat{\bm{\varepsilon}}_{\mathrm{s}}^{T},
\label{Eq:NH_gyration}
\end{equation}
where $\bm{g}$ is the gyration vector~\cite{Zvezdin1997}. The real and imaginary parts of these quantities have distinct physical consequences: $\mathrm{Re}\hat{\bm{\varepsilon}}_{\mathrm{s}}$ describes reciprocal refraction and birefringence, whereas $\mathrm{Im}\hat{\bm{\varepsilon}}_{\mathrm{s}}$ describes reciprocal absorption or gain; by contrast, $\mathrm{Re}\bm{g}$ gives rise to Faraday rotation and nonreciprocal phase shifts, whereas $\mathrm{Im} \bm{g}$ gives rise to magnetic circular dichroism~\cite{Zvezdin1997}. Thus, absorption and gain can make the material response non-Hermitian without breaking reciprocity, because they modify the symmetric part of the permittivity and preserve $\hat{\bm{\varepsilon}}=\hat{\bm{\varepsilon}}^{T}$ (Fig.~\ref{fig:nonhermitian_symmetry_concept}a). Under TR, a magnetization $\bm{M}$ and a drift or drive velocity $\bm{v}$~\cite{Landau1984,Caloz2018}, the Onsager--Casimir relations $\varepsilon_{ij}\!\left(\omega;\bm{M},\bm{v}\right)
=\varepsilon_{ji}\!\left(\omega;-\bm{M},-\bm{v}\right)$
~\cite{Onsager1931a,Onsager1931b,Casimir1945} makes the even piece symmetric and the odd piece antisymmetric. A $\mathcal{T}$-odd bias is therefore necessary for $\bm{g}\neq0$~\cite{Asadchy2020,Caloz2018}, absorption does not qualify. 

In coupled-mode theory, the amplitudes $\bm{a}$ of a set of interacting modes obey $\dot{\bm{a}}=-\mathrm{i}H_{\mathrm{eff}}\bm{a}$. Non-Hermiticity corresponds to $H_{\mathrm{eff}}\neq H_{\mathrm{eff}}^{\dagger}$. Reciprocity, by contrast, imposes a distinct constraint on the coupling amplitudes. Individual coupling phases depend on the phase convention chosen for the modal
amplitudes, whereas the reciprocity condition itself is gauge
invariant~\cite{Suh2004,Ranzani2015}. Its violation produces a nonreciprocal
phase, a nonreciprocal amplitude, or both~\cite{Ranzani2015}. Asymmetric coupling constitute a distinct and stronger form of directional coupling, characterized by 
\begin{equation}
\left|\left(H_{\mathrm{eff}}\right)_{mn}\right|
\neq
\left|\left(H_{\mathrm{eff}}\right)_{nm}\right|
\label{Eq:NH_Heff}
\end{equation}
which is incompatible with Hermiticity, since $H_{\mathrm{eff}} = H_{\mathrm{eff}}^{\dagger}$ requires $\left(H_{\mathrm{eff}}\right)_{nm} = \left(H_{\mathrm{eff}}\right)_{mn}^{*}$ and hence equal coupling magnitudes.

\subsection{Dissipation engineering}
\label{Subsec:Dissipation_gain_loss_engineering}
 
In \emph{dissipative reservoir engineering}, a lossy auxiliary degree of
freedom mediates an effective coupling between two optical
modes~\cite{Poyatos1996,Gardiner1993,Carmichael1993}.  Metelmann and Clerk showed that matching such a dissipative channel to a coherent one renders the transfer directional, generalising the structure of cascaded quantum systems to a wide class of interactions~\cite{Metelmann2015}. For two modes the condition follows from Eq.~(\ref{Eq:NH_Heff}). Writing the coherent coupling as $Je^{i\phi}$ and the
reservoir-mediated coupling as $-\tfrac{i}{2}\Gamma e^{i\theta}$,
\begin{equation}
\left|\left(H_{\mathrm{eff}}\right)_{12}\right|^{2}
-\left|\left(H_{\mathrm{eff}}\right)_{21}\right|^{2}
=-2J\Gamma\sin\!\left(\phi-\theta\right),
\label{Eq:NH_directionality}
\end{equation}
a two-mode form of the general result~\cite{Ranzani2014,Ranzani2015,Metelmann2015}. Directionality therefore requires all three of a coherent path ($J\neq0$), a dissipative path ($\Gamma\neq0$) and a nonzero relative phase $\phi-\theta$, which is the gauge-invariant combination~\cite{Ranzani2015}. The phase is the $\mathcal{T}$-odd bias and the dissipation is what converts it into an amplitude asymmetry. At $\Gamma=2J$ and $\phi-\theta=\pi/2$ the element $(H_{\mathrm{eff}})_{12}$ vanishes identically, extinguishing transfer from mode~2 to mode~1~\cite{Metelmann2015}.
 
\begin{figure*}[t!]
\centering
\includegraphics[width=170mm]{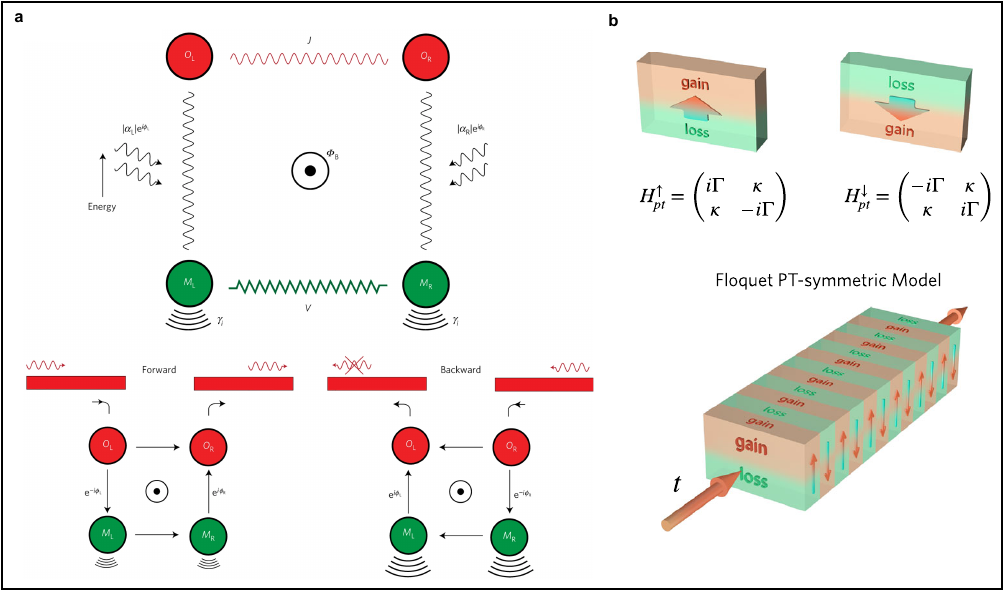}
\caption{\textbf{Non-Hermitian platforms in which an external bias is converted
into port-level contrast.}
\textbf{a} Optomechanical reservoir engineering. Two optical modes
($O_{\mathrm{L}}$, $O_{\mathrm{R}}$) are coupled directly at rate $J$ and
indirectly through two mechanical modes ($M_{\mathrm{L}}$, $M_{\mathrm{R}}$,
coupled at rate $V$ and damped at rate $\gamma_{i}$) by pump-enhanced
optomechanical interactions $|\alpha_{\mathrm{L,R}}|e^{i\phi_{\mathrm{L,R}}}$.
The pump phases impose a synthetic flux $\Phi_{\mathrm{B}}$ on the loop, so the
indirect path acquires $+(\phi_{\mathrm{R}}-\phi_{\mathrm{L}})$ in the forward
direction and $-(\phi_{\mathrm{R}}-\phi_{\mathrm{L}})$ in the backward
direction (lower panels). Mechanical damping makes the indirect path
dissipative, converting this phase asymmetry into an amplitude asymmetry:
forward transfer is allowed and backward transfer is cancelled (crossed arrow),
yielding $35\,\mathrm{dB}$ of isolation. Adapted with permission from
ref.~\cite{Fang2017}, Springer Nature.
\textbf{b} Floquet $\mathcal{PT}$ symmetry in an integrated waveguide pair.
Periodically flipping the gain/loss distribution alternates the two-mode
Hamiltonians $H_{pt}^{\uparrow}$ and $H_{pt}^{\downarrow}=(H_{pt}^{\uparrow})^{*}$,
with $\Gamma$ the gain/loss rate and $\kappa$ the inter-waveguide coupling;
both are symmetric, $H_{pt}=H_{pt}^{T}$, hence reciprocal. In the experiment
the modulation coordinate labelled $t$ is the propagation distance along the
waveguide, which emulates time; the structure is static and Lorentz reciprocal,
and the reported port dependence refers to which waveguide is excited.
Adapted from ref.~\cite{Liu2024} under CC~BY~4.0.}
\label{fig:nonhermitian_representative_platforms}
\end{figure*}

This mechanism was implemented in an optomechanical circuit by Fang \textit{et al.}~\cite{Fang2017}. Two optical cavities were coupled both directly and through a pair of mechanical modes, and phase-correlated optical pumps imposed a synthetic flux $\Phi_{\mathrm{B}}$ on the resulting loop (Fig.~\ref{fig:nonhermitian_representative_platforms}a). Mechanical damping made the indirect path non-conservative, supplying the $\Gamma$ of Eq.~(\ref{Eq:NH_directionality}); without it the synthetic phase alone yields a nonreciprocal phase and reciprocal transmission as in the photonic Aharonov--Bohm effect~\cite{Fang2012}. The experiment achieved $35\,\mathrm{dB}$ of optical isolation, with $12\,\mathrm{dB}$ of directional amplification in the forward direction under blue-detuned pumping~\cite{Fang2017}. 
 
A related strategy places loss selectively in multi-mode or multi-path structures~\cite{Huang2021,Huang2023}. Uniform absorption cannot break reciprocity~\cite{Jalas2013,Caloz2018}, but non-uniform loss reshapes the interference between modal pathways and can make the effective coupling between \emph{internal} modes direction dependent, in the sense of Eq.~(\ref{Eq:NH_Heff})~\cite{Huang2021,Huang2023}. The distinction that matters for isolation is that a passive, linear, time-invariant structure still satisfies $\hat{\mathbf{S}}=\hat{\mathbf{S}}^{T}$ at its ports~\cite{Jalas2013,Caloz2018,Asadchy2020}; what loss engineering delivers is mode selective contrast and suppression of a rejected internal channel, which becomes port-level isolation only when combined with a genuine bias.

A third route modulates gain and loss in time. Song \textit{et al.} analysed a waveguide carrying a traveling-wave modulation of $\mathrm{Im}\,\varepsilon$~\cite{Song2019}. Because the modulation carries both a frequency and a momentum, $(\Omega,\bm{q})$, it is phase matched to forward modes and mismatched to backward ones~\cite{Sounas2017Non-reciprocal}, and the effective Floquet Hamiltonian becomes direction dependent~\cite{Song2019,koutserimpas2018nonreciprocal}. The forward $\mathcal{PT}$ transition is then thresholdless while the backward one retains a finite threshold, giving direction-dependent amplification that can be converted into isolation by filtering or dissipating the unwanted channel~\cite{Song2019}. Here the medium is genuinely time varying, so $(\Omega,\bm{q})$ is a true $\mathcal{T}$-odd bias.

%The contrast with the on-chip Floquet $\mathcal{PT}$ platform of Liu \textit{et al.} is instructive~\cite{Liu2024}. In that experiment the gain/loss distribution is flipped periodically along the waveguide, with the propagation coordinate playing the role of time (Fig.~\ref{fig:nonhermitian_representative_platforms}b)~\cite{Liu2024}. Spontaneous $\mathcal{PT}$ breaking was observed at very small gain/loss levels, and amplification or suppression could be selected through the excitation port --- that is, through \emph{which waveguide} is excited~\cite{Liu2024}. The structure itself is static, so it is Lorentz reciprocal~\cite{Jalas2013,Caloz2018}. Floquet engineering therefore relaxes the gain required to reach the broken phase; it does not break reciprocity. %Isolation requires the modulation to run in real time, as in Ref.~\cite{Song2019}.

\subsection{EP-assisted nonreciprocity}
\label{Subsec:EP_assisted}
 
EPs are non-Hermitian degeneracies at which two or more eigenvalues \emph{and} their eigenvectors coalesce, so that the effective operator becomes defective~\cite{Kato1966Perturbation,Heiss2012,Ozdemir2019,miri2019exceptional}. Near an EP of order $N$, a generic perturbation $\epsilon$ splits the eigenvalues as $\epsilon^{1/N}$~\cite{Kato1966Perturbation,Heiss2012}, so a second-order EP responds to a small detuning, loss imbalance or coupling change with a square-root enhancement~\cite{ding2022non,Hodaei2017}. Applied to isolation, this is a gain in \emph{sensitivity}, not in symmetry: the enhancement is useful only when the forward and backward waves already see different parameters, and that difference must be supplied by a bias. Three routes have been proposed.

\begin{figure*}[t!]
\centering
\includegraphics[width=170mm]{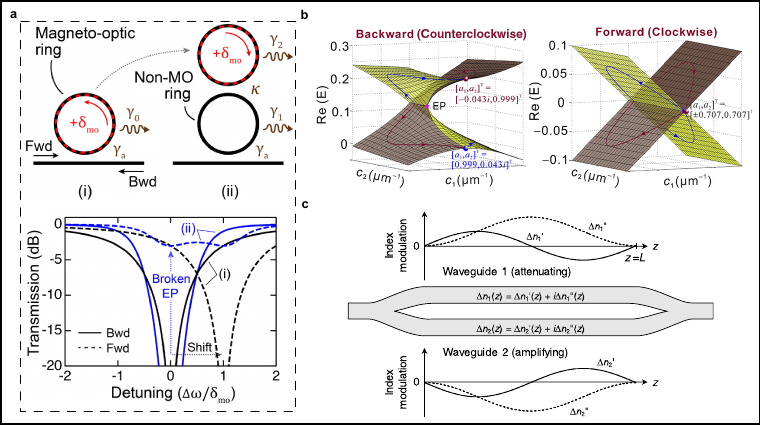}
\caption{\textbf{Exceptional-point-assisted routes to nonreciprocity. All
panels show calculated results.}
\textbf{a} EP-enhanced MO ring isolator. A single MO ring (i), whose resonance
is split by the MO shift $\delta_{\mathrm{mo}}$, is compared with an MO ring
coupled at rate $\kappa$ to a passive ring (ii); $\gamma_{0,1,2}$ are the
intrinsic loss rates and $\gamma_{a}$ the bus coupling rate. Calculated
forward (dashed) and backward (solid) transmission versus normalized detuning
$\Delta\omega/\delta_{\mathrm{mo}}$ shows that operation at the EP broadens the
isolation band, which scales as $\sqrt{\delta_{\mathrm{mo}}}$ rather than
linearly.
\textbf{b} Direction-dependent EP encircling in a complex-modulated waveguide.
Real parts of the two quasi-energy branches over the modulation-amplitude plane
$(c_{1},c_{2})$ for backward (left) and forward (right) propagation, which
satisfy different phase-matching conditions. The parameter loop (red and blue
trajectories) encloses the EP only in the backward case; the coalescing
eigenvectors $[a_{1},a_{2}]^{T}$ are indicated. The resulting asymmetry is in
even/odd \emph{mode conversion}, not in port transmission of a fixed mode.
\textbf{c} Nonlinear EP encircling in coupled waveguides. The complex index
modulations $\Delta n_{1,2}(z)=\Delta n'_{1,2}(z)+i\Delta n''_{1,2}(z)$ make
one arm attenuating and the other amplifying, tracing a loop around an EP along
$z$. \textbf{a}  Reproduced with permission from ref.~\cite{Grede2021} under CC~BY~4.0. \textbf{b}  Reproduced with permission from ref.~\cite{liu2023nonreciprocal},
\textcopyright{} Optica Publishing Group. \textbf{c} Reproduced with permission from ref.~\cite{choi2017extremely} under CC~BY~4.0.}
\label{fig:exceptional_points}
\end{figure*}

The first combines an EP with MO nonreciprocity~\cite{Grede2021}. In resonant integrated isolators the NRPS is small while the MO section adds absorption, which sets a tradeoff between isolation bandwidth and insertion loss~\cite{Li2024}. Grede \textit{et al.} showed that coupled-ring isolators operated at a second-order EP relax this tradeoff, because the isolation bandwidth then scales as the square root of the NRPS rather than linearly (Fig.~\ref{fig:exceptional_points}a)~\cite{Grede2021,Ruan2025}. The same square-root response, however, also amplifies detuning, fabrication disorder and noise, and the excess noise (Petermann) factor diverges at the EP, so the enhanced bandwidth is not obtained without a cost ~\cite{Wiersig2020,LauClerk2018,Wang2020Petermann}. Related work reports large transmission contrast near an EP in a gain--loss structure containing magnetic layers~\cite{thomas2016giant}. A distinct MO route replaces the NRPS with circular dichroism: because absorption then differs for the two circulations, the $\mathcal{PT}$ threshold itself becomes direction dependent in a coupled-ring geometry~\cite{kawaguchi2022non}.

The second route uses spatiotemporal modulation. The modulation supplies
frequency and momentum shifts, so counter-propagating waves satisfy different
phase-matching conditions~\cite{Sounas2017Non-reciprocal}, and the effective Floquet
Hamiltonian can host an EP for one direction
only~\cite{li2022nonreciprocal,koutserimpas2018nonreciprocal}. A closed loop in
parameter space can then encircle the EP in one direction and avoid it in the
other (Fig.~\ref{fig:exceptional_points}b), making the resulting state transfer
direction dependent~\cite{liu2023nonreciprocal}. Two qualifications are
essential. The predicted asymmetry is between \emph{modes} --- forward
even-to-odd conversion exceeding backward odd-to-even conversion --- rather
than between ports for a fixed mode~\cite{liu2023nonreciprocal}. An EP
encircling without a $\mathcal{T}$-odd bias is reciprocal: dynamically
encircling an EP along a static waveguide produces asymmetric mode
switching~\cite{Doppler2016} in a structure whose linear response remains
reciprocal~\cite{choi2017extremely,Khurgin2021}.

The third route relies on nonlinearity, which makes the effective detuning,
gain or loss depend on intensity~\cite{Chang2014}. When forward and backward
inputs build different intracavity fields, the system approaches the EP
differently in the two directions. In coupled whispering-gallery resonators the linear response is reciprocal, and direction-dependent transmission appears only once $\mathcal{PT}$-broken field localization enhances the nonlinear
response~\cite{peng2014parity}. Choi \textit{et al.} proposed the broadband
version of this idea: encircling an EP in coupled waveguides is reciprocal in the linear regime, but under gain saturation the same structure is predicted to
give more than $10\,\mathrm{dB}$ of contrast with near-unity forward
transmission over a bandwidth approaching $100\,\mathrm{THz}$
(Fig.~\ref{fig:exceptional_points}c)~\cite{choi2017extremely}. Related proposals use nonlinear EP encircling, nonlinear dissipation, or exceptional degeneracies
of Bloch modes~\cite{laha2020exceptional,li2021nonlinear,landers2025robust}. All
inherit the constraint of dynamic reciprocity discussed in
Section~\ref{Sec:Nonlinear_Systems}.

\section{Fundamental limits} \label{Sec:Fundamental_limits}
TR symmetry imposes strict limits on AT. Sounas and Alù showed that in any TR invariant asymmetric structures the ratio of internal field intensities for opposite excitations is bounded by the port reflection coefficient $R$~\cite{Sounas2017Time-reversal,ma2025time}
\begin{equation}
\frac{1 - R}{1 + R} \;\le\; \frac{|E_2|^2}{|E_1|^2} \;\le\; \frac{1 + R}{1 - R}.
\label{Eq:fundamental_limit}
\end{equation}
In the case of full transmission  $R=0$ (impedance-matched ports), the ratio approaches unity, implying that the field distributions under forward and backward excitation become identical throughout the structure. This result highlights that any significant internal field asymmetry necessarily relies on the presence of reflections or impedance mismatch. Equivalent bounds have been derived in different formalisms, including temporal coupled-mode theory, where they constrain the ratio of decay rates and modal amplitudes in resonant systems~\cite{Wang2024}. 

Nonlinear Fano-resonator-based isolators obey a fundamental trade-off between forward transmission and nonreciprocity: unitary transmission can only be achieved in the limit of vanishing nonreciprocity~\cite{Sounas2018Fundamental}. More broadly, symmetry considerations impose additional restrictions on the observable asymmetries. For instance, the extinction symmetry of reciprocal scatterers, i.e., the invariance of total extinction under reversal of the excitation direction, cannot be lifted unless all relevant symmetries are simultaneously broken. Even in systems lacking geometric or TR symmetry, a combination of nonreciprocal response, spatial asymmetry, and appropriate polarization conditions is required to achieve different extinction for opposite incidences~\cite{Yla2025}.

Geometric operations such as inversion, reflection, or rotation cannot by themselves break reciprocity. If  a $\mathcal{T}$-odd bias is absent every part of $\hat{\bm{\varepsilon}}$ is even and hence symmetric at every point. So the asymmetric transmission a geometrically asymmetric structure can display is reciprocal~\cite{Jalas2013,Caloz2018}. What broken spatial symmetry does control is which magnetoelectric couplings are allowed. Bianisotropy of any kind requires a non-centrosymmetric medium, and the reciprocal chiral coupling responsible for natural optical activity requires a point group containing no improper operation at all~\cite{Barron2004,Asadchy2020}. Natural optical activity and Faraday rotation are the canonical contrast. The former is governed by a $\mathcal{P}$-odd, $\mathcal{T}$-even pseudoscalar, while the latter by the $\mathcal{P}$-even, $\mathcal{T}$-odd gyration vector $\bm{g}$ of Eq.~(\ref{Eq:NH_gyration}). Being $\mathcal{T}$-even, the first cancels on a round trip \cite{Barron2004}. Being $\mathcal{T}$-odd, the second accumulates~\cite{Landau1984}.

\section{Conclusion and outlook}\label{Sec:Conclusions}
In this review, we have examined on-chip nonreciprocity and optical isolation from the perspective of time-reversal-odd physics and the symmetry of the constitutive tensors. In linear time-invariant systems, nonreciprocity requires biasing the structure with a physical quantity that is odd under time reversal. A nonzero ferromagnetic order (magnetization) makes the permittivity tensor antisymmetric, while ferrotoroidal order produces a nonreciprocal (Tellegen-type) magnetoelectric coupling; either condition is sufficient for a nonreciprocal response in local media. In linear time-varying media, momentum bias, angular momentum bias, synthetic gauge fields or direction-dependent attenuation mechanisms lead to symmetry breaking between counterpropagating modes and nonreciprocity.

Monolithic garnet films on silicon nitride and laser-written waveguides in latched garnet now provide broadband, fully passive isolation (Table~\ref{tab:MOisolators}), albeit requiring materials and processing steps beyond those of standard foundry platforms~\cite{Yan2024,Lapointe2025}. Space--time modulation removes the need for magnetic materials and has achieved one of the highest experimentally realized isolation ratios reported on chip (Table~\ref{tab:space_time_modulation})~\cite{Yu2023}, at the cost of a continuous microwave or acoustic drive~\cite{Yu2023,Han2026,Kim2026}, a millimetre- to centimetre-scale footprint, and the associated electrical-power and thermal budgets~\cite{Yu2023,Kittlaus2021}. Nonlinear approaches can offer compact implementations and have been hybrid-integrated with semiconductor lasers~\cite{White2023}, but their operation can require substantial optical power or an external pump, depending on the underlying mechanism. Moreover, nonlinear diodes generally exhibit asymmetric transmission only over restricted excitation conditions and power ranges~\cite{Fan2013Silicon}. Topological and non-Hermitian systems do not break Lorentz reciprocity by themselves; their value lies in providing unidirectional waveguide transport or in strengthening an asymmetry supplied elsewhere~\cite{Grede2021,Ruan2025}.

Several electromagnetic degrees of freedom remain essentially unused in integrated isolation and define concrete directions for future work. Most magneto-optical isolators to date break reciprocity through the permittivity tensor alone. First, magneto-plasmonic metamaterials can engage the permeability tensor as well~\cite{Dmitriev2013}, and optical gyromagnetic responses have now been measured in such systems~\cite{Yang2022}; their translation into integrated isolators, however, remains largely unexplored.

Second, nonlinear magneto-optics offers a largely unexplored route to optical isolation that can overcome some limitations of conventional nonlinear approaches. In nonlinear dielectric schemes (Section~\ref{Sec:Nonlinear_Systems}), phase matching can constrain the attainable bandwidth~\cite{xu2023breaking,Tsakmakidis2017}. The induced third-order permittivity change [Eq.~(\ref{Eq:third_order_permittivity})] is symmetric and even under time reversal, so the response remains reciprocal under simultaneous bidirectional excitation~\cite{Shi2015}. A further practical limitation is the separation of the signal from the pump and idler (Fig.~\ref{Fig:Nonlinear_System_Fig_2}c)~\cite{Hua2016}. By contrast, nonlinear magneto-optics introduces magnetization as an additional time-reversal-odd degree of freedom. Expanding the permittivity of a magnetically ordered medium in the magnetization gives 
$\varepsilon_{ij}=\varepsilon^{(0)}_{ij}
+\mathcal{K}_{ijk}M_k
+\mathcal{G}_{ijkl}M_kM_l
+\mathcal{H}_{ijklm}M_kM_lM_m+\cdots$
~\cite{Silber2026}. Consistent with the Onsager relation $\varepsilon_{ij}(\mathbf{M})=\varepsilon_{ji}(-\mathbf{M})$, terms odd in $\mathbf{M}$ contribute to the time-reversal-odd response, whereas even-order terms are time-reversal even~\cite{Visnovsky1986,Hamrlova2016}. The linear tensor $\mathcal{K}_{ijk}$ describes the conventional magneto-optic response, including Faraday rotation and nonreciprocal phase shifts, whereas the quadratic tensor $\mathcal{G}_{ijkl}$ contributes to reciprocal Cotton--Mouton birefringence. The cubic tensor $\mathcal{H}_{ijklm}$ therefore provides a higher-order magnetization-dependent nonreciprocal channel beyond the conventional linear magneto-optic response. This contribution has recently been identified in ferromagnetic thin films, with an angular dependence distinct from that of the linear effect, but has yet to be exploited in an integrated waveguide~\cite{Silber2026}. Nonlinear magneto-optical responses can also remain pronounced in ultrathin films and monolayers, where they can exceed their linear counterparts, making them attractive for two-dimensional magnetic structures~\cite{matsubara2012giant,hubner1995electronic}.

Third, magneto-electric (Tellegen) coupling introduces a distinct constitutive channel for nonreciprocal response that remains unexploited in integrated isolation. The magneto-electric couplings can be expanded to second order in the bias fields as $\xi_{ij}=\alpha_{ij}+\beta_{ijk}H_k+\gamma_{ikj}E_k+\cdots$ and $\zeta_{ij}=\alpha_{ji}+\beta_{kij}H_k+\gamma_{ijk}E_k+\cdots$~\cite{de2022aspects,de2025light}, where $\alpha_{ij}$ is the linear magneto-electric coefficient and $\beta_{ijk}$, $\gamma_{ikj}$ the second-order coefficients coupling to the magnetic and electric fields. Nonreciprocity resides in the Tellegen response when $\hat{\mathsf{\xi}}=+\hat{\zeta}^{\mathsf{T}}$. Metamaterials with spontaneous magnetization exhibit effective Tellegen parameters orders of magnitude beyond natural values and it have been measured at microwave frequencies~\cite{Safaei2024,yang2025gigantic}. The optical Tellegen effect has been recently demonstrated in a resonant metasurface at $830$~nm, reaching an effective parameter two order of magnitudes higher thant any natural material from spontaneous magnetization alone~\cite{SafaeiJazi2026}, although translation into waveguide geometries remains open. Nonlinear magneto-electric coupling has been predicted to make Faraday rotation electrically tunable~\cite{Chatterjee2026}. These developments position magneto-electric coupling as a promising route towards electrically tunable nonreciprocity. 

Fourth, a two-dimensional van der Waals multiferroic integrated with silicon microresonators has demonstrated room-temperature nonreciprocal operation~\cite{Dushaq2025}, offering a route that relaxes the lattice-matching and thermal-processing constraints associated with conventional integration of magneto-optic garnets. Extending such resonator-level demonstrations to fully integrated optical isolators remains an important next step. Finally, self-induced nonreciprocity, in which circularly polarized light generates a local spin polarization that acts as a time-reversal-odd bias, has enabled isolation exceeding 60~dB in atomic vapour, including in the presence of simultaneous counterpropagating excitation~\cite{Wang2025}. Translating this mechanism to an integrated platform remains an open challenge.

These mechanisms are increasingly being combined in hybrid architectures rather than pursued in isolation. We therefore expect future integrated isolators to combine linear and nonlinear magneto-optical and magneto-electric responses with topological and non-Hermitian design. The central challenge is therefore not merely to break reciprocity, but to translate that asymmetry into low-loss, broadband and fabrication-tolerant isolation on chip. In the near term, passive garnet devices based on remanent magnetization~\cite{Lapointe2025} and electro-optic isolators based on thin-film lithium niobate~\cite{Yu2023} offer particularly promising routes towards integration with on-chip lasers. If the emerging material platforms discussed above mature, on-chip isolation could become a standard library component of photonic integrated circuits, alleviating a long-standing constraint on integrated lasers, scalable optical interconnects, photonic processors and quantum photonic technologies.

\backmatter
\bmhead{Methodological note}
This review is based on a selective survey of the literature published primarily between 2010 and 2026, with emphasis on experimentally demonstrated devices and influential theoretical proposals. The tables summarize representative state-of-the-art metrics but are not exhaustive. Performance comparisons may differ due to measurement conditions, material platforms, and reporting standards.
\bmhead{Data Availability}
All the data needed to evaluate the conclusions of this work are present in the main text.

\bmhead{Acknowledgements}
This work is supported by the Russian Science Foundation (project \#23-72-10059-$\Pi$) . A.B. acknowledges the National Natural Science Foundation of China (Grant No. W2532010)  and the Academic leadership program Priority 2030. 

\section*{Declarations}

\bibliography{sn-bibliography}

\end{document}